\documentclass[journal]{IEEEtran}
\usepackage{graphicx}
\graphicspath{ {./img/} }
\DeclareGraphicsExtensions{.pdf,.png}

\usepackage{amsmath, amsfonts, amsthm, amssymb, mathtools, cases}
\usepackage{array}
\usepackage{subfigure}
\usepackage{textcomp}
\usepackage{stfloats}
\usepackage{url}
\usepackage{verbatim}
\usepackage{siunitx}

\usepackage[%
colorinlistoftodos,%
prependcaption,%
]{todonotes}
\setuptodonotes{inline}

\usepackage[%
style = ieee,%
citestyle = numeric-comp,%
bibencoding = utf8,%
datamodel = software,%
backend = biber,%
giveninits = true,%
abbreviate = true,
maxbibnames = 8,%
maxcitenames = 2,%
sortcites = true,%
doi = false,%
isbn = false,%
url = false,%
eprint = false,%
]{biblatex}

\usepackage{pgfplots}
\pgfplotsset{compat=newest}
\usepgflibrary{plotmarks}
\usetikzlibrary{
  tikzmark,
  calc,
  arrows,
  decorations.pathmorphing,
  shapes,
  matrix
}

\usepackage{algorithm, algpseudocode}
\makeatletter
\newcommand{\algmargin}{\the\ALG@thistlm}
\makeatother
\newlength{\whilewidth}
\usepackage{graphicx}
\graphicspath{ {./figs/} }
\DeclareGraphicsExtensions{.png,.pdf}
\usepackage{xcolor}
\usepackage{url}

\definecolor{cadetblue}{rgb}{0.37, 0.62, 0.63}
\definecolor{burntorange}{rgb}{0.8, 0.33, 0.0}
\definecolor{americanrose}{rgb}{1.0, 0.01, 0.24}
\definecolor{applegreen}{rgb}{0.55, 0.71, 0.0}
\definecolor{darkmagenta}{rgb}{0.55, 0.0, 0.55}
\definecolor{peru}{rgb}{0.80, 0.52, 0.25}
\definecolor{navy}{rgb}{0.0, 0.0, 0.5}
\definecolor{maroon}{rgb}{0.5, 0.0, 0.0}
\definecolor{gold}{rgb}{1.0, 0.84, 0.0}
\definecolor{crimson}{rgb}{0.86, 0.08, 0.24}

\usepackage{newfloat}

\usepackage[inline]{enumitem}
\usepackage{etoolbox}

\newcommand{\IntegerP}{\mathbb{N}}
\newcommand{\IntegerPP}{\mathbb{N}_*}
\newcommand{\Real}{\mathbb{R}}
\newcommand{\RealP}{\mathbb{R}_{+}}
\newcommand{\RealPP}{\mathbb{R}_{++}}

\newcommand\given{{\mathbin{}\mid\mathbin{}}}
\newcommand\vect[1]{\mathbf{#1}}
\newcommand\vectgr[1]{\boldsymbol{#1}}

\providecommand\given{} 
\newcommand\SetSymbol[1][]{
  \nonscript\,#1\vert \allowbreak \nonscript\,\mathopen{}}
\DeclarePairedDelimiterX\Set[1]{\lbrace}{\rbrace}%
{ \renewcommand\given{\SetSymbol[\delimsize]} #1 }
\DeclarePairedDelimiterX\innerp[2]{\langle}{\rangle}{#1
  \mathop{}\delimsize\vert\mathop{} #2}
\DeclarePairedDelimiterX\norm[1]\lVert\rVert{\ifblank{#1}{\:\cdot\:}{#1}}

\DeclareMathOperator{\trace}{trace}

\DeclareMathOperator{\sign}{sgn}

\DeclareMathOperator{\Fix}{Fix}

\DeclareMathOperator{\expect}{\mathbb{E}}
\DeclareMathOperator{\Prob}{\mathbb{P}}
\DeclareMathOperator*{\Plim}{\mathbb{P}-lim}

\DeclareMathOperator{\prox}{Prox}

\DeclareMathOperator{\Id}{Id}

\DeclareMathOperator{\dist}{dist}

\DeclareMathOperator{\vecinf}{\textbf{inf}}

\DeclareMathOperator{\argmin}{argmin}

\usepackage{stackengine}
\stackMath

\usepackage{thmtools}
\usepackage[unq]{unique}

\declaretheoremstyle[%
headfont=\normalfont\bfseries,
notefont=\mdseries,
notebraces={(}{)},
bodyfont=\normalfont,
postheadspace=1ex
]{mystyle}

\declaretheorem[style=mystyle,
name=Theorem,
refname={theorem,theorems},
Refname={Theorem,Theorems},
]{thm}

\declaretheorem[style=mystyle,
name=Lemma,
refname={lemma,lemmata},
Refname={Lemma,Lemmata},
numberlike=thm,
]{lemma}

\declaretheorem[style=mystyle,
name=Assumptions,
refname={assumptions},
Refname={Assumptions},
numberlike=thm,
]{assumptions}

\declaretheorem[style=mystyle,
name=Proposition,
refname={proposition,propositions},
Refname={proposition,propositions},
numberlike=thm,
]{prop}

\newlist{thmlist}{enumerate}{1}
\setlist[thmlist]{label=\textbf{(\roman{*})}, ref=\thethm(\roman{*}), noitemsep}

\newlist{lemlist}{enumerate}{1}
\setlist[lemlist]{label=\textbf{(\roman{*})}, ref=\thelemma(\roman{*}), noitemsep}

\newlist{exlist}{enumerate}{1}
\setlist[exlist]{label=\textbf{(\roman{*})}, ref=\theexample(\roman{*}), noitemsep}

\newlist{factlist}{enumerate}{1}
\setlist[factlist]{label=\textbf{(\roman{*})}, ref=\thefact(\roman{*}), noitemsep}

\newlist{proplist}{enumerate}{1}
\setlist[proplist]{label=\textbf{(\roman{*})}, ref=\theprop(\roman{*}), noitemsep}

\newlist{asslist}{enumerate}{1}
\setlist[asslist]{label=\textbf{(\roman{*})}, ref=\theassumption(\roman{*}), noitemsep}

\newlist{assslist}{enumerate}{1}
\setlist[assslist]{label=\textbf{(\roman{*})}, ref=\theassumptions(\roman{*}), noitemsep}

\newlist{deflist}{enumerate}{1}
\setlist[deflist]{label=\textbf{(\roman{*})}, ref=\thedefinition(\roman{*}), noitemsep}

\newlist{algolist}{enumerate}{1}
\setlist[algolist]{label=\textbf{(\roman{*})}, ref=\thealgo(\roman{*}), noitemsep}

\newlist{claimlist}{enumerate}{1}
\setlist[claimlist]{label=\textbf{(\roman{*})}, ref=\theclaim(\roman{*}), noitemsep}

\newlist{applist}{enumerate}{1}
\setlist[applist]{label=\textbf{(\roman{*})},
  ref=\thesection(\roman{*}), noitemsep}

\newlist{contriblist}{enumerate}{1}
\setlist[contriblist]{label=\textbf{(C\arabic{*})},
  ref=(C\arabic{*}), noitemsep}

\newlist{MyEnumSec}{enumerate}{1}
\setlist[MyEnumSec]{label=\textbf{\thesection(\roman{*})},
  ref=Item~\thesection(\roman{*}), noitemsep}

\newlist{MyEnumSubSec}{enumerate}{1}
\setlist[MyEnumSubSec]{label=\textbf{\thesubsection(\roman{*})},
  ref=Item~\thesubsection(\roman{*}), noitemsep, wide = 0pt, leftmargin = *}

\usepackage[capitalize]{cleveref}

\crefname{thm}{Theorem}{Theorems}
\crefname{prop}{Proposition}{Propositions}
\crefname{assumption}{Assumption}{Assumptions}
\crefname{assumptions}{Assumptions}{Assumptions}
\crefname{lemma}{Lemma}{Lemmata}
\crefname{definition}{Definition}{Definitions}
\crefname{example}{Example}{Examples}
\crefname{algo}{Algorithm}{Algorithms}
\crefname{fact}{Fact}{Facts}
\crefname{claim}{Claim}{Claims}
\crefname{appendix}{Appendix}{Appendices}
\crefname{coroll}{Corollary}{Corollaries}
\crefname{figure}{Figure}{Figures}
\crefname{section}{Section}{Sections}

\crefname{thmlisti}{Theorem}{Theorems}
\crefname{lemlisti}{Lemma}{Lemmata}
\crefname{proplisti}{Proposition}{Propositions}
\crefname{asslisti}{Assumption}{Assumptions}
\crefname{assslisti}{Assumption}{Assumptions}
\crefname{deflisti}{Definition}{Definitions}
\crefname{exlisti}{Example}{Examples}
\crefname{algolisti}{Algorithm}{Algorithms}
\crefname{factlisti}{Fact}{Facts}
\crefname{claimlisti}{Claim}{Claims}
\crefname{applisti}{Appendix}{Appendices}
\crefname{MyEnumSeci}{}{}
\crefname{MyEnumSubSeci}{}{}

\makeatletter

\newcommand*{\ie}{%
  \@ifnextchar{,}%
  {\textit{i.e.}}%
  {\textit{i.e.,}\@\xspace}%
}
\newcommand*{\eg}{%
  \@ifnextchar{,}%
  {\textit{e.g.}}%
  {\textit{e.g.,}\@\xspace}%
}
\newcommand*{\etc}{%
  \@ifnextchar{.}%
  {\textit{etc}}%
  {\textit{etc.}\@\xspace}%
}
\newcommand*{\etal}{%
  \@ifnextchar{.}%
  {\textit{et al}}%
  {\textit{et al.}\@\xspace}%
}
\newcommand*{\cf}{%
  \@ifnextchar{.}%
  {\textit{cf}}%
  {\textit{cf.}\@\xspace}%
}
\newcommand*{\aka}{%
  \@ifnextchar{,}%
  {\textit{a.k.a.}}%
  {\textit{a.k.a.}\@\xspace}%
}
\makeatother

\makeatletter
\newlength{\negph@wd}
\DeclareRobustCommand{\negphantom}[1]{%
  \ifmmode
    \mathpalette\negph@math{#1}%
  \else
    \negph@do{#1}%
  \fi
}
\newcommand{\negph@math}[2]{\negph@do{$\m@th#1#2$}}
\newcommand{\negph@do}[1]{%
  \settowidth{\negph@wd}{#1}%
  \hspace*{-\negph@wd}%
}
\makeatother

\allowdisplaybreaks

\DeclareUnicodeCharacter{0301}{\'{e}}

\title{Nonparametric Bellman Mappings\\ for Reinforcement
  Learning:\\ Application to Robust Adaptive Filtering}

\author{%
  Yuki~Akiyama, Minh~Vu, and Konstantinos~Slavakis%
  \IEEEauthorrefmark{1}%
  \thanks{\IEEEauthorrefmark{1}Y.~Akiyama, M.~Vu, and
    K.~Slavakis are with Tokyo Institute of Technology,
    Department of Information and Communications
    Engineering, 4259-G2-4 Nagatsuta-Cho, Midori-Ku,
    Yokohama, Kanagawa, 226-8502 Japan. Emails:
    \texttt{\{akiyama.y.am, vu.d.aa,
      slavakis.k.aa\}@m.titech.ac.jp}.%
  }%
}

\begin{document}
\sloppy
\maketitle

\begin{abstract}
  This paper designs novel nonparametric Bellman mappings in
  reproducing kernel Hilbert spaces (RKHSs) for
  reinforcement learning (RL). The proposed mappings benefit
  from the rich approximating properties of RKHSs, adopt no
  assumptions on the statistics of the data owing to their
  nonparametric nature, require no knowledge on transition
  probabilities of Markov decision processes, and may
  operate without any training data. Moreover, they allow
  for sampling on-the-fly via the design of trajectory
  samples, re-use past test data via experience replay,
  effect dimensionality reduction by random Fourier
  features, and enable computationally lightweight
  operations to fit into efficient online or time-adaptive
  learning. The paper offers also a variational framework to
  design the free parameters of the proposed Bellman
  mappings, and shows that appropriate choices of those
  parameters yield several popular Bellman-mapping
  designs. As an application, the proposed mappings are
  employed to offer a novel solution to the problem of
  countering outliers in adaptive filtering. More
  specifically, with no prior information on the statistics
  of the outliers and no training data, a policy-iteration
  algorithm is introduced to select online, per time
  instance, the ``optimal'' coefficient $p$ in the
  least-mean-$p$-power-error method. Numerical tests on
  synthetic data showcase, in most of the cases, the
  superior performance of the proposed solution over several
  RL and non-RL schemes.
\end{abstract}


\section{Introduction}\label{sec:intro}

\subsection{Motivation: Adaptive filters against outliers}%
\label{sec:RobAdaFilt}

The least-squares (LS) error/loss plays a pivotal role in
signal processing, e.g., adaptive filtering
(AdaFilt)~\cite{sayed2011adaptive}, and machine
learning~\cite{theodoridis.book:ml}. Notwithstanding, the LS
loss is notoriously sensitive to the presence of
outliers~\cite{rousseeuw1987}, where outliers are defined as
contaminating data that do not adhere to a nominal
data-generation model, and are often viewed as random
variables (RVs) with non-Gaussian heavy tailed
distributions, \eg, $\alpha$-stable
ones~\cite{shao1993signal, miotto2016pylevy}. To counter
outliers in AdaFilt, non-LS losses, such as the $p$-norm
($2 > p \in \RealPP$)~\cite{pei1994p-power,
  chambers1997robust, xiao1999adaptive, kuruoglu:02,
  gentile:03, vazquez2012, chen2015smoothed,
  slavakis2021outlier} and correntropy~\cite{singh.mcc:09,
  huang2017adaptive} have been studied (henceforth,
$\RealPP$ will denote the set of all positive real numbers).

Consider the classical linear data-generation model in
AdaFilt:
$y_n = \vectgr{\theta}_*^{\intercal} \vect{x}_n + o_n$,
where $n\in\IntegerP$ denotes discrete time ($\IntegerP$ is
the set of all non-negative integers),
$\vectgr{\theta}_* \in \Real^{L}$ is the $L\times 1$
vector/system with real-valued entries that needs to be
estimated (estimandum), $o_n$ is the real-valued RV which
models outliers/noise, $(\vect{x}_n, y_n)$ stands for the
input-output pair of available data, where
$\vect{x}_n \in \Real^{L}$ and $y_n \in \Real$, and
$\intercal$ denotes vector/matrix transposition. The
online-learning setting is considered, that is, data
$(\vect{x}_n, y_n)_{n \in \IntegerP}$ appear to the
user/agent in a streaming fashion, a pair
$(\vect{x}_n, y_n)$ per time index $n$, while no training
data are available. All operations in the following
discussion are performed online, so that the time index $n$
coincides with the iteration index of the proposed
reinforcement-learning (RL) algorithm.

The least-mean-$p$-power-error (LMP)
method~\cite{pei1994p-power} counters outliers by applying
the classical stochastic-gradient-descent (SGD) iteration to
the $p$-power error/loss
$\lvert y_n - \vect{x}_n^{\intercal} \vectgr{\theta}
\rvert^p$, $p \in [1,2]$, to generate the following sequence
of estimates $(\vectgr{\theta}_n)_{n\in\mathbb{N}}$ of the
estimandum $\vectgr{\theta}_*$: for an arbitrarily fixed
$\vectgr{\theta}_0 \in \Real^L$,
\begin{align}
  \vectgr{\theta}_{n+1} \coloneqq \vectgr{\theta}_n + \rho p
  \lvert e_n \rvert^{p-1} \sign(e_n)\, \vect{x}_n
  \,, \label{LMP}
\end{align}
where
$e_n \coloneqq y_n - \vect{x}_n^{\intercal}
\vectgr{\theta}_n = \vect{x}_n^{\intercal} (
\vectgr{\theta}_* - \vectgr{\theta}_n ) + o_n$ is the
classical a-priori error~\cite{sayed2011adaptive}, $\rho$ is
the learning rate (step size), and
$\sign(\cdot) \colon \Real \to \{ \pm 1\}$ provides the sign
of a real number. The $p$-power loss is a convex function of
$\vectgr{\theta}$ for $p \in [1,2]$. If $p = 1$ and $2$,
then~\eqref{LMP} boils down to the classical sign-LMS and
LMS, respectively~\cite{sayed2011adaptive}. The $p$-power
loss remains convex even if $p > 2$, but such values of $p$
may amplify large-variance outliers $o_n$ via
$\lvert e_n \rvert^{p-1}$ and inflict instabilities
on~\eqref{LMP}.

Intuition suggests that the choice of $p$ in~\eqref{LMP}
should be based on the probability density function (PDF) of
the RV $o_n$. Indeed, if $o_n$ obeys a Gaussian PDF, then
$p = 2$ yields the $2$-power loss which agrees with the
maximum-likelihood criterion. Nevertheless, having prior
knowledge on the statistics of the outliers is usually
infeasible in practice, as in cases where no training data
are available, and in dynamic environments where the
statistics of the outliers may be time varying.

Combinations of LMP filters, with different $p$-power
losses~\cite{chambers1997robust} as well as forgetting
factors~\cite{vazquez2012}, have been proposed to surmount
the problem of pinpointing the ``best'' $p$, but still, the
problem remains and translates to that of pinpointing the
``best'' combination, which again depends on the underlying
outlier PDF. A \textit{data-driven}\/ solution to the
problem of \textit{dynamically}\/ selecting $p$, per time
instance $n$, from streaming data with \textit{no prior
  knowledge}\/ on the statistics of $o_n$ and \textit{no
  training data}\/ seems to be missing from the AdaFilt
literature.

\subsection{Contributions}\label{sec:contribs}

Building on its short preliminary
version~\cite{minh:icassp23}, this manuscript offers a
solution to the aforementioned AdaFilt problem by
reinforcement learning
(RL)~\cite{bertsekas2019reinforcement,
  bertsekas1996neuro}. In RL, an agent takes a
decision/action based on feedback provided by the
surrounding environment on the agent's past actions. RL is a
sequential-decision-making framework with the goal of
minimizing the long-term loss/price $Q$ (a.k.a.\ Q-function)
to be paid by the agent for its own decisions. Central to an
RL design are the \textit{Bellman mappings (B-Maps)}\/ which
operate on the Q-functions, have deep roots in dynamic
programming~\cite{bertsekas2019reinforcement,
  bellman2003dp}, and a far-reaching range of applications
which extend from autonomous navigation, robotics, resource
planning, sensor networks, biomedical imaging, and can reach
even to gaming~\cite{bertsekas2019reinforcement}.

Rather than adopting a popular off-the-shelf RL method, this
manuscript designs a novel family of B-Maps
(\cref{sec:proposed.Bellman.maps}) to solve the AdaFilt
problem at hand. The contributions of this work are
summarized as follows.

\begin{contriblist} 

\item In contrast with the majority of existing B-Maps,
  which are defined in Banach spaces (no inner product
  available), the proposed B-Maps, as well as the
  Q-functions, are specifically defined in reproducing
  kernel Hilbert spaces (RKHSs) to take advantage of the
  rich approximating properties of
  RKHSs~\cite{aronszajn1950, scholkopf2002learning} and the
  flexibility an RKHS inner product brings into the design
  of loss functions and constraints.

\item The proposed B-Maps possess ample degrees of freedom;
  indeed, \cref{prop:T.yields.LSPE+BR} offers a variational
  framework to identify their free parameters, and shows
  that by appropriately designing those parameters, several
  popular B-Maps fall as special cases under the umbrella of
  the proposed design. \cref{sec:prior.art.Bellman} provides
  a thorough literature review on the prior art of B-Maps.

\item Owing to the kernel functions, the proposed B-Maps are
  rendered nonparametric, with no need for statistical
  priors and assumptions on the data, in an effort to reduce
  as much as possible the bias inflicted on data modeling by
  the user~\cite{Gyorfi:DistrFree:10}. The price to be paid
  for this distribution-free approach is that the dimensions
  of the Q-function estimates scale with the number of
  observed data. To surmount this ``curse of
  dimensionality,'' a dimensionality-reduction strategy
  based on random Fourier features is offered in
  \cref{sec:RFF}.

\item\label{contrib:no.training.data} The proposed B-Maps
  allow for sampling on-the-fly via the design of trajectory
  samples in \cref{sec:trajectories}, do not require any
  knowledge on transition probabilities of Markov decision
  processes, and enable computationally lightweight
  operations to fit into the online or time-adaptive
  learning required by the AdaFilt problem at hand.

\item For the first time in the literature, this manuscript
  and its short preliminary version~\cite{minh:icassp23}
  offer an RL-based solution (\cref{algo}) to the problem of
  countering outliers in AdaFilt.

\end{contriblist}

With regards to \ref{contrib:no.training.data}, it is worth
noting here that the recently popular deep learning (DeepL),
\eg, \cite{ddqn}, offers an alternative parametric way of
designing rich approximating spaces for
Q-functions. However, this is achieved at the price of
requiring learning from training data prior to the online
mode of operation (test-stage), or even re-training during
the test-stage to address the often met scenario of facing
test data with different statistics than those of the
training data (dynamic environments). Such modes of learning
raise computational-complexity issues and discourage the
application of DeepL solutions to online modes of operation
where a small computational-complexity footprint is desired.

The proposed RL solution is built on a continuous state
space, because of the nature of $(\mathbf{x}_n, y_n)$. In
contrast with \cite{minh:icassp23}, where the state space is
the high-dimensional $\Real^{2L+1}$, this study confines the
state space to the low-dimensional $\Real^4$. The action
space is considered to be discrete; an action is a value of
$p$ taken from a finite grid of the interval $[1, 2]$. The
well-known policy-iteration (PI)
strategy~\cite{bertsekas2019reinforcement} is adopted in
\cref{algo}, because of its well-documented merits (\eg,
\cite{ormoneit2002kernel, ormoneit:autom:02, xu2007klspi}),
especially for continuous state spaces, over the popular
strategies of temporal-difference (TD) and
Q-learning~\cite{bertsekas2019reinforcement}. To keep the
discussion simple, a quadratic loss on Q-functions is
defined via the proposed B-Maps, and the classical SGD rule
is applied to update the $Q$ estimates
(\cref{sec:T.mu.n}). To promote the use of past data,
experience replay~\cite{experiencereplay} is employed in
\cref{sec:exp.replay}, while \cite{minh:icassp23} uses
rollout~\cite{bertsekas2019reinforcement}.

Properties of the proposed B-Maps are established in
\Cref{thm:new.Bellman.maps.nonexp,%
  thm:new.Bellman.maps.consistency}, and a performance
analysis of \cref{algo} is provided in
\cref{sec:performance.analysis}. Numerical tests on
synthetic data in \cref{sec:tests} support the theoretical
findings and demonstrate that the advocated framework
outperforms, in most of the cases, several RL and non-RL
schemes. Due to lack of space, all proofs are
included in the appendices of the manuscript.

\section{Nonparametric Bellman Mappings for RL}%
\label{sec:new.Bellman.maps}

\subsection{Notation and preliminaries}\label{sec:notation}

\begin{figure}[t]
  \centering
  \includegraphics[width=\linewidth]{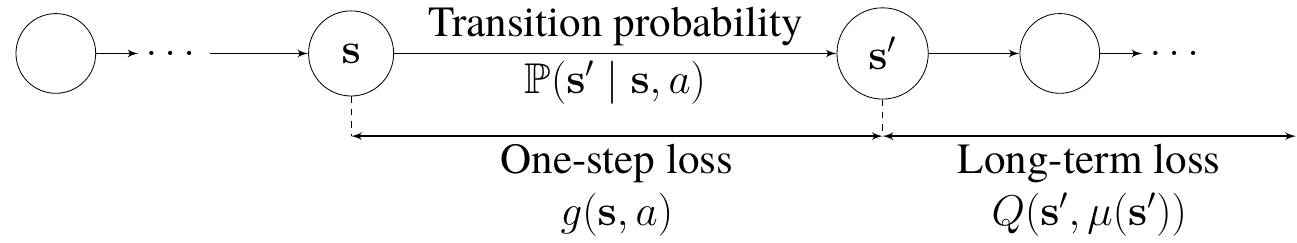}
  \caption{RL as a sequential-decision-making framework:
    Identify the agent's policy $\mu(\cdot)$ (a decision- or
    action-making function) which minimizes the total loss
    ($=$ one-step loss $+$ long-term loss) to be paid by the
    agent for its sequence of
    decisions/actions.}\label{fig:MDP}
\end{figure}

A \textit{continuous}\/ state space
$\mathfrak{S} \subset \Real^D$ is considered, with state
vector $\vect{s} \in \mathfrak{S}$, for some
$D\in \IntegerPP$ ($\IntegerPP$ is the set of all positive
integers). The action space is denoted by $\mathfrak{A}$,
with action $a\in \mathfrak{A}$. For convenience, the
state-action tuple is defined as
$\vect{z} \coloneqq (\vect{s}, a) \in \mathfrak{Z} \coloneqq
\mathfrak{S} \times \mathfrak{A}$. Moreover, let all
mappings
$\mathcal{M} \coloneqq \Set{ \mu(\cdot) \given \mu(\cdot)
  \colon \mathfrak{S} \to \mathfrak{A} : \vect{s} \mapsto
  \mu(\vect{s})}$, and define a policy
$\pi \in \Pi \coloneqq \mathcal{M}^{\IntegerP} \coloneqq
\Set{ (\mu_0, \mu_1, \ldots, \mu_n, \ldots) \given \mu_n\in
  \mathcal{M}, n\in \IntegerP}$. Given $\mu\in \mathcal{M}$,
the stationary policy $\pi_{\mu} \in \Pi$ is defined as
$\pi_{\mu} \coloneqq ( \mu, \mu, \ldots, \mu, \ldots)$. By
abuse of notation, $\mu$ will hereafter denote also the
stationary policy $\pi_{\mu}$.

RL can be viewed as a sequential-decision framework; see
\cref{fig:MDP}. In short, an agent, currently at state
$\vect{s}\in \mathfrak{S}$, takes an action/decision
$a\in \mathfrak{A}$ and transitions to a new state
$\vect{s}^{\prime} \in \mathfrak{S}$ with transition
(conditional) probability
$\Prob(\vect{s}^{\prime} \given \vect{s}, a)$ at the price
of the \textit{one-step}\/ loss $g(\vect{s}, a)$. Quantity
$Q(\vect{s}^{\prime}, \mu(\vect{s}^{\prime}))$ denotes the
\textit{long-term}\/ loss, or, the price to be paid if the
agent continues to take actions, from the state
$\vect{s}^{\prime}$ and on, according to the stationary
policy $\mu(\cdot)$. Typically,
$g(\cdot) \colon \mathfrak{Z}\to\Real$ and
$Q(\cdot) \colon \mathfrak{Z}\to\Real$ are considered points
of the functional Banach space $\mathcal{B}$ of all
(essentially) bounded functions, equipped with the
$\mathcal{L}_{\infty}$-norm~\cite{bertsekas2019reinforcement}. Recall
that by definition a Banach space $\mathcal{B}$ is not
equipped with an inner product.

Departing from standard RL routes which revolve around
Banach spaces $\mathcal{B}$, this study considers a
reproducing kernel Hilbert space (RKHS)
$\mathcal{H}$~\cite{aronszajn1950, scholkopf2002learning} as
the ambient space where $g$ and $Q$ belong to. The RKHS
$\mathcal{H}$ is a Hilbert space with inner product
$\innerp{\cdot}{\cdot}_{\mathcal{H}}$, norm
$\norm{\cdot}_{\mathcal{H}} \coloneqq
\innerp{\cdot}{\cdot}_{\mathcal{H}}^{1/2}$, and a
reproducing kernel
$\kappa(\cdot, \cdot): \mathfrak{Z}\times \mathfrak{Z} \to
\Real$ such that $\kappa(\vect{z}, \cdot)\in \mathcal{H}$,
$\forall \vect{z}\in \mathfrak{Z}$, and the
\textit{reproducing property}\/ holds true:
$Q(\vect{z}) = \innerp{Q}{\kappa(\vect{z},
  \cdot)}_{\mathcal{H}}$, $\forall Q\in \mathcal{H}$,
$\forall \vect{z}\in \mathfrak{Z}$. Space $\mathcal{H}$ may
be infinite dimensional; \eg, the case where
$\kappa(\cdot, \cdot)$ is a Gaussian
kernel~\cite{aronszajn1950, scholkopf2002learning}. For
compact notations, let the feature mapping
$\varphi(\vect{z}) \coloneqq \kappa(\vect{z},\cdot)$ and
$Q^{\intercal} Q^{\prime} \coloneqq
\innerp{Q}{Q^{\prime}}_{\mathcal{H}}$.

Finally, notation
$\mathcal{T}_{N} \coloneqq \{ (\vect{s}_i, a_i,
\vect{s}_i^{\prime}) \}_{i=1}^N \subset \mathfrak{S} \times
\mathfrak{A} \times \mathfrak{S}$, for $N \in \IntegerPP$,
will be used hereafter to denote a collection of trajectory
samples, with $\vect{s}_i^{\prime}$ being a potential
subsequent state of $\vect{s}_i$ after the agent takes
action $a_i$. Moreover, $g_i$ will stand for either
$g(\vect{s}_i, a_i, \vect{s}_i^{\prime})$ or
$g(\vect{s}_i, a_i)$, depending on the context of
discussion. For a reproducing kernel $\kappa(\cdot, \cdot)$
and its feature mapping $\varphi(\cdot)$, let for
convenience
$\vectgr{\Phi}_{\mathcal{T}_N} \coloneqq
[\varphi(\vect{z}_1), \ldots, \varphi(\vect{z}_N)]$, where
$\vect{z}_i \coloneqq (\vect{s}_i, a_i)$, and define then
$\vect{K}_{\mathcal{T}_N} \coloneqq
\vectgr{\Phi}_{\mathcal{T}_N}^{\intercal}
\vectgr{\Phi}_{\mathcal{T}_N}$ as the $N \times N$ kernel
matrix whose $(i,i^{\prime})$th entry is
$\kappa( \vect{z}_i, \vect{z}_{i^{\prime}})$. Moreover, let
$\vectgr{\Phi}_{\mu}^{\prime} \coloneqq [ \varphi(
\vect{s}_1^{\prime}, \mu( \vect{s}_1^{\prime} ) ), \ldots,
\varphi( \vect{s}_N^{\prime}, \mu( \vect{s}_N^{\prime} ) )
]$, and
$\vect{g} \coloneqq [ g(\vect{s}_1, a_1), \ldots,
g(\vect{s}_N, a_N) ]^{\intercal}$.

\subsection{Prior art of Bellman mappings}%
\label{sec:prior.art.Bellman}

The classical B-Maps are defined in a Banach space
$\mathcal{B}$ and quantify the total loss (= one-step loss +
expected long-term loss) to be paid by the agent when taking
action $a$ at state $\vect{s}$~\cite{bellemare:16}. More
specifically,
$T_{\mu}^{\diamond}, T^{\diamond}\colon \mathcal{B} \to
\mathcal{B}$, where $\forall Q\in \mathcal{B}$,
\begin{subequations}\label{classical.Bellman.maps}
  \begin{align}
    (T_{\mu}^{\diamond} Q)(\vect{s}, a)
    & \coloneqq g( \vect{s}, a ) + \alpha
      \expect_{\vect{s}^{\prime} \given (\vect{s}, a)} \{
      Q(\vect{s}^{\prime}, \mu(\vect{s}^{\prime})) \}
      \,, \label{classical.Bellman.mu} \\
    (T^{\diamond} Q)(\vect{s}, a)
    & \coloneqq g( \vect{s}, a ) + \alpha
      \expect_{\vect{s}^{\prime} \given (\vect{s}, a)}
      \{ \inf_{a^{\prime}\in \mathfrak{A}
      }Q(\vect{s}^{\prime}, a^{\prime})
      \}\,, \label{classical.Bellman.inf}
  \end{align}
  and where
  $\expect_{\vect{s}^{\prime} \given (\vect{s}, a)}\{\cdot
  \}$ stands for the conditional
  expectation~\cite{Williams:probability:91} over all
  possible subsequent states $\vect{s}^{\prime}$ of
  $\vect{s}$, conditioned on $(\vect{s}, a)$, and $\alpha$
  is the discount factor with typical values in
  $(0,1)$. Mapping~\eqref{classical.Bellman.mu} refers to
  the case where the agent takes actions according to the
  stationary policy $\mu(\cdot)$,
  while~\eqref{classical.Bellman.inf} serves as a greedy
  variation of~\eqref{classical.Bellman.mu}. Note
  that~\eqref{classical.Bellman.inf} can be recast in the
  form of \eqref{classical.Bellman.mu} whenever the $\inf$
  in~\eqref{classical.Bellman.inf} is achievable:
  \begin{alignat}{2}
    (T^{\diamond} Q)(\vect{s}, a)
    & {} \coloneqq {}
      && g( \vect{s}, a ) + \alpha
      \expect_{\vect{s}^{\prime} \given (\vect{s}, a)}
      \{ Q(\vect{s}^{\prime}, \mu_Q(\vect{s}^{\prime}))
      \} \label{classical.Bellman.min} \\
    & = && (T^{\diamond}_{\mu_Q} Q)(\vect{s}, a) \,, \notag
  \end{alignat}
\end{subequations}
where, given $Q$, the stationary policy $\mu_Q(\cdot)$ is
defined as
$\mu_Q(\vect{s}^{\prime}) \coloneqq \arg \min_{a^{\prime}\in
  \mathfrak{A}} Q(\vect{s}^{\prime}, a^{\prime})$, so that
$Q(\vect{s}^{\prime}, \mu_Q(\vect{s}^{\prime})) =
\min_{a^{\prime}\in \mathfrak{A}} Q(\vect{s}^{\prime},
a^{\prime})$.

Given a mapping $T\colon \mathcal{B} \to \mathcal{B}$, its
fixed-point set is defined as
$\Fix T \coloneqq \{ Q\in \mathcal{B} \given TQ = Q\}$. It
is well known that $\Fix T_{\mu}^{\diamond}$ and
$\Fix T^{\diamond}$ play a central role in identifying the
policy which minimizes the total
loss~\cite{bertsekas2019reinforcement}. Typically, the
discount factor $\alpha \in (0,1)$ to render
$T_{\mu}^{\diamond}, T^{\diamond}$
contractions~\cite{bertsekas2019reinforcement, hb.plc.book},
so that $\Fix T_{\mu}^{\diamond}$ and $\Fix T^{\diamond}$
become singletons~\cite{hb.plc.book}.

Motivated by the Nadaraya-Watson kernel
estimate~\cite{Gyorfi:DistrFree:10}, and for a non-negative
and not necessarily reproducing kernel function
$\chi(\cdot, \cdot)\colon \mathfrak{S} \times \mathfrak{S}
\to \Real$, kernel-based (KB)RL~\cite{ormoneit2002kernel,
  ormoneit:autom:02} is built on the following B-Maps;
$\forall (\vect{s}, a) \in \mathfrak{Z}$,
\begin{subequations}\label{KBRL.maps}
  \begin{align}
    T_{\textnormal{KBRL}, \mu}(Q) (\vect{s},a)
    & \coloneqq \smashoperator{\sum_{(\vect{s}_i, a_i,
      \vect{s}_i^{\prime}) \in
      \mathcal{T}_N^a}} \chi(\vect{s}, \vect{s}_i) \bigl(g_i
      + \alpha Q(\vect{s}_i^{\prime},
      \mu(\vect{s}_i^{\prime}) ) \bigr) \,, \label{KBRL.mu}
    \\
    T_{\textnormal{KBRL}}(Q) (\vect{s},a)
    & \coloneqq \smashoperator{\sum_{(\vect{s}_i, a_i,
      \vect{s}_i^{\prime}) \in
      \mathcal{T}_N^a}} \chi(\vect{s}, \vect{s}_i) \bigl(g_i
      + \alpha
      \inf_{a^\prime\in\mathfrak{A}}Q(\vect{s}_i^{\prime},
      a^{\prime}) \bigr) \,, \label{KBRL.inf}
  \end{align}
\end{subequations}
where
$\mathcal{T}_N^a \coloneqq \Set{ (\vect{s}_i, a_i,
  \vect{s}_i^{\prime})\in \mathcal{T}_N \given a_i = a }$,
$\mathcal{T}_N$ is typically considered to comprise
``historical'' (training) trajectory data,
$g_i \coloneqq g(\vect{s}_i, a_i, \vect{s}_i^{\prime})$, and
$\chi$ needs to satisfy
$\sum_{(\vect{s}_i, a_i, \vect{s}_i^{\prime}) \in
  \mathcal{T}_N^a} \chi(\vect{s}, \vect{s}_i) =
1$. Following~\cite{ormoneit2002kernel}, a simple way to
enforce the previous constraint on $\chi$ for every
$\vect{s}\in \mathfrak{S}$ is via another non-negative
``mother'' kernel function $\zeta(\cdot, \cdot)$:
\begin{align}
  \chi(\vect{s}, \vect{s}_i) \coloneqq \frac{\zeta(\vect{s},
  \vect{s}_i)} {\sum_{(\vect{s}_j,
  a_j, \vect{s}_j^{\prime}) \in \mathcal{T}_N^a } \zeta
  (\vect{s}, \vect{s}_j)} \,. \label{average.function.KBRL}
\end{align}
Even if $\zeta$ is a reproducing kernel of an RKHS
$\mathcal{H}$ and $Q\in \mathcal{H}$, due to the denominator
of~\eqref{average.function.KBRL}, there is no guarantee, in
general, that $\chi$, $T_{\textnormal{KBRL}, \mu}(Q)$, and
$T_{\textnormal{KBRL}}(Q)$ belong to
$\mathcal{H}$~\cite{ormoneit2002kernel}. For such a reason,
the discussion in~\cite{ormoneit2002kernel,
  ormoneit:autom:02} stays in a Banach space, with no use of
an RKHS inner product. The KBRL mappings~\eqref{KBRL.maps}
have been also adopted in~\cite{barreto:nips:11,
  barreto:nips:12, kveton_theocharous_2013,
  kveton_theocharous_2021}.

RKHSs were used as approximating spaces for conditional
expectations via distribution embeddings
in~\cite{Song:ICML:09, Song:ICML:10,
  grunewalder2012modelling}, where, for example, the
existence of an $h_{(\vect{s}, a)}^{\mu} \in\mathcal{H}$
such that
$\innerp{Q} {h_{(\vect{s}, a)}^{\mu} }_{\mathcal{H}} =
\expect_{\vect{s}^{\prime} \given (\vect{s}, a)} \{
Q(\vect{s}^{\prime}, \mu(\vect{s}^{\prime})) \}$,
$\forall (\vect{s}, a) \in \mathfrak{Z}$, is demonstrated
under certain conditions. By tailoring the arguments
of~\cite{grunewalder2012modelling} to the current context,
the B-Map
$\mathcal{H} \ni Q \mapsto T_{\textnormal{emb}, \mu}(Q)
(\vect{s}, a) \coloneqq g(\vect{s}, a) + \alpha
\innerp{Q}{\hat{h}_{(\vect{s}, a)}^{\mu} }_{\mathcal{H}}$
can be defined, with
\begin{align*}
  \hat{h}_{(\vect{s}, a)}^{\mu}
  \coloneqq \sum\nolimits_{i=1}^{N} \frac{c_{i}(\vect{s},
  a)}{\sum_{j = 1}^{N} \lvert
  c_{j}(\vect{s}, a) \rvert }
  \varphi(\vect{s}_i^\prime, \mu(\vect{s}_i^\prime) ) \in
  \mathcal{H} \,, 
\end{align*}
serving as a substitute for the unknown
$h_{(\vect{s}, a)}^{\mu}$, where
$\vect{c}(\vect{s}, a) \coloneqq [ c_1 (\vect{s}, a),
c_2(\vect{s}, a), \ldots, c_N(\vect{s}, a)]^{\intercal}
\coloneqq ( \vect{K}_{{\mathcal{T}_N}} + \sigma \vect{I}_N
)^{-1} \vectgr{\Phi}_{\mathcal{T}_N}^\intercal \varphi
(\vect{s}, a)$. Similarly to the discussion
following~\eqref{KBRL.maps}, $T_{\textnormal{emb}, \mu}(Q)$
is not guaranteed in general to belong to $\mathcal{H}$,
even if $Q\in\mathcal{H}$, and as such,
$T_{\textnormal{emb}, \mu}(Q)$ is treated as an element of a
Banach space in~\cite{grunewalder2012modelling}. Moreover,
notice that $\vect{c}(\vect{s}, a)$ needs to be computed at
each point $(\vect{s}, a)\in \mathfrak{Z}$, which poses
computational obstacles in cases where $\mathfrak{Z}$ is
either continuous or of massive cardinality. Further, the
previous ideas to approximate conditional expectations does
not seem to be straightforward in the case
of~\eqref{classical.Bellman.inf}, because the existence of
an $h_{(\vect{s}, a)}$ that satisfies
$\innerp{Q}{ h_{(\vect{s}, a)} }_{\mathcal{H}} =
\expect_{\vect{s}^{\prime} \given (\vect{s}, a)} \{
\inf_{a^{\prime}\in \mathfrak{A} } Q(\vect{s}^{\prime},
a^{\prime} ) \}$, and the linearity of the inner product,
would suggest that the previous conditional expectation is a
linear function of $Q$; however, this is not true in
general.

In quest of an inner product and more standard formulations,
the least-squares policy evaluation (LSPE) method was
introduced in Euclidean spaces in~\cite{Nedic:LSPE:03,
  bertsekas2004improved}, and extended into potentially
infinite dimensional RKHSs
in~\cite{jung2007kernelizing}. The basic LSPE
iteration~\cite{jung2007kernelizing} can be recast as a
Bellman mapping by the following variational form: for fixed
$\mu \in \mathcal{M}$, $\sigma \in \RealP$, and
$\forall Q\in \mathcal{H}$,
\begin{alignat}{2}
  T_{\textnormal{LSPE}, \mu} (Q)
  & {} \coloneqq {}
  && \arg\min\limits_{ Q^{\prime} \in
     \mathcal{H} }
     \sum\nolimits_{i=1}^{N} \bigl[ Q^{\prime}(\vect{z}_i) -
     g_i \notag \\
  &&& \hphantom{\arg\min\limits_{ Q^{\prime} \in \mathcal{H}
      } \sum\nolimits_{i=1}^{N} \bigl[}
      - \alpha Q(\vect{s}_i^{\prime}, \mu
      (\vect{s}_i^{\prime})) \bigr]^2 \notag \\
  &&& \hphantom{\arg\min\limits_{ Q^{\prime} \in
      \mathcal{H}}}  + \sigma \norm{Q^{\prime} -
      Q}_{\mathcal{H}}^2   \,. \label{LSPE.map}
\end{alignat}
The explicit form of $T_{\textnormal{LSPE}, \mu} (Q)$ can be
found in~\cref{prop:T.yields.LSPE}. LSPE shows strong
connections with the classical temporal-difference (TD)
learning~\cite{Sutton1988, bertsekas1996neuro, td-lambda,
  bertsekas2004improved, Bae:kerneltd1:11}, whose recursion
is an SGD step on the loss in~\eqref{LSPE.map} for $N=1$ and
$\sigma = 0$. Motivated by~\eqref{LSPE.map}, the popular LS
temporal difference (LSTD) method~\cite{Bradtke1996,
  Boyan2002, lagoudakis2003lspi, xu2007klspi,
  regularizedpi:16} computes a fixed point
$Q_{\textnormal{LSTD}, \mu}\in \Fix T_{\textnormal{LSPE},
  \mu}$, where
\begin{align}
  & \Fix T_{\textnormal{LSPE}, \mu} \notag \\
  & = \Set*{ Q\in\mathcal{H} \given (
    \vectgr{\Phi}_{\mathcal{T}_N}
    \vectgr{\Phi}_{\mathcal{T}_N}^{\intercal} - \alpha
    \vectgr{\Phi}_{\mathcal{T}_N}
    \vectgr{\Phi}_{\mu}^{\prime \intercal} ) Q =
    \vectgr{\Phi}_{\mathcal{T}_N} \vect{g} }
    \,; \label{fix.LSPE}
\end{align}
see \cref{app:prop:new.Bellman.maps.reproduce.LSTD} for a
proof of~\eqref{fix.LSPE}. That fixed point becomes unique
whenever
$\vect{K}_{\mathcal{T}_N} - \alpha
\vectgr{\Phi}_{\mu}^{\prime \intercal}
\vectgr{\Phi}_{\mathcal{T}_N}$ is invertible:
\begin{align}
  Q_{\textnormal{LSTD}, \mu} = \vectgr{\Phi}_{\mathcal{T}_N}
  (\vect{K}_{\mathcal{T}_N} - \alpha
  \vectgr{\Phi}_{\mu}^{\prime \intercal}
  \vectgr{\Phi}_{\mathcal{T}_N})^{-1} \vect{g}
  \,, \label{fixed.point.LSPE}
\end{align}
where~\eqref{fixed.point.LSPE} follows easily
from~\eqref{fix.LSPE} after using
$( \vectgr{\Phi}_{\mathcal{T}_N}
\vectgr{\Phi}_{\mathcal{T}_N}^{\intercal} - \alpha
\vectgr{\Phi}_{\mathcal{T}_N} \vectgr{\Phi}_{\mu}^{\prime
  \intercal} )^{-1} \vectgr{\Phi}_{\mathcal{T}_N} =
\vectgr{\Phi}_{\mathcal{T}_N} (\vect{K}_{\mathcal{T}_N} -
\alpha \vectgr{\Phi}_{\mu}^{\prime \intercal}
\vectgr{\Phi}_{\mathcal{T}_N})^{-1}$. Interestingly, it has
been demonstrated that, in general, LSPE/LSTD perform better
than TD in numerical tests~\cite{lagoudakis2003lspi}.

The Bellman-residual (BR) approach~\cite{Schweitzer1985,
  onlinebrloss:16} originates from a variational task, which
can be readily recast into the following form to fit the
current RKHS context:
\begin{alignat}{3}
  T_{\textnormal{BR}, \mu}(Q)
  & {} \coloneqq {}
  && \arg\min\limits_{ Q^{\prime} \in \mathcal{H} }
  && \sum\nolimits_{i=1}^{N} \bigl[ Q^{\prime}(\vect{z}_i) -
     g_i - \alpha Q^{\prime}(\vect{s}_i^{\prime}, \mu
     (\vect{s}_i^{\prime})) \bigr]^2 \notag \\
  &&&&& + \sigma \norm{Q^{\prime} - Q}_{\mathcal{H}}^2
        \,. \label{BR.map}
\end{alignat}
The explicit form of \eqref{BR.map} can be found in
\cref{prop:T.yields.BR}. Interestingly,
$T_{\textnormal{BR}, \mu}(Q) = \prox_{f/(2\sigma)}(Q)$,
where the popular proximal mapping is defined as
$\prox_{f/(2\sigma)}(Q) \coloneqq \arg\min_{ Q^{\prime} \in
  \mathcal{H} } f(Q^{\prime}) + 2\sigma \cdot (1/2)\norm{Q -
  Q^{\prime}}_{\mathcal{H}}^2$~\cite{hb.plc.book}, with
$f(Q^{\prime}) \coloneqq \sum\nolimits_{i=1}^{N} [
Q^{\prime}(\vect{z}_i) - g_i - \alpha
Q^{\prime}(\vect{s}_i^{\prime}, \mu (\vect{s}_i^{\prime}))
]^2$. Extensions to cases where the loss in~\eqref{BR.map}
is further regularized by additional convex functions, such
as $\ell_1$-norm loss, for example, to impose structure onto
the desired solutions, can be found
in~\cite{Qin:SparseRL:14, Mahadevan:PrimalDual:14,
  Liu:ProxGTD:18}.

\subsection{New Bellman mappings in RKHSs}%
\label{sec:proposed.Bellman.maps}

Hereafter, losses $g, Q$ are assumed to belong to an RKHS
$\mathcal{H}$. For some
$N_{\textnormal{av}} \in \IntegerPP$, consider the
state-space vectors
$\mathcal{S}_{\textnormal{av}} \coloneqq \Set{
  \vect{s}_i^{\textnormal{av}} }_{i=1}^{N_{\textnormal{av}}}
\subset \mathfrak{S}$, chosen by the user to enable sampling
of trajectory samples on-the-fly to approximate the
conditional expectation
in~\eqref{classical.Bellman.maps}. Define also, for
convenience in notation and for a $\mu \in \mathcal{M}$,
$\vectgr{\Phi}_{\mu}^{\textnormal{av}} \coloneqq [
\varphi(\vect{s}_1^{\textnormal{av}},
\mu(\vect{s}_1^{\textnormal{av}})), \ldots,
\varphi(\vect{s}_{N_{\textnormal{av}}}^{\textnormal{av}},
\mu(\vect{s}_{N_{\textnormal{av}}}^{\textnormal{av}}))
]$. Consider also the user-defined
$\{\psi_i\}_{i=1}^{N_{\textnormal{av}}} \subset
\mathcal{H}$, with
$\vectgr{\Psi} \coloneqq [ \psi_1, \ldots,
\psi_{N_{\textnormal{av}}} ]$. Then, the proposed B-Maps
$T_{\mu}, T\colon \mathcal{H} \to \mathcal{H}$ are defined
as follows: $\forall Q\in \mathcal{H}$,
$\forall \mu\in \mathcal{M}$,
\begin{subequations}\label{new.Bellman.maps}
  \begin{alignat}{2}
    T_{\mu} (Q)
    & {} \coloneqq {}
    && g + \alpha \sum\nolimits_{i=1}^{N_{\textnormal{av}}}
       Q (\vect{s}^{\textnormal{av}}_i,
       \mu(\vect{s}^{\textnormal{av}}_i) ) \cdot \psi_i
       \notag\\
    & =
    && g + \alpha \vectgr{\Psi}
       \vectgr{\Phi}_{\mu}^{\textnormal{av} \intercal} Q
       \,, \label{new.Bellman.mu}\\
    T(Q)
    & \coloneqq
    && g + \alpha
       \sum\nolimits_{i=1}^{N_{\textnormal{av}}}
       \inf\nolimits_{a_i \in \mathfrak{A}} Q
       (\vect{s}^{\textnormal{av}}_i, a_i) \cdot \psi_i
       \notag\\
    & =
    && g + \alpha \vectgr{\Psi}
       \vecinf_{\mathcal{S}_{\textnormal{av}}}(Q)
       \,, \label{new.Bellman.inf}
  \end{alignat}
\end{subequations}
where $\vecinf_{\mathcal{S}_{\textnormal{av}}} (Q)$ is
defined as the $N_{\textnormal{av}} \times 1$ vector whose
$i$th entry is
$\inf \nolimits_{ a \in \mathfrak{A} } Q
(\vect{s}^{\textnormal{av}}_i, a )$. The reproducing
property of the inner product in $\mathcal{H}$ was used to
obtain the latter formulation in~\eqref{new.Bellman.mu}. Let
also the $N_{\textnormal{av}} \times N_{\textnormal{av}}$
kernel matrices
$\vect{K}_{\Psi} \coloneqq \vectgr{\Psi}^{\intercal}
\vectgr{\Psi}$ and
$\vect{K}_{\mu}^{\textnormal{av}} \coloneqq
\vectgr{\Phi}_{\mu}^{\textnormal{av}\intercal}
\vectgr{\Phi}_{\mu}^{\textnormal{av}}$. Further, it is worth
noting that vectors $\mathcal{S}_{\textnormal{av}}$ may be
also used to incorporate training data
in~\eqref{new.Bellman.maps}.

In contrast with several of the popular B-Map designs in
\cref{sec:proposed.Bellman.maps} which are defined in Banach
spaces, the proposed~\eqref{new.Bellman.maps} are defined
directly in an RKHS $\mathcal{H}$. To highlight the ample
degrees of freedom and flexibility offered by the
user-defined $\{\psi_i\}_i$, the following proposition
demonstrates that by appropriately tuning the $\{\psi_i\}_i$
through a variational framework, the proposed
B-Maps~\eqref{new.Bellman.maps} yield the popular
designs~\eqref{LSPE.map}, \eqref{fixed.point.LSPE},
\eqref{BR.map}, as well as \cite[(7)]{Song:AISTATS:10} and
\cite[(3)]{grunewalder2012modelling} as special cases.

\begin{prop}\label{prop:T.yields.LSPE+BR}
  \textbf{(Variational framework for B-Maps)} Consider the
  user-defined loss function
  $\mathcal{L}\colon \Real^N \times \Real^{ N\times
    N_{\textnormal{av}} } \to \Real \colon (\vectgr{\gamma},
  \vectgr{\Upsilon}) \mapsto \mathcal{L}(\vectgr{\gamma},
  \vectgr{\Upsilon})$ and the regularizing function
  $\mathcal{R}\colon \Real^N \times \Real^{ N\times
    N_{\textnormal{av}} } \to \Real \colon (\vectgr{\gamma},
  \vectgr{\Upsilon}) \mapsto \mathcal{R}(\vectgr{\gamma},
  \vectgr{\Upsilon})$, and let
  $(\vectgr{\gamma}_{\star}, \vectgr{\Upsilon}_{\star})$
  stand for the minimizers of the following variational
  problem:
  \begin{alignat}{2}
    (\vectgr{\gamma}_{\star}, \vectgr{\Upsilon}_{\star}) \in
    \arg\min\limits_{\vectgr{\gamma}
    \in \Real^N, \vectgr{\Upsilon}\in \Real^{ N \times
    N_{\textnormal{av}} } }
    \mathcal{L}( \vectgr{\gamma},
    \vectgr{\Upsilon} ) + \sigma \mathcal{R}(
    \vectgr{\gamma}, \vectgr{\Upsilon} )
    \,, \label{T.yields.several.maps.task}
  \end{alignat}
  where $\sigma \in \RealP$.

  \begin{proplist}

  \item\label{prop:T.yields.LSPE} Consider the stationary
    policy $\mu$ in \eqref{LSPE.map}. Let
    $\mu_{\star} \in\mathcal{M}$ be a stationary policy
    s.t.\ $\mu_{\star}( \vect{s}_i ) \coloneqq a_i$ and
    $\mu_{\star}( \vect{s}_i^{\prime} ) \coloneqq \mu (
    \vect{s}_i^{\prime} )$,
    $\forall i\in \Set{1, \ldots, N}$. Define also
    $\vect{s}_i^{\textnormal{av}} \coloneqq \vect{s}_i$,
    $\forall i\in \Set{1, \ldots, N}$, and
    $\vect{s}_i^{\textnormal{av}} \coloneqq
    \vect{s}_{i-N}^{\prime}$,
    $\forall i\in \Set{N+1, \ldots, 2N}$, so that
    $\vectgr{\Phi}_{\mu_{\star}}^{\textnormal{av}} = [
    \vectgr{\Phi}_{\mathcal{T}_N},
    \vectgr{\Phi}_{\mu}^{\prime} ]$ and
    $N_{\textnormal{av}} = 2N$. Then,
    $T_{\textnormal{LSPE}, \mu}(Q)$ in~\eqref{LSPE.map}, and
    thus~\eqref{fixed.point.LSPE}, are reproduced by
    $T_{\mu_{\star}}(Q)$ in~\eqref{new.Bellman.mu} with
    \begin{align*}
      g \coloneqq \vectgr{\Phi}_{\mathcal{T}_N}
      \vectgr{\gamma}_{\star} \,, \qquad
      \vectgr{\Psi} \coloneqq \vectgr{\Phi}_{\mathcal{T}_N}
      \vectgr{\Upsilon}_{\star} \,,
    \end{align*}
    where the
    \begin{subequations}\label{gamma.Upsilon.LSPE}
      \begin{align}
        \vectgr{\gamma}_{\star}
        & \coloneqq ( \vect{K}_{\mathcal{T}_N} + \sigma
          \vect{I}_N )^{-1} \vect{g}
          \,, \label{gamma.LSPE}\\
        \vectgr{\Upsilon}_{\star}
        & \coloneqq ( \vect{K}_{\mathcal{T}_N} + \sigma
          \vect{I}_N )^{-1} [ (\sigma/\alpha)
          \vect{K}_{\mathcal{T}_N}^{\dagger} \,, \vect{I}_N
          ] \,, \label{Upsilon.LSPE}
      \end{align}
    \end{subequations}
    satisfy~\eqref{T.yields.several.maps.task} with
    \begin{subequations}\label{proposed.coeff.vec.LSPE}
      \begin{alignat}{2}
        \mathcal{L}(\vectgr{\gamma}, \vectgr{\Upsilon})
        & {} \coloneqq {}
        && \norm{ \vect{K}_{\mathcal{T}_N}( \vectgr{\gamma}
           + \alpha \vectgr{\Upsilon}
           \vectgr{\Phi}_{\mu_{\star}}^{\textnormal{av}
           \intercal} Q )  - \vect{g} - \alpha
           \vectgr{\Phi}_{\mu}^{\prime\intercal}Q
           }_{\Real^N}^2
           \,, \label{proposed.coeff.vec.loss.LSPE} \\
        \mathcal{R}(\vectgr{\gamma}, \vectgr{\Upsilon})
        & \coloneqq
        && ( \vectgr{\gamma} + \alpha \vectgr{\Upsilon}
           \vectgr{\Phi}_{\mu_{\star}}^{\textnormal{av}\intercal}
           Q - \vect{K}_{\mathcal{T}_N}^{\dagger}
           \vectgr{\Phi}_{\mathcal{T}_N}^{\intercal}Q )^{\intercal}
           \vect{K}_{\mathcal{T}_N} \notag\\
        &&& \cdot ( \vectgr{\gamma} + \alpha \vectgr{\Upsilon}
            \vectgr{\Phi}_{\mu_{\star}}^{\textnormal{av} \intercal}
            Q - \vect{K}_{\mathcal{T}_N}^{\dagger}
            \vectgr{\Phi}_{\mathcal{T}_N}^{\intercal}Q )
            \,, \label{proposed.coeff.vec.reg.LSPE}
      \end{alignat}
    \end{subequations}
    and $\vect{K}_{\mathcal{T}_N}^{\dagger}$ is the
    Moore-Penrose pseudoinverse of
    $\vect{K}_{\mathcal{T}_N}$.

  \item\label{prop:T.yields.BR} Consider the stationary
    policy $\mu$ in \eqref{BR.map}, and define $\mu_{\star}$
    and $\vectgr{\Phi}_{\mu_{\star}}^{\textnormal{av}}$ as
    in \cref{prop:T.yields.LSPE}. Define also the
    temporal-difference (TD) feature vectors
    $\vectgr{\Phi}_{\textnormal{TD}} \coloneqq
    \vectgr{\Phi}_{\mathcal{T}_N} - \alpha
    \vectgr{\Phi}_{\mu_{\star}}^{\prime}$ and
    $\vect{K}_{\textnormal{TD}} \coloneqq
    \vectgr{\Phi}_{\textnormal{TD}}^{\intercal}
    \vectgr{\Phi}_{\textnormal{TD}}$. Then,
    $T_{\textnormal{BR}, \mu}(Q)$ in~\eqref{BR.map} is
    reproduced by $T_{\mu_{\star}}(Q)$
    in~\eqref{new.Bellman.mu} with
    \begin{align*}
      g\coloneqq \vectgr{\Phi}_{\textnormal{TD}}
      \vectgr{\gamma}_{\star} \,, \qquad
      \vectgr{\Psi} \coloneqq \vectgr{\Phi}_{\textnormal{TD}}
      \vectgr{\Upsilon}_{\star} \,,
    \end{align*}
    where the
    \begin{subequations}\label{gamma.Upsilon.BR}
      \begin{align}
        \vectgr{\gamma}_{\star}
        & \coloneqq ( \vect{K}_{\textnormal{TD}} + \sigma
          \vect{I}_N )^{-1} \vect{g} \,, \label{gamma.BR}\\
        \vectgr{\Upsilon}_{\star}
        & \coloneqq ( \vect{K}_{\textnormal{TD}} + \sigma
          \vect{I}_N )^{-1}
          \vect{K}_{\textnormal{TD}}^{\dagger} [
          (\sigma/\alpha) \vect{I}_N \,, -\sigma \vect{I}_N
          ] \,, \label{Upsilon.BR}
      \end{align}
    \end{subequations}
    satisfy~\eqref{T.yields.several.maps.task} with
    \begin{subequations}\label{proposed.coeff.vec.BR}
      \begin{alignat}{2}
        \mathcal{L}(\vectgr{\gamma}, \vectgr{\Upsilon})
        & {} \coloneqq {}
        && \norm{ \vect{K}_{\textnormal{TD}} (
           \vectgr{\gamma} +
           \alpha \vectgr{\Upsilon}
           \vectgr{\Phi}_{\mu_{\star}}^{\textnormal{av}
           \intercal} Q ) - \vect{g} }_{\Real^N}^2
           \,, \label{proposed.coeff.vec.loss.BR}\\
        \mathcal{R}(\vectgr{\gamma}, \vectgr{\Upsilon})
        & \coloneqq
        && ( \vectgr{\gamma} + \alpha \vectgr{\Upsilon}
           \vectgr{\Phi}_{\mu_{\star}}^{\textnormal{av}
           \intercal} Q - \vect{K}_{\textnormal{TD}}^{\dagger}
           \vectgr{\Phi}_{\textnormal{TD}}^{\intercal}Q
           )^{\intercal}
           \vect{K}_{\textnormal{TD}} \notag\\
        &&& \cdot ( \vectgr{\gamma} + \alpha
            \vectgr{\Upsilon}
            \vectgr{\Phi}_{\mu_{\star}}^{\textnormal{av}
            \intercal} Q
            - \vect{K}_{\textnormal{TD}}^{\dagger}
            \vectgr{\Phi}_{\textnormal{TD}}^{\intercal}Q
            ) \,. \label{proposed.coeff.vec.reg.BR}
      \end{alignat}
    \end{subequations}

  \item\label{prop:T.inthispaper} Given a stationary policy
    $\mu \in \mathcal{M}$, let $\mu_{\star} \coloneqq \mu$,
    $\vect{s}_i^{\textnormal{av}} \coloneqq
    \vect{s}_i^{\prime}$, $\forall i\in \Set{1, \ldots, N}$,
    and
    $\vectgr{\Phi}_{ \mu_{\star} }^{\textnormal{av}}
    \coloneqq \vectgr{\Phi}_{\mu}^{\prime}$. For
    \begin{align*}
      g \coloneqq \vectgr{\Phi}_{\mathcal{T}_N}
      \vectgr{\gamma}_{\star} \,, \qquad
      \vectgr{\Psi} \coloneqq \vectgr{\Phi}_{\mathcal{T}_N}
      \vectgr{\Upsilon}_{\star} \,,
    \end{align*}
    where the
    \begin{subequations}\label{gamma.inthispaper}
      \begin{align}
        \vectgr{\gamma}_{\star}
        & \coloneqq ( \vect{K}_{\mathcal{T}_N} + \sigma
          \vect{I}_N )^{-1} \vect{g} \,, \label{gamma}\\
        \vectgr{\Upsilon}_{\star}
        & \coloneqq ( \vect{K}_{\mathcal{T}_N} + \sigma
          \vect{I}_N )^{-1} \,, \label{Upsilon}
      \end{align}
    \end{subequations}
    satisfy~\eqref{T.yields.several.maps.task} with
    \begin{subequations}\label{proposed.coeff.vec}
      \begin{alignat}{2}
        \mathcal{L}(\vectgr{\gamma}, \vectgr{\Upsilon})
        & {} \coloneqq {}
        && \norm{ \vect{K}_{\mathcal{T}_N}( \vectgr{\gamma}
           + \alpha \vectgr{\Upsilon}
           \vectgr{\Phi}_{\mu_{\star}}^{\textnormal{av}
           \intercal} Q )  - \vect{g} - \alpha
           \vectgr{\Phi}_{\mu}^{\prime\intercal}Q
           }_{\Real^N}^2 \,, \label{proposed.coeff.vec.loss} \\
        \mathcal{R}(\vectgr{\gamma}, \vectgr{\Upsilon})
        & \coloneqq
        && ( \vectgr{\gamma} + \alpha \vectgr{\Upsilon}
           \vectgr{\Phi}_{\mu_{\star}}^{\textnormal{av}\intercal}
           Q )^{\intercal} \vect{K}_{\mathcal{T}_N} (
           \vectgr{\gamma} + \alpha
           \vectgr{\Upsilon}
           \vectgr{\Phi}_{\mu_{\star}}^{\textnormal{av}
           \intercal} Q ) \,, \label{proposed.coeff.vec.reg}
      \end{alignat}
    \end{subequations}
    the B-Map $T_{ \mu_{\star} }(Q)$ in~\eqref{new.Bellman.mu}
    takes the form
    \begin{alignat*}{2}
      T_{ \mu_{\star} }(Q)
      & {} = {}
      && \vectgr{\Phi}_{\mathcal{T}_N} (
         \vect{K}_{\mathcal{T}_N} + \sigma \vect{I}_N )^{-1}
         \vect{g} \\
      &&& + \alpha \vectgr{\Phi}_{\mathcal{T}_N} (
          \vect{K}_{\mathcal{T}_N} + \sigma \vect{I}_N )^{-1}
          \vectgr{\Phi}_{\mu}^{\textnormal{av} \intercal}
          Q.
    \end{alignat*}
    Notice that operator
    $\vectgr{\Phi}_{\mathcal{T}_N} (
    \vect{K}_{\mathcal{T}_N} + \sigma \vect{I}_N )^{-1}
    \vectgr{\Phi}_{\mu}^{\textnormal{av} \intercal}$
    appears in \cite[(7)]{Song:AISTATS:10} and
    \cite[(3)]{grunewalder2012modelling}. However, the B-Map
    which is based on the previous operator and introduced
    in~\cite[(6), (7)]{grunewalder2012modelling} is defined
    in a Banach space, and not a Hilbert one.

  \end{proplist}

\end{prop}

\begin{IEEEproof}
  See \cref{app:prop:new.Bellman.maps.reproduce.LSTD}.
\end{IEEEproof}

More variations of~\eqref{new.Bellman.maps} can be generated
from~\eqref{T.yields.several.maps.task} by tuning the loss
functions $\mathcal{L}, \mathcal{R}$ appropriately. For
example, robust B-Map designs against outliers in sampling
can be obtained by letting the $\ell_1$-norm take the place
of the quadratic one in \eqref{proposed.coeff.vec.loss.LSPE}
and
\eqref{proposed.coeff.vec.loss.BR}. Task~\eqref{T.yields.several.maps.task}
for general (non)smooth convex $\mathcal{L}$ and
$\mathcal{R}$ can be handled efficiently
by~\cite{Slavakis:FMHSDM:18}. Such designs are deferred to
future publications.

\begin{thm}\label{thm:new.Bellman.maps.nonexp}
  \textbf{(Lipschitz continuity)}
  Mappings~\eqref{new.Bellman.maps} are Lipschitz
  continuous: $\forall Q_1, Q_2 \in \mathcal{H}$,%
  \begin{subequations}\label{Lipschitz.inequality}
    \begin{alignat}{2}
      & \norm{ T_{\mu}(Q_1) - T_{\mu} (Q_2) }_{\mathcal{H}}
      && \leq \beta \norm{Q_1 - Q_2}_{\mathcal{H}}
         \,, \label{Lipschitz.mu}\\
      & \norm{T(Q_1) - T(Q_2)}_{\mathcal{H}}
      && \leq \beta \norm{Q_1 - Q_2}_{\mathcal{H}}
         \,, \label{Lipschitz.optimal}
    \end{alignat}
  \end{subequations}
  where
  \begin{align}
    \beta \coloneqq \alpha \left( \norm{ \vect{K}_{\Psi} }_2
    \, \sup\nolimits_{\mu^{\prime}\in\mathcal{M}}
    \norm{\vect{K}^{\textnormal{av}}_{\mu^{\prime}}}_2
    \right)^{{1}/{2}} \,, \label{beta}
  \end{align}
  and $\norm{\cdot}_2$ stands for the spectral norm of a
  matrix. Hence, if $\beta = 1$,
  mappings~\eqref{new.Bellman.maps} are
  nonexpansive~\cite{hb.plc.book}, whereas, if
  $\beta < 1$, they are contractions~\cite{kreyszig:91} in
  $(\mathcal{H}, \innerp{\cdot}{\cdot}_{\mathcal{H}})$.
\end{thm}

\begin{IEEEproof}
  See \cref{app:thm:new.Bellman.maps.nonexp}.
\end{IEEEproof}

To state the following
\cref{thm:new.Bellman.maps.consistency}, a probability space
$(\Omega, \mathcal{F}, \Prob)$ is necessary, with sample
space $\Omega$, $\sigma$-algebra $\mathcal{F}$ of events,
and probability measure
$\Prob$~\cite{Williams:probability:91}. A statement
$(\ldots)$ will be said to hold true almost surely (a.s.),
if $(\ldots)$ holds true on an event
$\mathcal{E} \in \mathcal{F}$ with
$\Prob( \mathcal{E} ) = 1$. Moreover, by a slight abuse of
terminology, a bounded linear and self-adjoint mapping
$\mathcal{A} \colon \mathcal{H} \to \mathcal{H}$ will be
called positive definite, if its minimum spectral value
$\sigma_{\min} ( \mathcal{A} ) > 0$, where
$\sigma_{\min}(\cdot)$ is defined by~\eqref{sigma.min.def}.

\begin{assumptions}\label{ass:consistency.fix}\mbox{}
  \begin{assslist}

  \item\label{ass:HS} The RKHS $\mathcal{H}$ is separable,
    for a stationary policy $\mu(\cdot) \in \mathcal{M}$
    operators
    $\Sigma_{zz}, \Sigma_{s^{\prime} z}^{\mu}, \Sigma_{
      s^{\prime} \given z }^{\mu}$, defined
    by~\eqref{Sigma.def}, are bounded linear, $\Sigma_{zz}$
    of~\eqref{Sigma.zz} is positive definite, and
    $\Sigma_{zz}^{-3/2} \Sigma_{ s^{\prime}z }^{\mu}$
    of~\eqref{Sigma.conditional.map} is Hilbert-Schmidt (see
    \cref{thm:consistency.Gamma} in the appendices).

  \item Let $\vect{s}, a, \vect{s}^{\prime}$ be random
    variables (RVs) on the probability space
    $(\Omega, \mathcal{F}, \Prob)$, and assume that
    trajectory points
    $\{ (\vect{s}_i, a_i, \vect{s}_i^{\prime}) \}_{i=1}^N$
    are also RVs, but independent and identically
    distributed (IID) copies of
    $(\vect{s}, a, \vect{s}^{\prime})$.

  \item\label{ass:T.inthispaper} Motivated by
    \cref{prop:T.inthispaper}
    and~\cite[(7)]{Song:AISTATS:10}, let
    $N_{\textnormal{av}} = N$, set
    $\vect{s}_i^{\textnormal{av}} \coloneqq
    \vect{s}_i^{\prime}$, $\forall i\in \Set{1, \ldots, N}$,
    and define
    $\vectgr{\Psi} \coloneqq \vectgr{\Phi}_{\mathcal{T}_N} (
    \vect{K}_{\mathcal{T}_N} + N \sigma^{\prime}_N
    \vect{I}_N )^{-1}$ so that \eqref{new.Bellman.maps}
    become: $\forall Q\in \mathcal{H}$,
    \begin{subequations}\label{new.Bellman.maps.special}
      \begin{align}
        T_{\mu} (Q)
        & = g + \alpha \vectgr{\Phi}_{\mathcal{T}_N} (
          \vect{K}_{\mathcal{T}_N} + N \sigma^{\prime}_N
          \vect{I}_N )^{-1}
          \vectgr{\Phi}_{\mu}^{\textnormal{av} \intercal} Q
          \,, \label{new.Bellman.mu.special} \\
        T(Q)
        & = g + \alpha \vectgr{\Phi}_{\mathcal{T}_N} (
          \vect{K}_{\mathcal{T}_N} + N \sigma^{\prime}_N
          \vect{I}_N )^{-1}
          \vecinf_{\mathcal{S}_{\textnormal{av}}}(Q)
          \,, \label{new.Bellman.inf.special}
      \end{align}
    \end{subequations}
    where $\sigma^{\prime}_N \in \RealPP$ is a
    regularization coefficient, dependent on $N$.

  \item\label{ass:reg.coeff.sigma}
    $\lim_{N \to \infty} \sigma^{\prime}_N = 0$ and
    $\lim_{N \to \infty} N \sigma^{\prime\, 3}_N = +\infty$.

  \item The $\inf$ operators in \eqref{classical.Bellman.inf}
    and \eqref{new.Bellman.inf.special} are achievable.

  \item\label{ass:beta.inf} There exists
    $\beta_{\infty} \in (0,1)$ s.t.\ $\beta = \beta(N)$ in
    \eqref{beta} satisfies $\beta(N) \leq \beta_{\infty}$,
    $\forall N$, a.s.

  \end{assslist}
\end{assumptions}

\begin{thm}\label{thm:new.Bellman.maps.consistency}
  \textbf{(Consistency of fixed points)} Under
  \cref{ass:consistency.fix},
  $T_{\mu}^{\diamond}, T^{\diamond}$
  in~\eqref{classical.Bellman.maps} and $T_{\mu}, T$
  in~\eqref{new.Bellman.maps.special} are contractions in
  the Hilbert space $\mathcal{H}$, and thus possess unique
  fixed points
  $Q_{\mu}^{\diamond}, Q_*^{\diamond}, Q_{\mu}, Q_*$,
  respectively, a.s. Notice that $Q_{\mu}, Q_*$ depend on
  $N$, \ie, $Q_{\mu} = Q_{\mu}(N)$ and $Q_* =
  Q_*(N)$. Furthermore,
  \begin{subequations}
    \begin{alignat}{2}
      & \Plim\nolimits_{N\to \infty}
        \norm{Q_{\mu}^{\diamond} -
        Q_{\mu}(N)}_{\mathcal{H}}
      && = 0 \,, \label{consistency.mu}\\
      & \Plim\nolimits_{N\to \infty} \norm{Q_*^{\diamond}
        - Q_*(N)}_{\mathcal{H}}
      && = 0 \,, \label{consistency.*}
    \end{alignat}
  \end{subequations}
  where $\Plim$ stands for convergence in
  probability~\cite{Williams:probability:91}.
\end{thm}

\begin{IEEEproof}
  See \cref{app:thm:new.Bellman.maps.consistency}.
\end{IEEEproof}

\section{Application to Robust Adaptive Filtering}%
\label{sec:AdaFilt}

The following discussion applies the novel B-Maps of
\cref{sec:proposed.Bellman.maps} to the setting
of~\cref{sec:RobAdaFilt}. To abide by the online or
time-adaptive premise of~\cref{sec:RobAdaFilt}, the
arguments of \cref{sec:proposed.Bellman.maps} will be
equipped hereafter with a discrete time index
$n\in \IntegerP$. It is important to note that $n$ serves
also as the \textit{iteration index}\/ of the proposed
RL-based \cref{algo}. Index $n$ appears in the following
discussion in various forms, such as a sub-/super-script, or
as $[n]$. For example, $N$ of \cref{sec:prior.art.Bellman}
becomes $N[n]$ from now and on to highlight the fact that
$N$ depends on $n$.

Although \eqref{new.Bellman.maps} and
\cref{prop:T.yields.LSPE+BR} introduce considerable freedom
in designing B-Maps, this manuscript focuses on the setting
of \cref{prop:T.inthispaper} to avoid lengthy
discussions. The online variant $T_{\mu_n}^{(n)}$ of the
mapping in \cref{prop:T.inthispaper} will be presented in
\cref{sec:T.mu.n}. Other designs are deferred to future
publications.

\cref{algo} offers an RL way to robustify LMP~\eqref{LMP} by
letting the data themselves select the ``optimal'' $p_n$ per
time $n$ (see \cref{algo:update.theta} of \cref{algo}),
without any assumptions and prior knowledge on the
statistics of the outliers. More specifically, instead
of~\eqref{LMP},
\begin{align}
  \vectgr{\theta}_{n+1} \coloneqq \vectgr{\theta}_{n+1} (
  a_n ) \coloneqq \vectgr{\theta}_n + \rho\, p_n \lvert e_n
  \rvert^{p_n - 1} \sign(e_n)\, \vect{x}_n \,, \label{dLMP}
\end{align}
where $e_n$ is defined after~\eqref{LMP}. It is clear
by~\eqref{dLMP} and \cref{algo:update.theta} of \cref{algo}
that $\vectgr{\theta}_{n+1}$ depends on the action
$a_n = p_n$. To highlight this observation,
$\vectgr{\theta}_{n+1}( a_n )$ is used together with
$\vectgr{\theta}_{n+1}$ in~\eqref{dLMP}, as well as in the
following discussion.

\cref{algo} belongs to the class of \textit{policy-iteration
  (PI)}\/ algorithms of
RL~\cite{bertsekas2019reinforcement}. More precisely, it is
an \textit{approximate}\/ (A)PI algorithm, because the
expectation operators in~\eqref{classical.Bellman.maps} are
approximated by sample averaging in
\cref{prop:T.inthispaper}. Typically, (A)PI comprises two
major steps: policy improvement in
\cref{algo:policy.improvement} and policy evaluation in
\cref{algo:policy.evaluation}. The following discussion
details \cref{algo}.

\begin{algorithm}[t]
  \begin{algorithmic}[1]
    \renewcommand{\algorithmicindent}{1em}

    \State{Arbitrarily initialize $\vectgr{\theta}_0$,
      $Q_0$, and $\vect{s}_{-1}$.}

    \While{$n \in \IntegerP$}\label{line:iter}

      \State{Data $(\vect{x}_n, y_n)$ become available to
        the user/agent.}

      \State{New state $\vect{s}_{n}$ is defined
        by~\eqref{def.states}.}\label{algo:states}

      \State{\textbf{Policy improvement:} Let $a_{n}
        \coloneqq \mu_{n}(\vect{s}_{n})$
        by~\eqref{mu.n}.}\label{algo:policy.improvement}

      \State{Compute $\vectgr{\theta}_{n+1}$ by~\eqref{dLMP},
        with $p_{n} \coloneqq a_{n}$.}\label{algo:update.theta}

      \State{\textbf{Policy evaluation:} Compute $Q_{n+1}$
        by~\eqref{Q.n.to.n+half} and
        \eqref{Q.n+half.to.n+1}.}\label{algo:policy.evaluation}

      \State{Increase $n$ by one, and go to
        \cref{line:iter}.}

    \EndWhile
  \end{algorithmic}

  \caption{Approximate policy iteration for LMP}\label{algo}

\end{algorithm}

\subsection{State-action space and policy improvement}%
\label{sec:state.action.space}

This subsection refers to
\Cref{algo:states,algo:policy.improvement,algo:update.theta}
of \cref{algo}. Action space $\mathfrak{A}$ is defined as
any finite grid of the interval $[1,2]$, and it is the
domain $p_n$ in \eqref{dLMP} takes values from. State space
is defined as $\mathfrak{S} \coloneqq \Real^4$, with the
dimension of $\mathfrak{S}$ rendered independent of the
filter length $L$. In contrast, the state space
in~\cite{minh:icassp23} is $\Real^{2L+1}$, directly
dependent on $L$, potentially high-dimensional in the case
of long filters $\vectgr{\theta}_*\in \Real^L$, with the
likely unpleasant side-effect of hindering learning due to
the notorious ``curse of dimensionality.''  State-action
space is defined as
$\mathfrak{Z} \coloneqq \mathfrak{S}\times \mathfrak{A}
\coloneqq \Set{ \vect{z} \coloneqq (\vect{s}, a) \given
  \vect{s}\in \mathfrak{S}, a\in \mathfrak{A}}$.

At time $n$, available to the user are the state-action pair
$(\vect{s}_{n-1}, a_{n-1})$, data
$\mathfrak{D}_{( n-M_{ \textnormal{av} } ):n} \coloneqq
((\vect{x}_{\nu}, y_{\nu}))_{\nu = n-M_{ \textnormal{av}
  }}^{n}$, for some buffer length
$M_{ \textnormal{av} } \in \RealPP$, as well as estimates
$(\vectgr{\theta}_{n}, \vectgr{\theta}_{n-1})$. The new
state $\vect{s}_n$ is defined as
\begin{align}
  \vect{s}_n \coloneqq \vect{s}_{n-1}^{\prime} = [
  s_{n-1}^{\prime (1)}, s_{n-1}^{\prime (2)},
  s_{n-1}^{\prime (3)}, s_{n-1}^{\prime (4)}]^{\intercal}
  \,, \label{def.sn}
\end{align}
where $\vect{s}_{n-1}^{\prime}$ stands for the subsequent
state of $\vect{s}_{n-1}$, which depends on
$( \vect{s}_{n-1}, a_{n-1}, \mathfrak{D}_{( n-M_{
    \textnormal{av} } ):n}, \vectgr{\theta}_{n},
\vectgr{\theta}_{n-1} )$ and is defined by
\begin{subequations}\label{def.states}
  \begin{alignat}{2}
    s_{n-1}^{\prime (1)}
    & {} \coloneqq {}
      && \log\, \lvert e_n \rvert^2 \,, \label{state1}\\
    s_{n-1}^{\prime (2)}
    & \coloneqq
      && \tfrac{1}{M_{\textnormal{av}}}
      \sum_{m=1}^{M_{\textnormal{av}}}
      \log \frac{ \lvert y_{n-m} -
      \vectgr{\theta}_{n}^\intercal(a_{n-1})\,
      \vect{x}_{n-m} \rvert^2 } { \norm{\vect{x}_{n-m} }_2^2
         } \,, \label{state2}\\
    s_{n-1}^{\prime (3)}
    & \coloneqq
      && \log\, \norm{\vect{x}_n}_2 \,, \label{state3}\\
    s_{n-1}^{\prime (4)}
    & \coloneqq
      && \varpi s_{n-1}^{(4)} + (1 - \varpi) \log
         ( \tfrac{1}{\rho}
         \norm{\vectgr{\theta}_{n}(a_{n-1}) -
         \vectgr{\theta}_{n-1}}_2 ) \notag \\
    & =
      && \varpi s_{n-1}^{(4)} + (1 - \varpi)( p_{n-1} - 1 )
         \log \lvert e_{n-1} \rvert \notag \\
    &&& + (1 - \varpi) \log\, \norm{ \vect{x}_{n-1}
        }_2  \notag\\
    &&& + (1 - \varpi) \log p_{n-1} \,, \label{state4}
  \end{alignat}
\end{subequations}
with $\varpi \in (0,1)$ being a user-defined parameter, while
$\rho$ comes from \eqref{dLMP}. The $\log (\cdot)$ function
is employed to decrease the dynamic range of the positive
values in \eqref{def.states}. Any logarithmic function can
be used in~\eqref{def.states}; the $10$-base one is used in
\cref{sec:tests}. The classical prior loss in
AdaFilt~\cite{sayed2011adaptive} is used in \eqref{state1},
an $M_{\textnormal{av}}$-length sliding-window sampling
average of the posterior loss~\cite{sayed2011adaptive} is
provided in \eqref{state2}, normalized by the norm of the
input signal to remove as much as possible its effect on the
error, the instantaneous norm of the input signal in
\eqref{state3}, and a smoothing auto-regressive process in
\eqref{state4} to monitor the consecutive displacement of
the estimates $(\vectgr{\theta}_n)_{n\in\IntegerP}$. The
reason for including $\rho$ in \eqref{state4} is to remove
$\rho$'s effect from $s_4^{(n)}$.

With the estimate $Q_n$ available to the user/agent, policy
improvement in \cref{algo:policy.improvement} is achieved by
the standard greedy rule~\cite{bertsekas2019reinforcement}
\begin{align}
  \mu_{n}(\vect{s}) \coloneqq \arg\min\nolimits_{a\in
  \mathfrak{A}} Q_n( \vect{s}, a) \,, \quad \forall
  \vect{s}\in \mathfrak{S} \,. \label{mu.n}
\end{align}
More specifically, the next action $a_n$ for the agent is
identified by plugging $\vect{s}_n$ in the place of
$\vect{s}$ in~\eqref{mu.n}. Now that
$a_n \coloneqq \mu_{n}(\vect{s}_{n})$ is available,
recursion~\eqref{dLMP} is applied with $p_n \coloneqq a_n$
to \cref{algo:update.theta} of \cref{algo} to obtain the new
estimate $\vectgr{\theta}_{n+1}$.

\subsection{Defining $T_{\mu_n}^{(n)}$ and loss
  $\mathcal{L}_{\mu_n}^{(n)} [ \cdot ]$}%
\label{sec:T.mu.n}

This subsection defines the online version $T_{\mu_n}^{(n)}$
of the B-Map discussed in
\cref{prop:T.inthispaper}. State-action pair
$\vect{z}_{n} = (\vect{s}_{n}, a_{n})$ and stationary policy
$\mu_n$ are now available to the user/agent by the
discussion in \cref{sec:state.action.space}, and therefore,
trajectory samples
$\mathcal{T}_{N[n]}^{(n)} \coloneqq \{ (\vect{s}_i[n],
a_i[n], \vect{s}_i^{\prime}[n]) \}_{i=1}^{N[n]}$ can be
now defined according to \cref{sec:trajectories}.

Let $N_{\textnormal{av}}[n] \coloneqq N[n]$,
$\vect{s}_i^{\textnormal{av}}[n] \coloneqq
\vect{s}_i^{\prime}[n]$,
$\forall i\in \Set{1, \ldots, N[n]}$, and for some
$\sigma\in \RealP$,
\begin{alignat}{2}
  \vectgr{\Psi}[n]
  & {} \coloneqq {}
  && [ \psi_1[n], \ldots, \psi_{N[n]}[n] ]
     \notag\\
  & \coloneqq
  && \vectgr{\Phi}_{\mathcal{T}_{N[n]}^{(n)}} (
     \vect{K}_{\mathcal{T}_{N[n]}^{(n)}} +
     \sigma \vect{I}_{N[n]} )^{-1}
     \,, \label{psi.def-n} \\
  \vectgr{\Phi}_{\mathcal{T}_{N[n]}^{(n)}}
  & \coloneqq
  && [ \varphi(\vect{s}_1[n], a_1[n]), \ldots,
     \varphi(\vect{s}_{N[n]}[n],
     a_{N[n]}[n])] \,, \notag\\
  \boldsymbol{K}_{\mathcal{T}_{N[n]}^{(n)}}
  & \coloneqq
  && \vectgr{\Phi}_{\mathcal{T}_{N[n]}^{(n)}}^{\intercal}
     \vectgr{\Phi}_{\mathcal{T}_{N[n]}^{(n)}}
     \,, \notag\\
  \vectgr{\Phi}_{\mu_n}^{\textnormal{av}}[n]
  & \coloneqq
  && [ \varphi(\vect{s}_1^{\prime}[n],
     \mu_n(\vect{s}_1^{\prime}[n]) ), \notag \\
  &&& \hphantom{[ } \ldots, \varphi(\vect{s}_{N[n]}^{\prime}
      [n], \mu_n(\vect{s}_{N[n]}^{\prime}[n]) ) ] \,, \notag
\end{alignat}
where $\Set{ \mu_n(\vect{s}_i^{\prime}[n]) }_{i=1}^{N[n]}$
are computed by~\eqref{mu.n}. As such,
\eqref{new.Bellman.mu} takes the following form:
\begin{align}
  T_{\mu_n}^{(n)}(Q) \coloneqq g + \alpha \vectgr{\Psi}[n]
  \vectgr{\Phi}_{\mu_n}^{\textnormal{av}\intercal}[n] Q \,,
  \quad \forall Q\in \mathcal{H} \,.
  \label{new.Bellman.mu.online}
\end{align}
Computing $T_{\mu_n}^{(n)}(Q)$ for a given
$Q\in \mathcal{H}$ amounts to identifying
$(T_{\mu_n}^{(n)}(Q))(\vect{s}, a)$ for \textit{all}\/
$\vect{z} = (\vect{s}, a)\in \mathfrak{S} \times
\mathfrak{A}$, which is a computationally infeasible task
given that $\mathfrak{S}$ is the \textit{continuous}\/
$\Real^4$. To surmount this obstacle, this study uses the
point evaluation of $T_{\mu_n}^{(n)}(Q)$ at a
\textit{single}\/ state-action vector
$\vect{z}_{\nu_*} = (\vect{s}_{\nu_*}, a_{\nu_*})$, chosen
by the user from the history of state-action pairs
$\Set{ \vect{z}_{\nu} = (\vect{s}_{\nu}, a_{\nu})
}_{\nu=0}^{n-1}$, that is, $\nu_*\in \Set{ 0, \ldots, n-1}$,
to define the following superset of $\Fix T_{\mu_n}^{(n)}$:
\begin{subequations}\label{H_n}
  \begin{alignat}{2}
    &&& \negphantom{ {} \coloneqq {} }
        H_{\mu_n}^{(n)} [\vect{z}_{\nu_*}] \notag\\
    & {} \coloneqq {}
    && \Set{ Q\in\mathcal{H} \given
       (T_{\mu_n}^{(n)}(Q) - Q)
       (\vect{z}_{\nu_*}) = 0 } \notag \\
    & = && \Set{ Q\in\mathcal{H} \given
           \innerp{T_{\mu_n}^{(n)}(Q) -
           Q}{\varphi(\vect{z}_{\nu_*})}_{\mathcal{H}} = 0
           } \label{H_n.i}\\
    & = && \Set{ Q\in\mathcal{H} \given
           \innerp{Q}{h_{\mu_n}^{(n)} [\vect{z}_{\nu_*}]
           }_{\mathcal{H}} = g(\vect{z}_{\nu_*}) }
           \,, \label{H_n.ii}
  \end{alignat}
\end{subequations}
where~\eqref{H_n.i} follows by the reproducing property,
and~\eqref{H_n.ii} by
incorporating~\eqref{new.Bellman.mu.online} with
\begin{align}
  & h_{\mu_n}^{(n)} [\vect{z}_{\nu_*}] \notag \\
  & \coloneqq \varphi(\vect{z}_{\nu_*}) - \alpha
    \sum \nolimits_{i=1}^{N[n]} (\psi_i[n])
    (\vect{z}_{\nu_*})\, \varphi (\vect{s}_i^{\prime}[n],
    \mu_n (\vect{s}_i^{\prime}[n]) ) \label{def.vec.h}
\end{align}
in~\eqref{H_n.i}. Notice that
$H_{\mu_n}^{(n)} [\vect{z}_{\nu_*}]$ is a hyperplane of
$\mathcal{H}$, with $h_{\mu_n}^{(n)} [\vect{z}_{\nu_*}]$
being its normal vector,
$\varphi(\vect{z}_{\nu_*}) = \kappa(\vect{z}_{\nu_*},
\cdot)$, and $(\psi_i[n]) (\vect{z}_{\nu_*})$ stands for the
value of $\psi_i[n]$ at $\vect{z}_{\nu_*}$. Hyperplane
$H_{\mu_n}^{(n)} [\vect{z}_{\nu_*}]$ is well defined and
non-empty even if $\Fix T_{\mu_n}^{(n)} = \emptyset$. Recall
here that the Banach-Picard fixed-point
theorem~\cite{kreyszig:91} guarantees that
$\Fix T_{\mu_n}^{(n)}$ is non-empty and a singleton in the
case where $T_{\mu_n}^{(n)}$ is a contraction. It is also
worth stressing here that exact knowledge of the one-step
loss $g$, as in $g(\vect{z})$ for \textit{all}\/
$\vect{z}\in \mathfrak{S} \times \mathfrak{A}$, is no longer
necessary since the definition of
$H_{\mu_n}^{(n)} [\vect{z}_{\nu_*}]$ requires only a point
evaluation at $\vect{z}_{\nu_*}$, which, for the present
setting, is set to be
\begin{align}
  g(\vect{z}_{\nu_*}) = g(\vect{s}_{\nu_*}, a_{\nu_*})
  \coloneqq s^{\prime (2)}_{\nu_*} = s^{(2)}_{\nu_* + 1}
  \,, \label{one.step.loss.value}
\end{align}
where the rightmost equality in~\eqref{one.step.loss.value}
follows by~\eqref{def.sn}. Define also the quadratic loss
$\mathcal{L}_{\mu_n}^{(n)} [\vect{z}_{\nu_*}] (\cdot) \colon
\mathcal{H}\to \RealP$ as
\begin{alignat}{2}
  \mathcal{L}_{\mu_n}^{(n)} [ \vect{z}_{\nu_*}] (Q)
  & {} \coloneqq {}
  && \tfrac{1}{2} \innerp{T_{\mu_n}^{(n)}(Q) -
     Q}{\varphi(\vect{z}_{\nu_*})}_{\mathcal{H}}^2 \notag \\
  & = && \tfrac{1}{2} \left[
         \innerp{Q}{h_{\mu_n}^{(n)} [\vect{z}_{\nu_*}]
         }_{\mathcal{H}} - g(\vect{z}_{\nu_*}) \right]^2
         \,. \label{loss.hyperplane}
\end{alignat}
It can be verified by~\eqref{H_n} that
\begin{align*}
  H_{\mu_n}^{(n)} [\vect{z}_{\nu_*}] = \arg\min \nolimits_{Q
  \in \mathcal{H}} \mathcal{L}_{\mu_n}^{(n)}
  [\vect{z}_{\nu_*}] (Q) \,.
\end{align*}

\subsection{Trajectory samples and policy evaluation}%
\label{sec:trajectories}

This subsection details the way the trajectory samples
$\mathcal{T}_{N[n]}^{(n)} \coloneqq \{ (\vect{s}_i[n],
a_i[n], \vect{s}_i^{\prime}[n]) \}_{i=1}^{N[n]}$ are
constructed. Instrumental to the construction is the buffer
$\mathfrak{B}_n \coloneqq \Set{ \mathfrak{b}_j \coloneqq
  (\bar{\vect{s}}_j, \bar{a}_j, g( \bar{\vect{s}}_j,
  \bar{a}_j), \bar{\vect{s}}^{\prime}_j ) }_{j=1}^{\lvert
  \mathfrak{B}_n \rvert}$, where
$(\bar{\vect{s}}_j, \bar{\vect{s}}^{\prime}_j) \in
\mathfrak{S}^2$, $\bar{a}_j \in \mathfrak{A}$, and where
$\bar{\vect{s}}^{\prime}_j$ is determined by
$(\bar{\vect{s}}_j, \bar{a}_j)$. The way to update
$\mathfrak{B}_n$ is provided first, while the design of
trajectory samples follows next by utilizing the strategy of
experience replay~\cite{experiencereplay}.

\subsubsection{Updating $\mathfrak{B}_{n-1}$ to $\mathfrak{B}_n$}

At time $n$, buffer $\mathfrak{B}_{n-1}$ and tuple
$( \vect{s}_{n-1}, a_{n-1}, g(\vect{s}_{n-1}, a_{n-1}),
\vect{s}_n )$ are available to the user. Given a
user-defined distance function
$\dist_{\mathfrak{S}}(\cdot, \cdot) \colon \mathfrak{S}
\times \mathfrak{S} \to \RealP$, for example,
$\dist_{\mathfrak{S}}(\cdot, \cdot) \coloneqq 1 -
\kappa_{\textnormal{G}}(\cdot, \cdot)$, where
$\kappa_{\textnormal{G}}(\cdot, \cdot)$ is a Gaussian
kernel, and a threshold $\delta_{\mathfrak{S}}\in \RealPP$,
consider the following criterion:
\begin{align}
  \dist_{\mathfrak{S}} ( \vect{s}_{n-1}, \bar{\vect{s}}_j )
  {} & {} > \delta_{\mathfrak{S}} \,, \notag \\
  & \forall \mathfrak{b}_{j} = (\bar{\vect{s}}_j, \bar{a}_j,
    g( \bar{\vect{s}}_j, \bar{a}_j),
    \bar{\vect{s}}^{\prime}_j ) \in \mathfrak{B}_{n-1}
    \,. \label{buffer.criterion.i}
\end{align}
If~\eqref{buffer.criterion.i} is satisfied, $\vect{s}_{n-1}$
is considered to be ``different'' enough from \textit{all}\/
states $\bar{\vect{s}}_j$ which appear in the tuples
$\mathfrak{b}_{j}$ of $\mathfrak{B}_{n-1}$, and to carry
``sufficiently novel'' information to be included in
$\mathfrak{B}_{n}$. Consequently, generate all tuples
$\mathfrak{C}_{n-1} \coloneqq \Set{ ( \vect{s}_{n-1}, a,
  g(\vect{s}_{n-1}, a), \vect{s}_{n-1}^{\prime} (
  \vect{s}_{n-1}, a, \mathfrak{D}_{( n-M_{ \textnormal{av} }
    ):n}, \vectgr{\theta}_{n}, \vectgr{\theta}_{n-1} )
  \given a \in \mathfrak{A} }$, where
$\vect{s}_{n-1}^{\prime} ( \vect{s}_{n-1}, a,
\mathfrak{D}_{( n-M_{ \textnormal{av} } ):n},
\vectgr{\theta}_{n}, \vectgr{\theta}_{n-1} )$ is obtained by
replacing $a_{n-1}$ with $a$ in~\eqref{def.states}. Define
then
\begin{align*}
  \mathfrak{B}_n \coloneqq \mathfrak{B}_{n-1} \cup \Set{
  ( \vect{s}_{n-1}, a_{n-1}, g( \vect{s}_{n-1}, a_{n-1} ),
  \vect{s}_n)} \cup \mathfrak{C}_{n-1} \,.
\end{align*}
On the other hand, if~\eqref{buffer.criterion.i} is not
satisfied, then $\mathfrak{B}_n \coloneqq
\mathfrak{B}_{n-1}$.

\subsubsection{Experience replay}%
\label{sec:exp.replay}

Now that buffer $\mathfrak{B}_n$ has been updated and given
a user-defined distance function
$\dist_{\mathfrak{Z}}(\cdot, \cdot) \colon \mathfrak{Z}
\times \mathfrak{Z} \to \RealP$, for example,
$\dist_{\mathfrak{Z}}(\cdot, \cdot) \coloneqq 1 -
\kappa_{\textnormal{G}}(\cdot, \cdot)$, where
$\kappa_{\textnormal{G}}(\cdot, \cdot)$ is a Gaussian
kernel, and a threshold $\delta_{\mathfrak{Z}}\in \RealPP$,
define
\begin{align}
  \mathfrak{T}_{\vect{z}} \coloneqq \{ \mathfrak{b}_j =
  {} & (\bar{\vect{s}}_j, \bar{a}_j, g( \bar{\vect{s}}_j,
       \bar{a}_j), \bar{\vect{s}}^{\prime}_j
       ) \in \mathfrak{B}_n \notag \\
     & \given \dist_{\mathfrak{Z}} ( \vect{z},
       (\bar{\vect{s}}_j, \bar{a}_j ) ) \leq
       \delta_{\mathfrak{Z}} \} \,, \quad \forall\vect{z}\in
       \mathfrak{Z} \,. \label{mathfrakT.z}
\end{align}
In other words, $\mathfrak{T}_{\vect{z}}$ includes all
tuples $\mathfrak{b}_j$ of $\mathfrak{B}_n$ whose
state-action pairs $(\bar{\vect{s}}_j, \bar{a}_j )$ are
``sufficiently similar'' with $\vect{z}$. Identify then
$\mathfrak{T}_{\vect{z}_{n-1}}$ by~\eqref{mathfrakT.z}, and
define the trajectory samples
\begin{alignat*}{2}
  \mathcal{T}_{N[n]}^{(n)}
  & = && \Set{ ( \vect{s}_i[n], a_i[n],
         \vect{s}_i^{\prime}[n] ) }_{i=1}^{N[n]} \\
  & {} \coloneqq {}
      && \Set{ (\bar{\vect{s}}_j, \bar{a}_j,
         \bar{\vect{s}}_j^{\prime}) \given
         \mathfrak{b}_j = (\bar{\vect{s}}_j, \bar{a}_j, g(
         \bar{\vect{s}}_j, \bar{a}_j),
         \bar{\vect{s}}^{\prime}_j ) \in
         \mathfrak{T}_{\vect{z}_{n-1}} } \\
  &&& \cup \Set{ (\vect{s}_{n-1}, a_{n-1},
      \vect{s}_{n-1}^{\prime}) } \,,
\end{alignat*}
where $\vect{s}_{n-1}^{\prime}$ is defined
by~\eqref{def.states}. Consider also $T_{\mu_n}^{(n)}$ and
loss $\mathcal{L}_{\mu_n}^{(n)} [\vect{z}_{n-1}] (\cdot)$ as
in~\eqref{new.Bellman.mu.online}
and~\eqref{loss.hyperplane}, respectively, and apply the SGD
rule, with a learning rate $\eta\in \RealPP$, to form the
update:
\begin{alignat}{3}
  Q_{n+1/2}
  & {} \coloneqq {}
    && Q_n - \eta \nabla
  && \mathcal{L}_{\mu_n}^{(n)} [\vect{z}_{n-1}] (Q_n)
     \notag \\
  & =
    && Q_n - \eta \Bigl[
  && \!\!
     \innerp{Q_n}{h_{\mu_n}^{(n)} [\vect{z}_{n-1}]
     }_{\mathcal{H}} \notag \\
  &&&&& \!\! - g(\vect{z}_{n-1}) \Bigr] \cdot
        h_{\mu_n}^{(n)} [\vect{z}_{n-1}]
        \,, \label{Q.n.to.n+half}
\end{alignat}
where $g(\vect{z}_{n-1})$ is computed
by~\eqref{one.step.loss.value}, with $\vect{z}_{n-1}$ taking
the place of $\vect{z}_{\nu_*}$.

It is worth mentioning here that in the case where the set
$\mathcal{T}_{N[n]}^{(n)}$ of trajectory samples is a
singleton, more specifically,
$\mathcal{T}_{N[n]}^{(n)} = \{ (\vect{s}_{n-1}, a_{n-1},
\vect{s}_{n-1}^{\prime}) \}$, and whenever $\sigma = 0$
in~\eqref{psi.def-n}, then~\eqref{Q.n.to.n+half}
corresponds to~\cite{onlinebrloss:16}.

To exploit also state-action pairs other than
$\vect{z}_{n-1}$, the strategy of experience replay is
adopted to choose a
$\mathfrak{b}_{j_*} \in \mathfrak{B}_n \setminus \Set{
  (\bar{\vect{s}}_j, \bar{a}_j, g( \bar{\vect{s}}_j,
  \bar{a}_j), \bar{\vect{s}}^{\prime}_j ) \in \mathfrak{B}_n
  \given (\bar{\vect{s}}_j, \bar{a}_j) = \vect{z}_{n-1}}$
via a probability distribution, whose details are skipped
here but can be found in~\cite{experiencereplay}. Having
identified such a
$\mathfrak{b}_{j_*} = ( \bar{\vect{s}}_{j_*}, \bar{a}_{j_*},
g( \bar{\vect{s}}_{j_*}, \bar{a}_{j_*}),
\bar{\vect{s}}^{\prime}_{j_*} )$, let
$\bar{\vect{z}}_{j_*} \coloneqq ( \bar{\vect{s}}_{j_*},
\bar{a}_{j_*} )$ and define
$\mathfrak{T}_{\bar{\vect{z}}_{j_*}}$ via
\eqref{mathfrakT.z}. Next, let the trajectory samples
\begin{alignat*}{2}
  &&& \negphantom{ {} \coloneqq {} }
      \mathcal{T}_{N[n+1/2]}^{(n+1/2)} \\
  & = && \Set{ ( \vect{s}_i[n+1/2], a_i[n+1/2],
         \vect{s}_i^{\prime}[n+1/2] ) }_{i=1}^{N[n+1/2]} \\
  & {} \coloneqq {}
      && \Set{ (\bar{\vect{s}}_j, \bar{a}_j,
         \bar{\vect{s}}_j^{\prime}) \given
         \mathfrak{b}_j = (\bar{\vect{s}}_j, \bar{a}_j, g(
         \bar{\vect{s}}_j, \bar{a}_j),
         \bar{\vect{s}}^{\prime}_j ) \in
         \mathfrak{T}_{\bar{\vect{z}}_{j_*}} } \,.
\end{alignat*}
Define also $T_{\mu_n}^{(n+1/2)}$ and loss
$\mathcal{L}_{\mu_n}^{(n+1/2)} [\bar{\vect{z}}_{j_*}]
(\cdot)$ as in~\eqref{new.Bellman.mu.online}
and~\eqref{loss.hyperplane}, respectively, and apply again
the SGD rule to obtain the Q-function estimate of
\cref{algo:policy.evaluation} in \cref{algo}:
\begin{alignat}{3}
  Q_{n+1}
  & {} \coloneqq {}
  && Q_{n+1/2} - \eta
     \nabla
  && \mathcal{L}_{\mu_n}^{(n+1/2)}
     [\bar{\vect{z}}_{j_*}] (Q_{n+1/2}) \notag \\
  & =
  && Q_{n+1/2} - \eta \Bigl[
  && \!\!
     \innerp{Q_{n+1/2}}
     {h_{\mu_n}^{(n+1/2)} [\bar{\vect{z}}_{j_*}]
     }_{\mathcal{H}} \notag \\
  &&&&& \!\! - g(\bar{\vect{z}}_{j_*}) \Bigr] \cdot
        h_{\mu_n}^{(n+1/2)} [\bar{\vect{z}}_{j_*}]
        \,. \label{Q.n+half.to.n+1}
\end{alignat}

\subsection{Dimensionality reduction by random Fourier
  features}%
\label{sec:RFF}

At every time instance $n$, \cref{algo} adds new features
into the representation of $Q_{n+1}$ via \eqref{def.vec.h},
\eqref{Q.n.to.n+half}, and~\eqref{Q.n+half.to.n+1},
justifying the ``nonparametric'' characterization of the
proposed design. These new features contribute information
in $Q_{n+1}$ along novel dimensions of $\mathcal{H}$ which
may have not been explored prior to time $n$. Novel
dimensions are welcome since they lead into a ``rich''
kernel-based representation of the Q-function. However,
due to the potentially infinite dimensionality of
$\mathcal{H}$, the length of the Q-function representation
may grow unbounded as new dimensions/features are added up,
raising in turn hardware/computational obstacles due to the
need for large storage space and large number of
computations to process the long Q-function
representations (``curse of dimensionality''). The desire
for low hardware/computational footprints calls for
dimensionality reduction, which is achieved here by
employing random Fourier features (RFF)~\cite{rff} as
follows.

According to Bochner's theorem, there exist pairs
$(\kappa, \vect{v})$, where $\kappa$ is a real-valued
reproducing kernel and $\vect{v}$ an RV, s.t.\
$\kappa(\vect{z}, \vect{z}^{\prime}) =
\expect_{\vect{v}}\Set{ \cos
  [{\vect{v}^{\intercal}(\vect{z}-\vect{z}^{\prime})} ] }$,
$\forall \vect{z}, \vect{z}^{\prime}\in
\Real^D$~\cite{rff}. An example of such a pair is
$(\kappa_{\textnormal{G}}, \vect{v}_{\textnormal{G}})$,
where $\kappa_{\textnormal{G}}$ is the Gaussian kernel and
$\vect{v}_{\textnormal{G}}$ follows the Gaussian
distribution $\mathcal{N}(0, \vect{I}_D)$. It can be
verified that
$\expect_{\vect{v}, u} \{\cos [\vect{v}^\intercal (\vect{z}
+ \vect{z}^{\prime} ) + 2u] \} = 0$, where $u$ is an RV
uniformly distributed over $[0, 2\pi)$ and independent of
$\vect{v}$. Hence, Bochner's theorem yields:
\begin{alignat}{2}
  \kappa(\vect{z}, \vect{z}^{\prime})
  & {} = {}
  && \expect_{\vect{v}} \{ \cos [\vect{v}^{\intercal}
     (\vect{z} - \vect{z}^{\prime})] \} \notag\\
  & =
  && \expect_{\vect{v}, u}\{ \cos [
     (\vect{v}^{\intercal}\vect{z} + u) -
     (\vect{v}^{\intercal} \vect{z}^{\prime} + u)] \notag\\
  &&& + \cos [ (\vect{v}^{\intercal} \vect{z} + u) +
      (\vect{v}^{\intercal}\vect{z}^{\prime} +
      u) ] \} \notag\\
  & =
  && 2 \expect_{\vect{v}, u} \{ \cos
         (\vect{v}^{\intercal} \vect{z} + u) \cdot
         \cos (\vect{v}^{\intercal} \vect{z}^{\prime} + u)\}
         \notag\\
  & \approx
  && 2 \tfrac{1}{D_{\textnormal{RFF}}}
     \sum\nolimits_{i=1}^{D_{\textnormal{RFF}}}
     \cos (\vect{v}_i^{\intercal} \vect{z} +
     u_i) \cdot \cos (\vect{v}_i^{\intercal}
     \vect{z}^{\prime} + u_i) \notag\\
  & =
  && \varphi_{\textnormal{RFF}}^{\intercal}( \vect{z} )
         \,  \varphi_{\textnormal{RFF}} ( \vect{z}^{\prime}
     ) \,, \label{rff.innerp}
\end{alignat}
where $\approx$ in~\eqref{rff.innerp} holds true by the law
of large numbers~\cite{Ash:Probability} for a large
user-defined number $D_{\textnormal{RFF}}$ of IID copies
$\{ \vect{v}_i \}_{i=1}^{D_{\textnormal{RFF}}}$ and
$\{ u_i \}_{i=1}^{D_{\textnormal{RFF}}}$ of $\vect{v}$ and
$u$, respectively, and $\varphi_{\textnormal{RFF}}$ is the
feature mapping defined as
\begin{align}
  \varphi_{\textnormal{RFF}}
  & \colon \Real^D \to \Real^{D_{\textnormal{RFF}}} \notag\\
  & \colon \vect{z} \mapsto \sqrt{
    \tfrac{2}{D_{\textnormal{RFF}}} } [ \cos
    (\vect{v}_1^{\intercal} \vect{z} + u_1), \ldots, \cos
    (\vect{v}_{D_{\textnormal{RFF}}}^{\intercal} \vect{z} +
    u_{D_{\textnormal{RFF}}})]^{\intercal} \,. \label{rff.phi}
\end{align}
Mapping~\eqref{rff.phi} serves as a low dimensional
rendition of the feature mapping
$\varphi\colon \Real^D \to \mathcal{H}$ introduced in
\cref{sec:notation}, since $D_{\textnormal{RFF}}$ can be
made smaller than the typically large and potentially
infinite $\dim\mathcal{H}$. The feature-mapping
$\varphi_{\textnormal{RFF}}$ and inner-product
$\varphi_{\textnormal{RFF}}^{\intercal}( \vect{z} ) \,
\varphi_{\textnormal{RFF}} ( \vect{z}^{\prime} )$
approximations of $\varphi$ and
$\kappa(\vect{z}, \vect{z}^{\prime})$, respectively, are
used in \cref{algo} to bound the hardware/computational
complexity. The results of \cref{sec:tests} are based on
these approximations.

\section{Performance Analysis of \cref{algo}}%
\label{sec:performance.analysis}

In the following discussion, a statement $( \ldots (n) )$,
which depends on the iteration index $n\in \IntegerP$, will
be said to hold true ``for all sufficiently large $n$,'' if
there exists a large $n_0\in \IntegerPP$ s.t.\
$( \ldots (n) )$ holds true $\forall n \geq n_0$. Moreover,
within the context of a probability space
$(\Omega, \mathcal{F}, \Prob)$ (see the discussion before
\cref{ass:consistency.fix}), a statement $(\ldots)$ will be
said to hold true with high probability (w.h.p.), if
$(\ldots)$ holds true on an event
$\mathcal{E} \in \mathcal{F}$ with
$\Prob( \mathcal{E} ) \geq 1 - \varepsilon$, for a
sufficiently small $\varepsilon \in \RealPP$.

Central to the following discussion are the sequence of
estimates $(Q_n)_{n\in \IntegerP}$ generated by \cref{algo},
the classical B-Maps $T^{\diamond}, T_{\mu_n}^{\diamond}$
in~\eqref{classical.Bellman.maps}, as well as the newly
proposed $T_{\mu_n}^{(n)}$ one in~\eqref{new.Bellman.maps}
for a stationary policy
$\mu_n(\cdot) \colon \mathfrak{S} \to \mathfrak{A}$. Recall
also that $Q_*^{\diamond}$ stands for a fixed point of
mapping $T^{\diamond}$, \ie,
$Q_*^{\diamond} \in \Fix ( T^{\diamond} ) \Leftrightarrow
Q_*^{\diamond} = T^{\diamond} (Q_*^{\diamond})$, that
$Q_{\mu_n}^{\diamond} \in \Fix ( T_{\mu_n}^{\diamond} )$ and
$Q_{\mu_n} \in \Fix ( T_{\mu_n}^{(n)} )$.

\cref{thm:new.Bellman.maps.consistency} asserts that for an
arbitrarily fixed $\epsilon \in \RealPP$ and for any $n$,
$\norm{Q_{\mu_n}^{\diamond} - Q_{\mu_n}(N[n])}_{\mathcal{H}}
\leq \epsilon$ holds true w.h.p.\ for all sufficiently large
$N[n]$. By this result, it is expected that for sufficiently
large $N[n]$, the size of the event
\begin{align}
  E_{n,N[n]}^{(\epsilon)} \coloneqq \Set*{ \omega \in
  \Omega \given \norm{Q_{\mu_n}^{\diamond} -
  Q_{\mu_n}(N[n])}_{\mathcal{H}} \leq \epsilon }
  \,. \label{event.En.epsilon}
\end{align}
can be considered to be large. This observation serves as
the motivation behind the following \cref{ass:liminf.event}.

\begin{assumptions}\label{ass:PI-bound} \mbox{}

  \begin{assslist}

  \item\label{ass:liminf.event} Consider an
    $\epsilon\in \RealPP$ s.t.\
    \begin{align}
      E^{(\epsilon)} \coloneqq \liminf_{n\to \infty}
      \liminf_{N[n] \to \infty} E^{(\epsilon)}_{n, N[n]} \neq
      \emptyset \,, \label{event.E.epsilon}
    \end{align}
    where for events
    $( \mathcal{E}_{\nu})_{\nu \in \IntegerP}$,
    $\liminf_{\nu \to \infty} \mathcal{E}_{\nu} \coloneqq
    \cup_{\nu} \cap_{\nu^{\prime} \geq \nu}
    \mathcal{E}_{\nu^{\prime}}$~\cite{Williams:probability:91}.

  \item Presume \cref{ass:consistency.fix}.

  \item\label{ass:beta.infty.online} In the current online
    setting, \cref{ass:beta.inf} takes the following
    form. There exists $\beta_{\infty} \in (0,1)$ s.t.\
    $\beta_n = \beta_n(N[n]) \leq \beta_{\infty}$, for all
    sufficiently large $n$ and $N[n]$, a.s., where $\beta_n$
    is defined according to \eqref{beta} as
    \begin{align}
      \beta_n \coloneqq \alpha \left( \norm{
      \vect{K}_{\Psi_n} }_2 \,
      \sup\nolimits_{\mu^{\prime}\in\mathcal{M}}
      \norm{\vect{K}^{\textnormal{av}, n}_{\mu^{\prime}}}_2
      \right)^{{1}/{2}} \,. \label{beta.n}
    \end{align}

  \item\label{ass:Qn.close.Qmun} There exists
    $\Delta_0 \in \RealPP$ s.t.\
    $\norm{Q_{\mu_n} - Q_n}_\mathcal{H}\leq \Delta_0$, for
    all sufficiently large $n$, a.s.

  \item\label{ass:Tdiammu.minus.Tdiam} There exists
    $\Delta_1\in \RealPP$ s.t.\
    $\norm{T_{\mu_{n+1}}^{\diamond} (Q_n) - T^{\diamond}
      (Q_n)}_\mathcal{H} \leq \Delta_1$, for all
    sufficiently large $n$, a.s.

  \item\label{ass:Id-Tdiamond.bound} There exists
    $\Delta_2\in \RealPP$ s.t.\
    $\norm{ (T_{\mu_{n+1}}^{\diamond} -
      \Id)(Q_{\mu_n}^{\diamond}) }_\mathcal{H} \leq
    \Delta_2$, for all sufficiently large $n$, a.s.

  \end{assslist}
\end{assumptions}

The following theorem bounds the distance of the sequence of
estimates $(Q_n)_n$ from the fixed point $Q_*^{\diamond}$.

\begin{thm}\label{thm:PI-bound}
  Under \cref{ass:PI-bound}, for every $\omega \in
  E^{(\epsilon)}$ of~\eqref{event.E.epsilon},
  \begin{align*}
    \limsup\nolimits_{n\to\infty}\norm{ Q_n -
    Q_*^{\diamond} }_{\mathcal{H}} \leq \Delta_3 \,,
  \end{align*}
  where
  $\Delta_3 \coloneqq \epsilon + \Delta_0 + [ 2 \beta_\infty
  (\Delta_0 + \epsilon) + \Delta_1 +
  \Delta_2/(1-\beta_\infty) ] / (1-\beta_\infty)$.
\end{thm}

\begin{IEEEproof}
  See \cref{app:PI-bound}.
\end{IEEEproof}

If the size of $E^{(\epsilon)}$ is large, which is something
anticipated by the discussion around
\eqref{event.En.epsilon}, the assertion of
\cref{thm:PI-bound} holds true w.h.p. Instead of the
point-wise, sample-point-based analysis of
\cref{thm:PI-bound}, the following \cref{thm:evaluation}
provides a bound on the sequence $(Q_n)_n$ via the
expectation operator $\expect\{ \cdot \}$. To this end, the
following assumptions are necessary.

\begin{assumptions}\label{assumption:evaluation}\mbox{}
  \begin{assslist}

  \item\label{ass:no.exp.replay} For simplicity, policy
    evaluation in \cref{algo:policy.evaluation} of
    \cref{algo} is performed without considering experience
    replay~\eqref{Q.n+half.to.n+1}, that is,
    $Q_{n+1} \coloneqq Q_n - \eta \nabla
    \mathcal{L}_{\mu_n}^{(n)} [\vect{z}_{n-1}] (Q_{n})$.

  \item\label{ass:stationary.policy} There exists a policy
    $\mu \colon \mathfrak{S} \to \mathfrak{A}$ s.t.\
    $\mu_n = \mu$ for all sufficiently large $n$.

  \item\label{ass:T.mu.diamond.contraction} There exists
    $\beta_{\infty} \in (0,1)$ s.t.\ mapping
    $T_{\mu_n}^{\diamond}$ is a $\beta_{\infty}$-contraction
    for all sufficiently large $n$, a.s.

  \item\label{ass:independency} \textbf{(Independency)} The
    $\sigma$-algebra $\sigma( \{\vect{z}_n, \xi_{n+1}\} )$,
    generated by the state-action pair $\vect{z}_n$ and
    $\xi_{n+1}$~\eqref{xi.def}, is independent of the
    filtration $\mathcal{F}_n$ for all sufficiently large
    $n$, where
    $\mathcal{F}_n \coloneqq \sigma( \{ Q_{\nu} \}_{\nu=0}^n
    )$ is defined as the $\sigma$-algebra generated by the
    sequence of estimates
    $\{ Q_{\nu}
    \}_{\nu=0}^n$~\cite{Williams:probability:91}.

  \item\label{ass:stationary.moments} \textbf{(Stationary
      moment)} There exists
    $\mathfrak{m}_{\xi}^{(4)} \in \RealPP$ s.t.\
    $\mathfrak{m}_{\xi}^{(4)} = \expect \{
    \norm{\xi_n}_{\mathcal{H}}^4 \}$, for all sufficiently
    large $n$, where $\xi_n$ is defined by~\eqref{xi.def}.

  \item \textbf{(Stationary covariance operators)} There
    exist bounded linear operators
    $\Sigma_{zz}, \Sigma_{\xi z}, \Sigma_{\xi \xi} \colon
    \mathcal{H} \to \mathcal{H}$ s.t.\
    $\Sigma_{zz}^{(n)} = \Sigma_{zz}, \Sigma_{\xi z}^{(n)} =
    \Sigma_{\xi z}$, and
    $\Sigma_{\xi \xi}^{(n)} = \Sigma_{\xi \xi}$, for all
    sufficiently large $n$, where
    $\Sigma_{zz}^{(n)}, \Sigma_{\xi z}^{(n)}, \Sigma_{\xi
      \xi}^{(n)}$ are defined
    by~\eqref{prop:E.loss.covariace.maps}.

  \item\label{ass:positive.A.Sigma} \textbf{(Positive
      definite $\mathcal{A}_{\mu_n}, \Sigma_{zz}$)} For all
    sufficiently large $n$, the linear bounded and
    self-adjoint mappings
    $\mathcal{A}_{\mu_n}$~\eqref{def.map.A} and
    $\Sigma_{zz}$ are positive definite, \ie, their minimum
    spectral values
    $\sigma_{\min} ( \mathcal{A}_{\mu_n} ) > 0$ and
    $\sigma_{\min} ( \Sigma_{zz} ) > 0$, where
    $\sigma_{\min}(\cdot)$ is defined
    by~\eqref{sigma.min.def}.

  \item\textbf{(Bounded kernel)} There exists
    $B_{\kappa} \in \RealPP$ s.t.\
    $\kappa( \vect{z}, \vect{z} ) \leq B_{\kappa}$,
    $\forall \vect{z} \in \mathfrak{Z}$.

  \end{assslist}
\end{assumptions}

\begin{thm}\label{thm:evaluation}
  Under \cref{assumption:evaluation}, for any sufficiently
  small step size $\eta$, there exist
  $\Delta_4, \Delta_5 \in \RealPP$ s.t.\
  \begin{alignat*}{2}
    &&& \negphantom{ {} \leq {} } \limsup
        \nolimits_{n\to\infty} \expect \{ \norm{ Q_n -
        {Q}_*^{\diamond} }_{\mathcal{H}}^2 \} \\
    & {} \leq {}
      && \overbrace{ \Delta_4 \eta }^{ \textnormal{T}_1 } +
         \overbrace{ \Delta_5 \limsup\nolimits_{n\to \infty}
         \expect\{ \norm{ \Sigma_{ s^{\prime} \given z
         }^{\mu_n} - \hat{\Sigma}_{ s^{\prime} \given z
         }^{\mu_n}( N[n] ) }^2 \} }^{ \textnormal{T}_2 } \\
    &&& + \underbrace{ 2 \limsup \nolimits_{
        n\to\infty } \, \expect\{ \norm{
        Q_{\mu_n}^{\diamond} - Q_*^{\diamond}
        }_{\mathcal{H}}^2 \} }_{ \textnormal{T}_3 } \,,
  \end{alignat*}
  where operator
  $\Sigma_{ s^{\prime} \given z }^{\mu_n}$ is defined
  by~\eqref{Sigma.conditional.map} and $\hat{\Sigma}_{
    s^{\prime} \given z }^{\mu_n}( N[n] )$ is defined by
  \begin{alignat*}{2}
    \hat{\Sigma}_{
    s^{\prime} \given z }^{\mu_n}( N[n] )
    & {} \coloneqq {}
    && \tfrac{1}{ \sqrt{N[n]} } \vectgr{\Phi}_{
       \mathcal{T}_{N[n]} } ( \tfrac{1}{N[n]} \vect{K}_{
       \mathcal{T}_{N[n]} }
       + \sigma^{\prime}_{N[n]} \vect{I}_{N[n]} )^{-1} \\
    &&& \cdot \tfrac{1}{ \sqrt{N[n]} }
        \vectgr{\Phi}_{\mu}^{\textnormal{av}\intercal} \,;
  \end{alignat*}
  see also \eqref{hat.classical.Bellman.mu.alternative},
  \cref{prop:T.inthispaper}, and \cref{ass:T.inthispaper}.
\end{thm}

\begin{IEEEproof}
  See \cref{app:thm:evaluation}.
\end{IEEEproof}

Three terms
$\textnormal{T}_1, \textnormal{T}_2, \textnormal{T}_3$
contribute to the bound in \cref{thm:evaluation}:
$\textnormal{T}_1$ which stems from the
stochastic-gradient-descent recursion of
\cref{ass:no.exp.replay}, $\textnormal{T}_2$ because of the
error in approximating the expectation operator by
trajectory sampling (\cref{sec:trajectories}), and
$\textnormal{T}_3$ which quantifies the disagreement between
$\mu_n$ and the ``optimal'' policy through the lenses of the
classical B-Maps~\eqref{classical.Bellman.maps}. It is worth
mentioning here that under \cref{ass:consistency.fix} and
according to \cref{thm:consistency.Gamma} in
\cref{app:thm:new.Bellman.maps.consistency},
$\norm{ \Sigma_{ s^{\prime} \given z }^{\mu_n} -
  \hat{\Sigma}_{ s^{\prime} \given z }^{\mu_n}( N[n] ) }$
can be made arbitrarily small w.h.p.\ for sufficiently large
$N[n]$.

To establish bounds on the distance
$\norm{ \vectgr{\theta}_n - \vectgr{\theta}_* }$ between the
estimate $\vectgr{\theta}$ and the estimandum
$\vectgr{\theta}_*$ (see \cref{sec:RobAdaFilt}), standard
arguments from the analysis of stochastic gradient descent
(SGD) can be applied, because~\eqref{dLMP} is nothing but
SGD on the convex $p$-power loss ($p\in [1,2]$). For example,
the discussion of~\cite[\S8.2]{BertsekasConvexAnalysis} can
be followed by employing also the minimizer of the $p$-power
loss as an auxiliary quantity to facilitate the
analysis. Because of space limitations, rather than such a
straightforward but tedious performance analysis, this study
prefers to offer \cref{thm:optimal.regret} instead, which
showcases the connection between
$\norm{ \vectgr{\theta}_n - \vectgr{\theta}_* }$ and the
main mathematical object of this work, the Q-functions. To
this end, \cref{assumption:optimal.regret} are needed.

\begin{assumptions}\label{assumption:optimal.regret}\mbox{}

  \begin{assslist}

  \item Presume \cref{ass:stationary.policy}.

  \item Presume \cref{ass:T.mu.diamond.contraction}.

  \item Let $\alpha\in (0,1)$.

  \item \textbf{(Bounded errors)} There exist
    $\Delta_6, \Delta_7 \in \RealPP$ s.t.\
    \begin{align*}
      \Delta_6 < \lvert y_{n-m} -
      \vectgr{\theta}_{n}^{\intercal} \vect{x}_{n-m}
      \rvert < \Delta_7 \,,
    \end{align*}
    $\forall m\in \{ 0, \ldots, M_{\textnormal{av}}-1 \}$
    and for all sufficiently large $n$, a.s.

  \item\label{ass:independency.signals}
    \textbf{(Independency)} The input-signal $\vect{x}_n$ is
    independent of the outlier/noise $o_n$ for all
    sufficiently large $n$. Moreover, the $\sigma$-algebra
    generated by
    $\{ \vect{x}_m \given m\in \{ n - M_{\textnormal{av}} +
    1, \ldots, n\} \}$ and
    $\{ o_m \given m\in \{ n - M_{\textnormal{av}} + 1,
    \ldots, n\} \}$ is independent of the $\sigma$-algebra
    generated by $\vectgr{\theta}_n$ for all sufficiently
    large $n$.

  \item \textbf{(Stationary $o_n$)} There exists $\sigma_o
    \in \RealPP$ s.t.\ $\expect \{ o_n^2 \} = \sigma_o^2$
    and $\expect\{ o_n \} = 0$, for all sufficiently large
    $n$.

  \item\label{ass:stationary.x} \textbf{(Stationary
      $\vect{x}_n$)} There exists a positive definite matrix
    $\Sigma_{xx} \in \Real^{L\times L}$, with
    $\sigma_{\min}( \Sigma_{xx} ) = \lambda_{\min}(
    \Sigma_{xx} ) > 0$ and where
    $\lambda_{\min}( \Sigma_{xx} )$ stands for the minimum
    eigenvalue of $\Sigma_{xx}$, s.t.\
    $\expect\{ \vect{x}_n \vect{x}_n^{\intercal} \} =
    \Sigma_{xx}$, for all sufficiently large $n$. Moreover,
    $\liminf_{n\to\infty} \expect \{ \log\, \norm{
      \vect{x}_n }_2^2 \} > -\infty$.

  \item\label{ass:expect.Qmu.upper.bound} For the sequence
    of states $( \vect{s}_n )_n$ in \cref{algo},
    $\limsup_{n\to \infty} \lvert \expect \{
    Q_{\mu}^{\diamond} ( \vect{s}_n, \mu( \vect{s}_n ) ) \}
    \rvert < +\infty$.

  \end{assslist}

\end{assumptions}

\begin{thm}\label{thm:optimal.regret} Under
  \cref{assumption:optimal.regret}, there exist $\Delta_8\in
  \RealPP$ and $\Delta_9 \in \RealP$ s.t.\ for all
  sufficiently large $n$,
  \begin{alignat*}{2}
    &&& \negphantom{ {} \leq {} } \expect\{ \norm{
        \vectgr{\theta}_* - \vectgr{\theta}_{n} (\mu(
        \vect{s}_{n-1} )) }_2^2 \} \\
    & {} \leq {}
      && \frac{1 - \alpha}{\Delta_8 \lambda_{\min}
         (\Sigma_{xx})} \expect \{
         Q_{\mu}^{\diamond}( \vect{s}_{n},
         \mu(\vect{s}_{n}) ) \} \\
    &&& + \frac{1}{\Delta_8 \lambda_{\min}
        (\Sigma_{xx})} [ \log \trace ( \Sigma_{xx} )  -
        \Delta_8 \sigma_o^2 - \Delta_9 ] \,.
  \end{alignat*}
\end{thm}

\begin{IEEEproof}
  See \cref{app:optimal.regret}.
\end{IEEEproof}

Under the light of \cref{thm:optimal.regret}, the greedy
rule~\eqref{mu.n} makes now sense in the context of AdaFilt,
because, choosing a policy by the greedy rule
$\arg\min_{ \mu(\cdot) \in \mathcal{M} } \expect \{
Q_{\mu}^{\diamond}( \vect{s}_{n}, \mu(\vect{s}_{n}) ) \}$
pushes the upper bound of \cref{thm:optimal.regret} to lower
levels, and, thus, potentially forces the estimate
$\vectgr{\theta}_n$ to approach $\vectgr{\theta}_*$.

\section{Numerical Tests}\label{sec:tests}

\cref{algo} is tested on synthetic data against

\begin{enumerate}

\item \textbf{Non-RL-based methods:}

  \begin{enumerate} [label=\textbf{(\roman*)}]

  \item LMP~\eqref{LMP}, for values of
    $p \in \mathfrak{A} \coloneqq \{1, 1.25, 1.5, 1.75, 2\}$
    which are kept fixed throughout all iterations;

  \item \cite{vazquez2012}, which uses a combination of
    adaptive filters with different forgetting factors but
    with the same $p$-power loss;

  \item \cite{chambers1997robust}, where two LMP
    recursions~\eqref{LMP}, with different $p$, are combined
    to tackle outliers;

  \item the variable-kernel-width and correntropy-based
    VKW-MCC~\cite{huang2017adaptive};

  \end{enumerate}

\item \textbf{RL-based methods:}

  \begin{enumerate} [label=\textbf{(\roman*)}, start = 5]

  \item the kernel-based TD(0)~\cite{Bae:kerneltd1:11},
    equipped with RFF (\cref{sec:RFF});

  \item the popular kernel (K)LSPI~\cite{xu2007klspi}; and

  \item the predecessor~\cite{minh:icassp23} of this work
    which is based on RKHS arguments.

  \end{enumerate}

\end{enumerate}

Action space is defined as
$\mathfrak{A} \coloneqq \{1, 1.25, 1.5, 1.75, 2\}$ for all
employed RL methods, including \cref{algo}. Not only
\cref{algo}, but all tested RL methods other
than~\cite{minh:icassp23} define state vectors
by~\eqref{def.states} and are equipped with experience
replay (\cref{sec:exp.replay}). On the other hand,
\cite{minh:icassp23} defines the state space as
$\mathfrak{S}\subset\Real^{2L+1}$ and utilizes no experience
replay, but uses rollout
instead~\cite{bertsekas2019reinforcement}. Notice also
according to the discussion which
follows~\eqref{Q.n.to.n+half} that \cref{algo} with only one
trajectory sample, \ie, $N[n] = 1$ per $n$, corresponds
to~\cite{onlinebrloss:16}. Nevertheless, \cref{algo} is
equipped with RFF (\cref{sec:RFF}) which is not available
in~\cite{onlinebrloss:16}.

The performance metric is the normalized deviation of the
estimate $\vectgr{\theta}_n$ from the estimandum
$\vectgr{\theta}_*$; see the vertical axes in all
figures. The classical Gaussian
kernel~\cite{scholkopf2002learning} was employed to define
$\mathcal{H}$, approximated by RFF with
$D_{\textnormal{RFF}} = 500$ as described in
\cref{sec:RFF}. The dimension of
$\vect{x}_n, \vectgr{\theta}_*$ is set to $L = 100$, where
$\vect{x}_n$ and $\vectgr{\theta}_*$ are generated by the
Gaussian distribution $\mathcal{N}(\vect{0}, \vect{I}_{L})$,
with $(\vect{x}_n)_{n \in \IntegerP}$ designed to be
IID. The learning rate in~\eqref{dLMP} is $\rho =
10^{-3}$. Moreover, $M_{\textnormal{av}} = 300$ and
$\varpi = 0.3$ in~\eqref{def.states}, while $\eta = 0.1$ in
\eqref{Q.n.to.n+half} and
\eqref{Q.n+half.to.n+1}. In~\eqref{buffer.criterion.i},
$\delta_{\mathfrak{S}} = 1 - 0.99$.

To study the effect of sampling size $N[n]$, controlled by
$\delta_{\mathfrak{Z}}$ in \eqref{mathfrakT.z}, the
following sets of parameters were tested:
\begin{enumerate*}[label=\textbf{(\roman*)}]
\item $\delta_{\mathfrak{Z}} = 1 - 0.98$ in
  \eqref{mathfrakT.z}, which yields $N[n] \geq 1$,
  and $\sigma = 10^{-1}$ in \eqref{psi.def-n}; and
\item $\delta_{\mathfrak{Z}} = 0$, which forces $N[n] = 1$,
  and $\sigma = 0$, corresponding thus
  to~\cite{onlinebrloss:16}.
\end{enumerate*}
Additionally, to study the effect of the long-term loss,
$\alpha = 0$ was also tested. Note that $\alpha = 0$
suggests that \cref{algo} uses no trajectory samples.

In \cref{algo}, policy improvement and evaluation are
scheduled to be run at every iteration $n$. Nevertheless, to
promote stability and allow the policy-iteration step reach
a ``steady state'' between two consecutive invocations of
the policy-improvement step, a less greedy approach is
followed here and \cref{algo:policy.improvement} of
\cref{algo} is not run at every $n$, but it is invoked
periodically, every other $N_p = 500$ iterations, \ie, at
iterations $\Set{ n = N_pk \given k\in\IntegerP }$. Between
two consecutive policy-improvement steps, that is, during
iterations $\{ N_pk, \ldots, N_p(k+1) - 1 \}$, the policy
stays fixed to $\mu_{ N_pk } (\cdot)$. The same strategy is
also followed for KLSPI~\cite{xu2007klspi}.

KLSPI~\cite{xu2007klspi} was originally designed to generate
trajectory samples offline by using training data. However,
since this work considers the online setting, where no
training data are available and only test data are
considered, and to ensure fairness among all competing
methods, matrix $\vect{A}$ and vector $\vect{b}$, which
appear in~\cite{xu2007klspi} and are learned from training
data, are substituted by the following online versions
$\vect{A}_n$ and $\vect{b}_n$: upon defining
$\vect{k}_n \coloneqq \vectgr{\Phi}_{ \textnormal{KLSPI}
}^{(n) \intercal}\, \varphi ( \vect{s}_n, \mu_n(\vect{s}_n)
)$, where $\vectgr{\Phi}_{ \textnormal{KLSPI} }^{(n)}$ is a
time-varying basis of RKHS vectors, constructed online by an
approximate-linear-dependency criterion~\cite{xu2007klspi},
let
\begin{align*}
  \vect{A}_{n}
  & \coloneqq \vect{A}_{n-1} + \vect{k}_{n-1} (
    \vect{k}_{n-1} - \alpha \vect{k}_{n} )^{\intercal} \,, \\
  \vect{b}_{n}
  & \coloneqq \vect{b}_{n-1} + g( \vect{s}_n,
    \mu_n(\vect{s}_n) )\, \vect{k}_{n} \,,
\end{align*}
where
$\vect{A}_1 \coloneqq \vect{k}_0 ( \vect{k}_0 -
\alpha \vect{k}_1 )^{\intercal}$ and
$\vect{b}_0 \coloneqq g( \vect{s}_0, \mu_0(\vect{s}_0) )\,
\vect{k}_0$.


Two types of outliers are considered. Heavy-tailed
$\alpha$-stable outliers, generated
by~\cite{miotto2016pylevy} with parameters
$\alpha_{\textnormal{stable}} = 1$,
$\beta_{\textnormal{stable}} = 0.5$,
$\sigma_{\textnormal{stable}} = 1$, and ``sparse'' outliers
which appear in $10\%$ of the data, while Gaussian noise
with $\textnormal{SNR} = 30\textnormal{dB}$ is added to the
rest $90\%$. Sparse outliers are generated by the uniform
distribution and take values from the interval
$[-100, 100]$.

All methods were finely tuned to perform their best for
every considered setting of the environment. All curves in
the subsequent figures are the uniformly averaged results of
\num{100} independent tests.

\subsection{Scenario 1}\label{sec:sce1}

\begin{figure}[t]

  \centering

  \subfigure[$\alpha$-stable outliers]{ \includegraphics[
    width = .45\columnwidth]{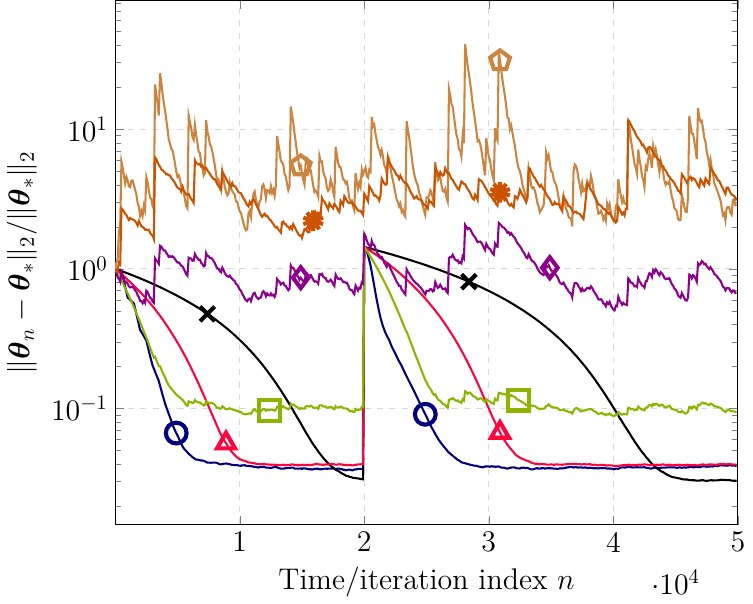}\label{1-1} }%
  \subfigure[Sparse outliers]{ \includegraphics[ width =
    .45\columnwidth]{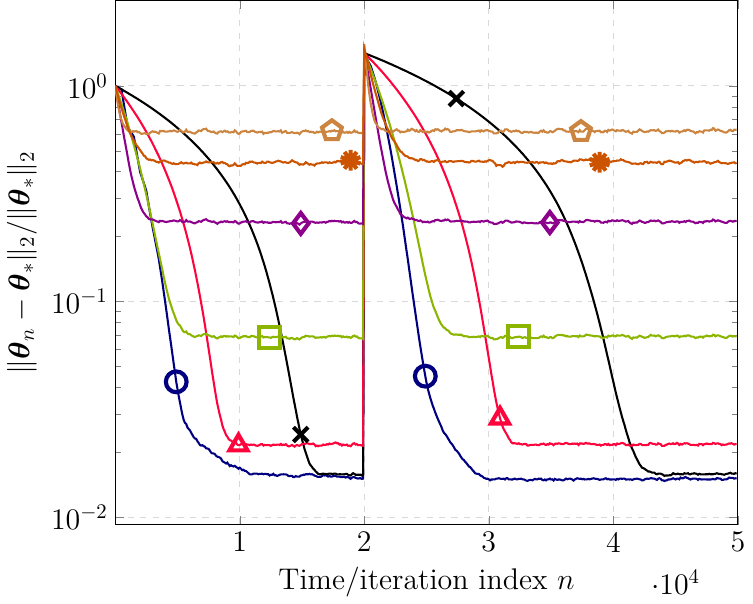}\label{1-2} }

  \caption{Scenario~1 (\cref{sec:sce1}): \cref{algo} against
    LMP. \protect\tikz[baseline = -0.5ex]{
      \protect\node[mark size = 3pt, color = navy, line
      width = .5pt ] {\protect\pgfuseplotmark{o}}}:
    \cref{algo} with $\alpha=0.9, N[n]\geq 1$. Marks
    \protect\tikz[baseline = -0.5ex]{ \protect\node[mark
      size = 3pt, color = black, line width = .5pt ]
      {\protect\pgfuseplotmark{x}}}, \protect\tikz[baseline
    = -0.5ex]{ \protect\node[mark size = 3pt, color =
      americanrose, line width = .5pt ]
      {\protect\pgfuseplotmark{triangle}}},
    \protect\tikz[baseline = -0.5ex]{ \protect\node[mark
      size = 3pt, color = applegreen, line width = .5pt ]
      {\protect\pgfuseplotmark{square}}},
    \protect\tikz[baseline = -0.5ex]{ \protect\node[mark
      size = 3pt, color = darkmagenta, line width = .5pt ]
      {\protect\pgfuseplotmark{diamond}}},
    \protect\tikz[baseline = -0.5ex]{ \protect\node[mark
      size = 3pt, color = peru, line width = .5pt ]
      {\protect\pgfuseplotmark{pentagon}}} correspond to
    \eqref{LMP} with $p=1, 1.25, 1.5, 1.75, 2$,
    respectively. Mark \protect\tikz[baseline = -0.5ex]{
      \protect\node[mark size = 3pt, color = burntorange,
      line width = .5pt ]
      {\protect\pgfuseplotmark{10-pointed star}}} denotes an
    algorithm which randomly chooses $p$, $\forall n$.}%
  \label{fig:vs.LMP}

\end{figure}

\begin{figure}[t]

  \centering

  \subfigure[$\alpha$-stable
  outliers]{ \includegraphics[width =
    .45\columnwidth]{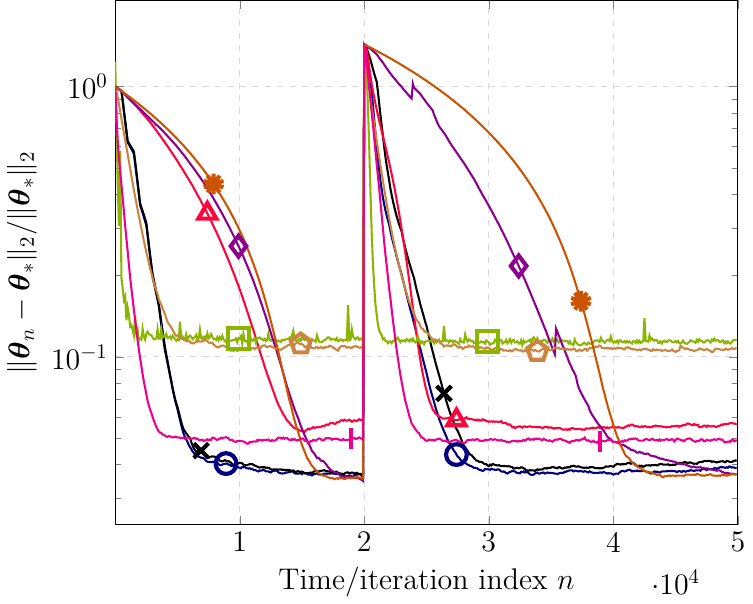}\label{3-1} }%
  \subfigure[Sparse outliers]{\includegraphics[width =
    .45\columnwidth]{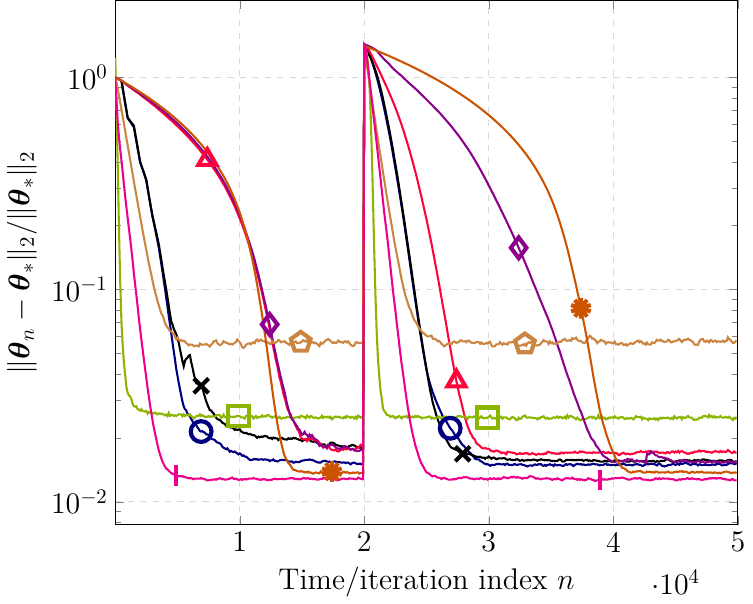}\label{3-2}}

  \caption{Scenario~1 (\cref{sec:sce1}): \cref{algo} against
    non-RL- and RL-based methods. \protect\tikz[baseline =
    -0.5ex]{ \protect\node[mark size = 3pt, color = navy,
      line width = .5pt ] {\protect\pgfuseplotmark{o}}}:
    \cref{algo} with $\alpha=0.9, N[n] \geq
    1$. \protect\tikz[baseline = -0.5ex]{ \protect\node[mark
      size = 3pt, color = black, line width = .5pt ]
      {\protect\pgfuseplotmark{x}}}: \cref{algo} with
    $\alpha=0.9, N[n] = 1$. \protect\tikz[baseline =
    -0.5ex]{ \protect\node[mark size = 3pt, color =
      americanrose, line width = .5pt ]
      {\protect\pgfuseplotmark{triangle}}}: Kernel-based
    TD(0) with $\alpha =
    0.9$~\cite{Bae:kerneltd1:11}. \protect\tikz[baseline =
    -0.5ex]{ \protect\node[mark size = 3pt, color =
      applegreen, line width = .5pt ]
      {\protect\pgfuseplotmark{square}}}: \cite{vazquez2012}
    with $p=1, \gamma_1 = 0.9, \gamma_2 =
    0.99$. \protect\tikz[baseline = -0.5ex]{
      \protect\node[mark size = 3pt, color = darkmagenta,
      line width = .5pt ]
      {\protect\pgfuseplotmark{diamond}}}: KLSPI with
    $\alpha=0.9$~\cite{xu2007klspi}. \protect\tikz[baseline
    = -0.5ex]{ \protect\node[mark size = 3pt, color = peru,
      line width = .5pt ]
      {\protect\pgfuseplotmark{pentagon}}}: mixed
    norm~\cite{chambers1997robust}. \protect\tikz[baseline =
    -0.5ex]{ \protect\node[mark size = 3pt, color =
      burntorange, line width = .5pt ]
      {\protect\pgfuseplotmark{10-pointed star}}}: the
    predecessor~\cite{minh:icassp23} of the current
    work. \protect\tikz[baseline = -0.5ex]{
      \protect\node[mark size = 3pt, color = magenta, line
      width = .5pt ] {\protect\pgfuseplotmark{|}}}:
    VKW-MCC~\cite{huang2017adaptive}.}%
  \label{fig:vs.TD.KLSPI}

\end{figure}

\begin{figure}[t]

  \centering

  \subfigure[$\alpha$-stable
  outliers]{\includegraphics[width =
    .45\columnwidth]{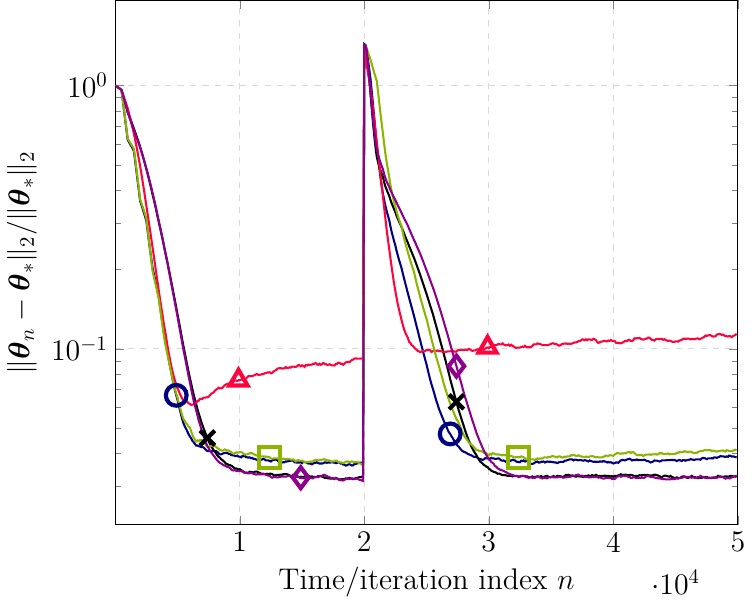}\label{2-1}}%
  \subfigure[Sparse outliers]{\includegraphics[width =
    .45\columnwidth]{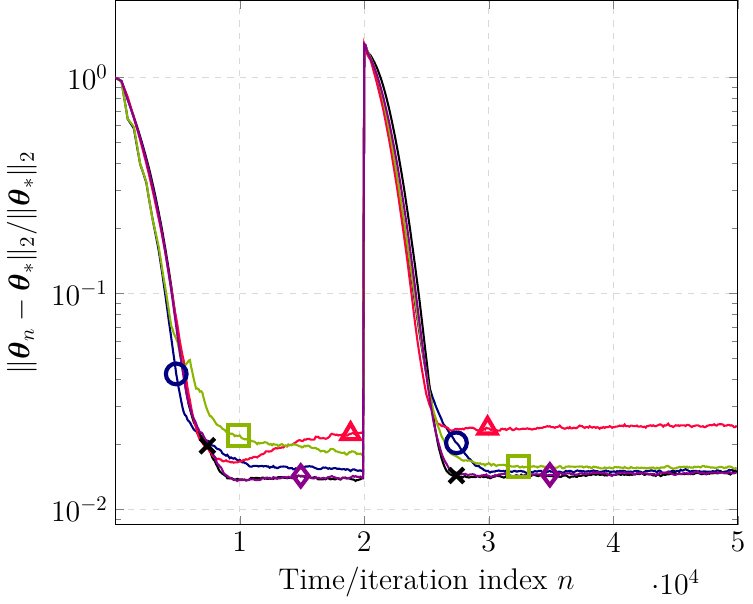}\label{2-2}}

  \caption{Scenario~1 (\cref{sec:sce1}): Versions of
    \cref{algo} under several parameter
    settings. \protect\tikz[baseline = -0.5ex]{
      \protect\node[mark size = 3pt, color = navy, line
      width = .5pt ] {\protect\pgfuseplotmark{o}}}:
    $\alpha = 0.9, N[n]\geq 1$. \protect\tikz[baseline =
    -0.5ex]{ \protect\node[mark size = 3pt, color = black,
      line width = .5pt ] {\protect\pgfuseplotmark{x}}}:
    $\alpha=0.75, N[n]\geq 1$. \protect\tikz[baseline =
    -0.5ex]{ \protect\node[mark size = 3pt, color =
      americanrose, line width = .5pt ]
      {\protect\pgfuseplotmark{triangle}}}: $\alpha = 0$.
    \protect\tikz[baseline = -0.5ex]{ \protect\node[mark
      size = 3pt, color = applegreen, line width = .5pt ]
      {\protect\pgfuseplotmark{square}}}:
    $\alpha=0.9, N[n]=1$.  \protect\tikz[baseline = -0.5ex]{
      \protect\node[mark size = 3pt, color = darkmagenta,
      line width = .5pt ]
      {\protect\pgfuseplotmark{diamond}}}:
    $\alpha=0.75, N[n]=1$.}\label{fig:vs.params}

\end{figure}

\Cref{fig:vs.LMP,fig:vs.TD.KLSPI,fig:vs.params} refer to the
scenario where the statistics (PDF) of the outliers stay
fixed throughout all iterations. Moreover, as it is
customary in the AdaFilt literature, system
$\vectgr{\theta}_*$ changes randomly at a specific time
index (here at time \#\num{20000}) to test the tracking
ability of all competing methods.

\Cref{fig:vs.LMP,fig:vs.TD.KLSPI,fig:vs.params} demonstrate
that \cref{algo} shows high estimation accuracy while
tracking swiftly the estimandum $\vectgr{\theta}_*$. In
\cref{fig:vs.LMP}, \cref{algo} reaches steady state faster
than the ``best'' versions of LMP ($p=1, 1.25$). An
inspection of \cref{fig:vs.LMP} suggests that \cref{algo}
selects large values of $p$, within
$\{1, 1.25, 1.5, 1.75, 2\}$, in the beginning state of
learning to speed up convergence and then changes to select
small values of $p$ to score high accuracy in the steady
state.

\cref{fig:vs.TD.KLSPI} shows that \cref{algo} outperforms
the kernel-based TD(0)~\cite{Bae:kerneltd1:11} and
KLSPI~\cite{xu2007klspi}. The
predecessor~\cite{minh:icassp23} scores an almost identical
steady-state performance with \cref{algo}, but with a slower
convergence speed. VKW-MCC~\cite{huang2017adaptive} shows
excellent performance under sparse outliers. However, under
$\alpha$-stable outliers, it is outperformed by \cref{algo}.

In \cref{fig:vs.params}, where several parameters of
\cref{algo} are validated, $\alpha = 0.75$ shows the best
estimation accuracy among $\alpha \in \{ 0, 0.75, 0.9
\}$. Among them, $\alpha = 0$ does not perform well, which
strongly suggests that the long-term loss is crucial in
choosing $p$. The number of samples $N[n]$ does not seem to
affect performance significantly in this specific AdaFilt
application.

\subsection{Scenario 2}\label{sec:sce2}

\begin{figure}[t]

  \centering

  \subfigure[$\alpha$-stable $\to$
  sparse]{ \includegraphics[width =
    .45\columnwidth]{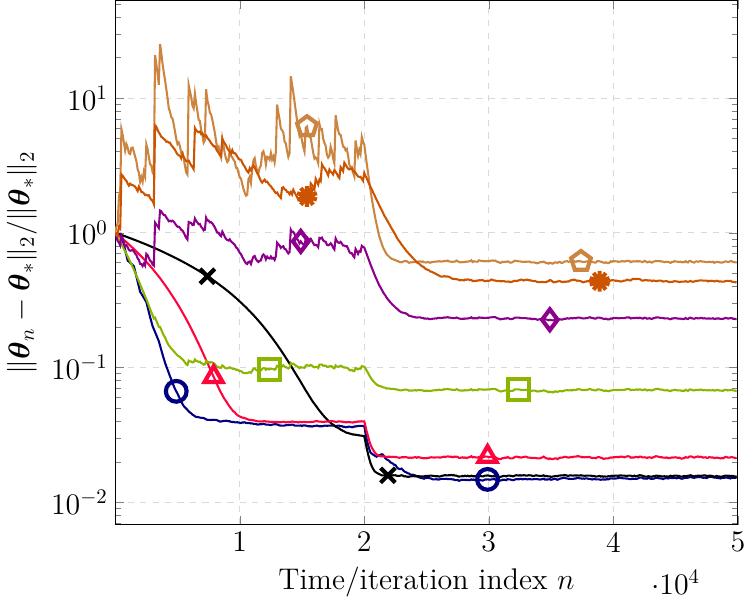}\label{4-1}}%
  \subfigure[Sparse $\to \alpha$-stable]{ \includegraphics[
    width = .45\columnwidth]{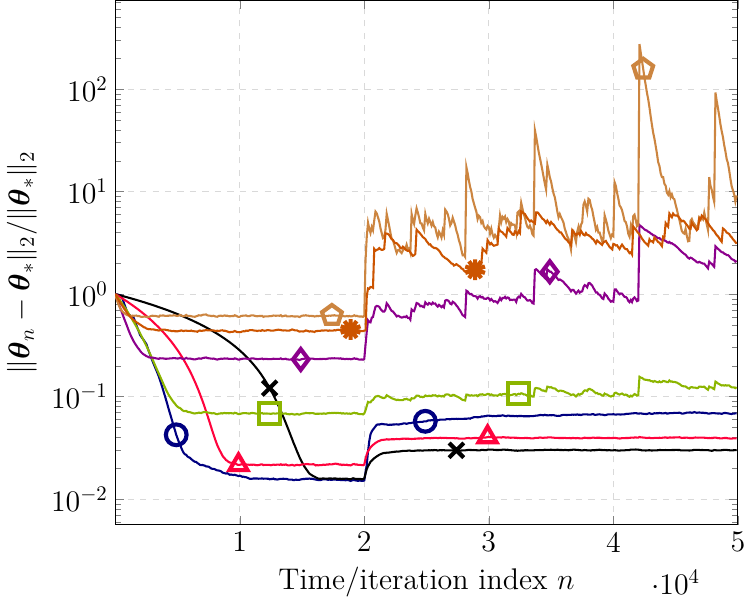}\label{4-2} }

  \caption{Scenario~2 (\cref{sec:sce2}): \cref{algo} against
    LMP. \protect\tikz[baseline = -0.5ex]{
      \protect\node[mark size = 3pt, color = navy, line
      width = .5pt ] {\protect\pgfuseplotmark{o}}}:
    \cref{algo} with $\alpha=0.9, N[n]\geq1$. Marks
    \protect\tikz[baseline = -0.5ex]{ \protect\node[mark
      size = 3pt, color = black, line width = .5pt ]
      {\protect\pgfuseplotmark{x}}}, \protect\tikz[baseline
    = -0.5ex]{ \protect\node[mark size = 3pt, color =
      americanrose, line width = .5pt ]
      {\protect\pgfuseplotmark{triangle}}},
    \protect\tikz[baseline = -0.5ex]{ \protect\node[mark
      size = 3pt, color = applegreen, line width = .5pt ]
      {\protect\pgfuseplotmark{square}}},
    \protect\tikz[baseline = -0.5ex]{ \protect\node[mark
      size = 3pt, color = darkmagenta, line width = .5pt ]
      {\protect\pgfuseplotmark{diamond}}},
    \protect\tikz[baseline = -0.5ex]{ \protect\node[mark
      size = 3pt, color = peru, line width = .5pt ]
      {\protect\pgfuseplotmark{pentagon}}} correspond to
    \eqref{LMP} with $p=1, 1.25, 1.5, 1.75, 2$,
    respectively. Mark \protect\tikz[baseline = -0.5ex]{
      \protect\node[mark size = 3pt, color = burntorange,
      line width = .5pt ]
      {\protect\pgfuseplotmark{10-pointed star}}} denotes an
    algorithm which randomly chooses $p$, $\forall n$.}%
  \label{fig:vs.LMP-2}

\end{figure}

\begin{figure}[t]

  \centering

  \subfigure[$\alpha$-stable $\to$
  sparse]{ \includegraphics[width =
    .45\columnwidth]{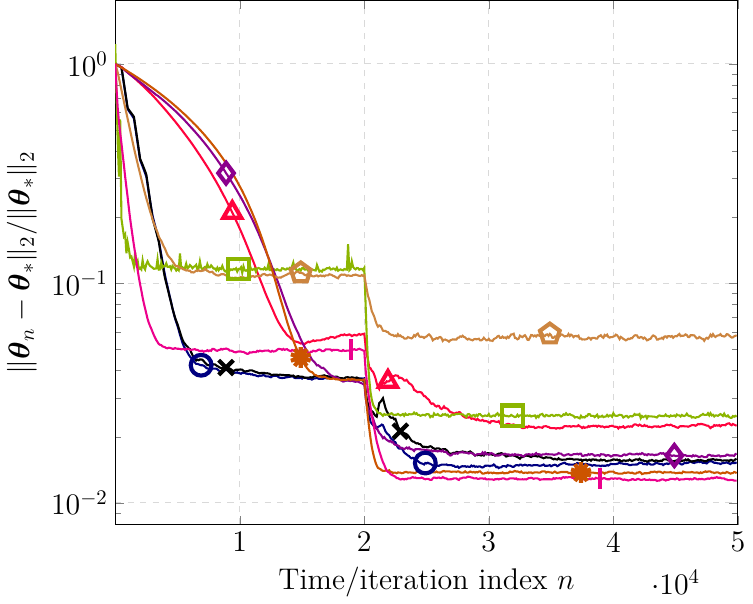}\label{6-1} }%
  \subfigure[Sparse
  $\to\alpha$-stable]{\includegraphics[width =
    .45\columnwidth]{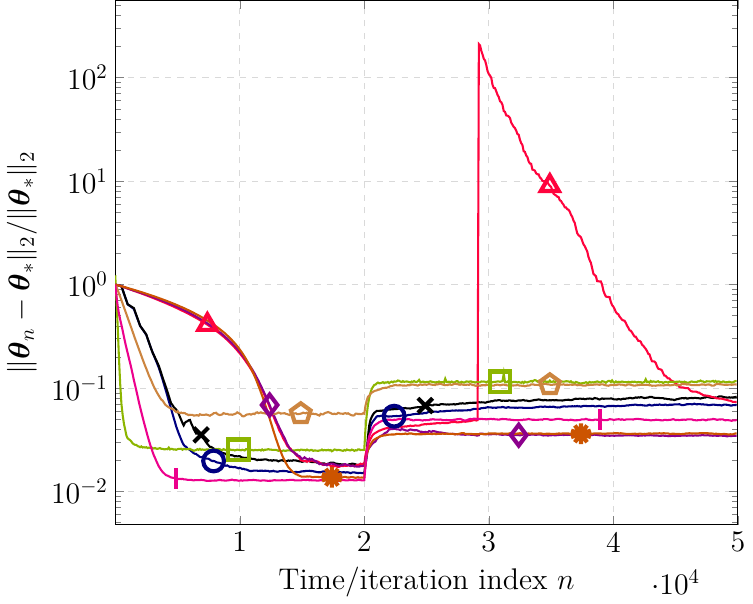}\label{6-2}}

  \caption{Scenario~2 (\cref{sec:sce2}): \cref{algo} against
    non-RL- and RL-based methods. \protect\tikz[baseline =
    -0.5ex]{ \protect\node[mark size = 3pt, color = navy,
      line width = .5pt ] {\protect\pgfuseplotmark{o}}}:
    \cref{algo} with $\alpha=0.9, N[n] \geq
    1$. \protect\tikz[baseline = -0.5ex]{ \protect\node[mark
      size = 3pt, color = black, line width = .5pt ]
      {\protect\pgfuseplotmark{x}}}: \cref{algo} with
    $\alpha=0.9, N[n]=1$. \protect\tikz[baseline = -0.5ex]{
      \protect\node[mark size = 3pt, color = americanrose,
      line width = .5pt ]
      {\protect\pgfuseplotmark{triangle}}}: Kernel-based
    TD(0) with $\alpha =
    0.9$~\cite{Bae:kerneltd1:11}. \protect\tikz[baseline =
    -0.5ex]{ \protect\node[mark size = 3pt, color =
      applegreen, line width = .5pt ]
      {\protect\pgfuseplotmark{square}}}: \cite{vazquez2012}
    with $p=1, \gamma_1 = 0.9, \gamma_2 =
    0.99$. \protect\tikz[baseline = -0.5ex]{
      \protect\node[mark size = 3pt, color = darkmagenta,
      line width = .5pt ]
      {\protect\pgfuseplotmark{diamond}}}: KLSPI with
    $\alpha=0.9$~\cite{xu2007klspi}. \protect\tikz[baseline
    = -0.5ex]{ \protect\node[mark size = 3pt, color = peru,
      line width = .5pt ]
      {\protect\pgfuseplotmark{pentagon}}}: mixed
    norm~\cite{chambers1997robust}. \protect\tikz[baseline =
    -0.5ex]{ \protect\node[mark size = 3pt, color =
      burntorange, line width = .5pt ]
      {\protect\pgfuseplotmark{10-pointed star}}}: the
    predecessor~\cite{minh:icassp23} of the current work.
    \protect\tikz[baseline = -0.5ex]{ \protect\node[mark
      size = 3pt, color = magenta, line width = .5pt ]
      {\protect\pgfuseplotmark{|}}}:
    VKW-MCC~\cite{huang2017adaptive}.}\label{fig:vs.TD.KLSPI-2}

\end{figure}

\begin{figure}[t]

  \centering

  \subfigure[$\alpha$-stable $\to$
  sparse]{\includegraphics[width =
    .45\columnwidth]{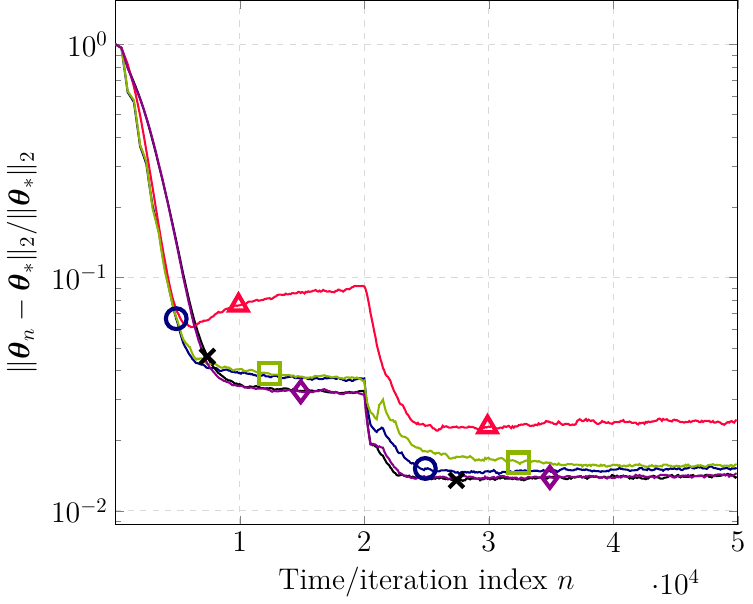}\label{5-1}}%
  \subfigure[Sparse
  $\to\alpha$-stable]{\includegraphics[width =
    .45\columnwidth]{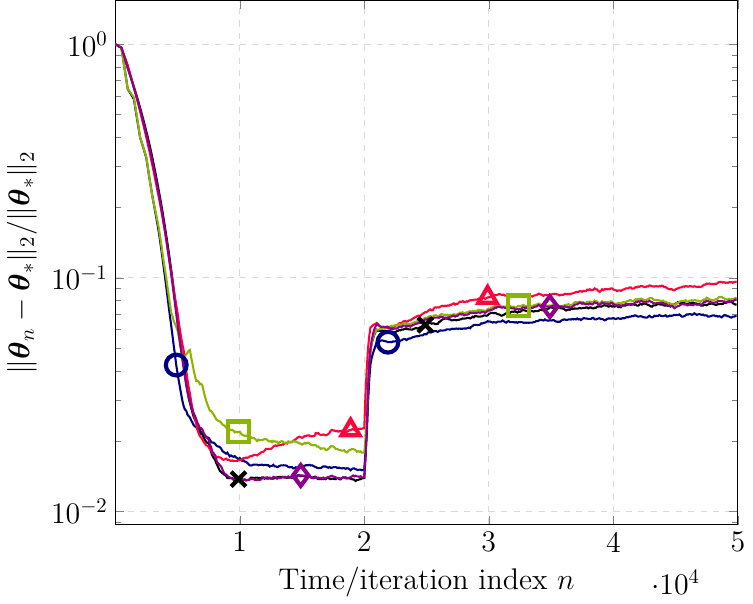}\label{5-2}}

  \caption{Scenario~2 (\cref{sec:sce2}): Versions of
    \cref{algo} under several parameter
    settings. \protect\tikz[baseline = -0.5ex]{
      \protect\node[mark size = 3pt, color = navy, line
      width = .5pt ] {\protect\pgfuseplotmark{o}}}:
    $\alpha = 0.9, N[n]\geq 1$.  \protect\tikz[baseline =
    -0.5ex]{ \protect\node[mark size = 3pt, color = black,
      line width = .5pt ] {\protect\pgfuseplotmark{x}}}:
    $\alpha = 0.75, N[n]\geq 1$.  \protect\tikz[baseline =
    -0.5ex]{ \protect\node[mark size = 3pt, color =
      americanrose, line width = .5pt ]
      {\protect\pgfuseplotmark{triangle}}}: $\alpha = 0$.
    \protect\tikz[baseline = -0.5ex]{ \protect\node[mark
      size = 3pt, color = applegreen, line width = .5pt ]
      {\protect\pgfuseplotmark{square}}}:
    $\alpha = 0.9, N[n]=1$.  \protect\tikz[baseline =
    -0.5ex]{ \protect\node[mark size = 3pt, color =
      darkmagenta, line width = .5pt ]
      {\protect\pgfuseplotmark{diamond}}}:
    $\alpha = 0.75, N[n]=1$.}\label{fig:vs.params-2}

\end{figure}

\Cref{fig:vs.LMP-2,fig:vs.TD.KLSPI-2,fig:vs.params-2} refer
to the scenario where the system $\vectgr{\theta}_*$ stays
fixed but the statistics (PDF) of the outliers change at a
specific time instance (here, at iteration
\#\num{20000}). Both sparse and $\alpha$-stable outliers are
considered in the following two dynamic sub-scenarios:
\begin{enumerate*} [label=\textbf{(\roman*)}]

\item $\alpha$-stable outliers appear at
  $n\in \{1, \ldots, \num{20000} \}$, followed by sparse
  ones at $n\in \{\num{20001}, \ldots, \num{50000} \}$; and

\item sparse outliers contaminate signals whenever
  $n\in \{1, \ldots, \num{20000} \}$, while $\alpha$-stable
  ones appear at
  $n\in \{ \num{20001}, \ldots, \num{50000} \}$.

\end{enumerate*}

\Cref{1-1,4-2} show different steady-state performances of
\cref{algo} after iteration \#\num{20000}, despite the fact
that the statistics of the $\alpha$-stable outliers are the
same. More specifically, in \cref{1-1}, the steady-state
performance level of \cref{algo} is almost identical to that
of LMP for $p = 1.25$, whereas LMP with $p = 1.25$ scores a
lower steady-state level than \cref{algo} after iteration
\#\num{20000} in \cref{4-2}. On the other hand, \cref{algo}
shows excellent performance in \cref{4-1} after the sudden
transition to sparse outliers. The previous discussion
concludes that \cref{algo} appears to underperform in cases
where there is a sudden change from light-tailed outliers to
the heavy-tailed $\alpha$-stable ones. In contrast, notice
the excellent performance of \cref{algo} under
$\alpha$-stable outliers in \cref{3-1}. Moreover, \cref{6-2}
demonstrates that kernel-based TD(0)~\cite{Bae:kerneltd1:11}
deteriorates significantly whenever the outlier PDF suddenly
changes to $\alpha$-stable outliers. The rest of the
RL-based methods exhibit more or less robust performance
against the abrupt change to $\alpha$-stable outliers in
\cref{6-2}.

Finally, notice that under the heavy-tailed $\alpha$-stable
outliers, versions of \cref{algo} with $\alpha > 0$ perform
better than version with $\alpha = 0$ in \cref{2-1}, whereas
versions $\alpha > 0$ and $\alpha = 0$ perform similarly
after iteration \#\num{20000} in \cref{5-2}.

\section{Conclusions}

This paper designed novel nonparametric Bellman mappings
(B-Maps) in reproducing kernel Hilbert spaces (RKHSs) for
reinforcement learning (RL). The new B-Maps exhibit several
desirable features (see \cref{sec:contribs}), with ample
degrees of freedom. To benefit from that freedom, a
variational framework (\cref{prop:T.yields.LSPE+BR}) was
provided to identify the free parameters of the B-Maps. As a
side effect, it was demonstrated that several
state-of-the-art designs become special cases of the
proposed B-Maps. Other non-trivial designs of B-Maps are
deferred to a future work. On the application front, the
manuscript considered the problem of selecting online, per
time instance, the ``optimal'' coefficient $p$ in the
least-mean-$p$-power method, with no prior information on
the outlier statistics and no training data. Interestingly,
the proposed design is general enough for the novel B-Maps
to be applied also to domains other than adaptive
filtering. Such application domains, together with their RL
designs, are currently under consideration and will be
presented soon at other publication venues.

\nocite{*}
\printbibliography
\clearpage
\newpage

\begin{center}
  \Large\textbf{Appendices}
\end{center}

\appendices

\section{Proof of \cref{prop:T.yields.LSPE+BR}}%
\label[appendix]{app:prop:new.Bellman.maps.reproduce.LSTD}

First, recall that the orthogonal projection mapping
$P_{\vectgr{\Phi}_{\mathcal{T}_N}}$ onto the linear span of
$\{ \varphi(\vect{z}_i)\}_{i=1}^N$ is provided by
$P_{\vectgr{\Phi}_{\mathcal{T}_N}} =
\vectgr{\Phi}_{\mathcal{T}_N}
\vect{K}_{\mathcal{T}_N}^{\dagger}
\vectgr{\Phi}_{\mathcal{T}_N}^{\intercal}$. Moreover, let
$P_{\vectgr{\Phi}_{\mathcal{T}_N}^{\perp}}$ denote the
orthogonal projection mapping onto the orthogonal complement
of the linear span of $\{
\varphi(\vect{z}_i)\}_{i=1}^N$. The Pythagoras theorem
states that
$\Id = P_{\vectgr{\Phi}_{\mathcal{T}_N}} +
P_{\vectgr{\Phi}_{\mathcal{T}_N}^{\perp}}$, where $\Id$ is
the identity operator in $\mathcal{H}$.

Define
$\vect{c}\coloneqq \vectgr{\gamma} + \alpha \vectgr{\Upsilon}
\vectgr{\Phi}_{\mu_{\star}}^{\textnormal{av} \intercal} Q$ for convenience, and observe that
\begin{align}
  & ( \vect{c} - \vect{K}_{\mathcal{T}_N}^{\dagger}
    \vectgr{\Phi}_{\mathcal{T}_N}^{\intercal}Q )^{\intercal}
    \vect{K}_{\mathcal{T}_N}
    ( \vect{c} - \vect{K}_{\mathcal{T}_N}^{\dagger}
    \vectgr{\Phi}_{\mathcal{T}_N}^{\intercal}Q ) \notag\\
  & = ( \vect{c} - \vect{K}_{\mathcal{T}_N}^{\dagger}
    \vectgr{\Phi}_{\mathcal{T}_N}^{\intercal}Q )^{\intercal}
    \vectgr{\Phi}_{\mathcal{T}_N}^{\intercal}
    \vectgr{\Phi}_{\mathcal{T}_N} ( \vect{c} -
    \vect{K}_{\mathcal{T}_N}^{\dagger}
    \vectgr{\Phi}_{\mathcal{T}_N}^{\intercal}Q ) \notag\\
  & = \innerp{ \vectgr{\Phi}_{\mathcal{T}_N} (\vect{c} -
    \vect{K}_{\mathcal{T}_N}^{\dagger}
    \vectgr{\Phi}_{\mathcal{T}_N}^{\intercal}Q)}{
    \vectgr{\Phi}_{\mathcal{T}_N}(\vect{c} -
    \vect{K}_{\mathcal{T}_N}^{\dagger}
    \vectgr{\Phi}_{\mathcal{T}_N}^{\intercal}Q)}_{\mathcal{H}}
    \notag\\
  & = \norm{ \vectgr{\Phi}_{\mathcal{T}_N} \vect{c} -
    \vectgr{\Phi}_{\mathcal{T}_N}
    \vect{K}_{\mathcal{T}_N}^{\dagger}
    \vectgr{\Phi}_{\mathcal{T}_N}^{\intercal} Q
    }^2_{\mathcal{H}} \notag\\
  & = \norm{ \vectgr{\Phi}_{\mathcal{T}_N} \vect{c} - P_{\vectgr{\Phi}_{\mathcal{T}_N}}
    Q  }^2_{\mathcal{H}} \,. \label{LSPE.regularizer}
\end{align}

A change of variables and the Pythagoras theorem suggest that
\begin{subequations}
  \begin{alignat}{3}
    &&& \negphantom{ {} = {} } T_{\textnormal{LSPE}, \mu}(Q)
        \notag\\
    & {} = {} && \arg\min_{ Q^{\prime} \in \mathcal{H} }
    && \norm{\vectgr{\Phi}_{\mathcal{T}_N}^{\intercal}
       Q^{\prime} - \vect{g} - \alpha
       \vectgr{\Phi}_{\mu}^{\prime\intercal}Q }_{\Real^N}^2
       \notag \\
    &&&&& + \sigma \norm{ Q^{\prime} - Q }^2_{\mathcal{H}}
          \notag\\
    & = && \arg\min_{ Q^{\prime} \in \mathcal{H} }
    && \norm{
       \vectgr{\Phi}_{\mathcal{T}_N}^{\intercal} (Q^{\prime} -
       P_{\vectgr{\Phi}_{\mathcal{T}_N}^{\perp}} Q) -
       \vect{g} - \alpha
       \vectgr{\Phi}_{\mu}^{\prime\intercal}Q }_{\Real^N}^2
       \notag\\
    &&&&& + \sigma \norm{ Q^{\prime} - Q }^2_{\mathcal{H}}
          \notag \\
    & = && \arg\min_{ Q^{\prime\prime} \in \mathcal{H} }
    && \norm{ \vectgr{\Phi}_{\mathcal{T}_N}^{\intercal}
       Q^{\prime\prime} - \vect{g} -
       \alpha \vectgr{\Phi}_{\mu}^{\prime\intercal}Q
       }_{\Real^N}^2 \notag\\
    &&&&& + \sigma \norm{ Q^{\prime\prime} +
          P_{\vectgr{\Phi}_{\mathcal{T}_N}^{\perp}} Q - Q
          }^2_{\mathcal{H}} \notag \\
    & = && \arg\min_{ Q^{\prime\prime} \in \mathcal{H} }
    && \norm{ \vectgr{\Phi}_{\mathcal{T}_N}^{\intercal}
       Q^{\prime\prime} - \vect{g} - \alpha
       \vectgr{\Phi}_{\mu}^{\prime\intercal}Q }_{\Real^N}^2
       \notag\\
    &&&&& + \sigma \norm{ Q^{\prime\prime} -
          P_{\vectgr{\Phi}_{\mathcal{T}_N}} Q
          }^2_{\mathcal{H}}  \label{argmin.LSPE.ii} \\
    & = && \vectgr{\Phi}_{\mathcal{T}_N} (
           \vect{K}_{\mathcal{T}_N}
    && + \sigma \vect{I}_N )^{-1} \vect{g} \notag\\
    &&&&& + \sigma \vectgr{\Phi}_{\mathcal{T}_N} (
          \vect{K}_{\mathcal{T}_N}
          + \sigma \vect{I}_N )^{-1}
          \vect{K}_{\mathcal{T}_N}^{\dagger}
          \vectgr{\Phi}_{\mathcal{T}_N}^{\intercal} Q
          \notag\\
    &&&&& + \alpha \vectgr{\Phi}_{\mathcal{T}_N} (
          \vect{K}_{\mathcal{T}_N}
          + \sigma \vect{I}_N )^{-1}
          \vectgr{\Phi}_{\mu}^{\prime \intercal} Q
          \,, \label{Tlspe.solution}
  \end{alignat}
\end{subequations}
where
$\vectgr{\Phi}_{\mathcal{T}_N} ( \vect{K}_{\mathcal{T}_N} +
\sigma \vect{I}_N )^{-1} = ( \vectgr{\Phi}_{\mathcal{T}_N}
\vectgr{\Phi}_{\mathcal{T}_N}^{\intercal} + \sigma\Id)^{-1}
\vectgr{\Phi}_{\mathcal{T}_N}$ was used in the last
equality.

Moreover, notice again by the Pythagoras theorem and
$P_{\vectgr{\Phi}_{\mathcal{T}_N}}^2 =
P_{\vectgr{\Phi}_{\mathcal{T}_N}}$ that
\begin{alignat*}{2}
  &&& \negphantom{ {} \geq {} } \norm{
      \vectgr{\Phi}_{\mathcal{T}_N}^{\intercal}
      Q^{\prime\prime} - \vect{g} - \alpha
      \vectgr{\Phi}_{\mu}^{\prime\intercal}Q }_{\Real^N}^2
      + \sigma \norm{ Q^{\prime\prime} -
      P_{\vectgr{\Phi}_{\mathcal{T}_N}} Q }^2_{\mathcal{H}}
  \\
  & {} \geq {}
    && \norm{\vectgr{\Phi}_{\mathcal{T}_N}^{\intercal}
       Q^{\prime\prime} - \vect{g} - \alpha
       \vectgr{\Phi}_{\mu}^{\prime\intercal}Q }_{\Real^N}^2
       + \sigma \norm{ P_{\vectgr{\Phi}_{\mathcal{T}_N}}(
       Q^{\prime\prime} - P_{\vectgr{\Phi}_{\mathcal{T}_N}}
       Q) }^2_{\mathcal{H}} \\
  & =
    && \norm{ \vectgr{\Phi}_{\mathcal{T}_N}^{\intercal}
       P_{\vectgr{\Phi}_{\mathcal{T}_N}} Q^{\prime\prime} -
       \vect{g} - \alpha
       \vectgr{\Phi}_{\mu}^{\prime\intercal} Q }_{\Real^N}^2
  \\
  &&& + \sigma \norm{ P_{\vectgr{\Phi}_{\mathcal{T}_N}}
      Q^{\prime\prime} - P_{\vectgr{\Phi}_{\mathcal{T}_N}} Q
      }^2_{\mathcal{H}} \,,
\end{alignat*}
which clearly suggests that the minimizer in
\eqref{argmin.LSPE.ii} lies in the linear span of
$\{ \varphi(\vect{z}_i)\}_{i=1}^N \Leftrightarrow (
Q^{\prime\prime} = \vectgr{\Phi}_{\mathcal{T}_N} \vect{c},\
\text{for}\ \exists \vect{c}\in \Real^N) \Leftrightarrow (
Q^{\prime\prime} = P_{\vectgr{\Phi}_{\mathcal{T}_N}}
Q^{\prime\prime} )$. Hence,
\begin{alignat*}{2}
  &&& \negphantom{ {} = {} } T_{\textnormal{LSPE}, \mu}Q \\
  & {} = {}
    && \arg\min_{ Q^{\prime\prime} \in \mathcal{H}
       \,\given\, Q^{\prime\prime} =
       P_{\vectgr{\Phi}_{\mathcal{T}_N}} Q^{\prime\prime} }
       \norm{ \vectgr{\Phi}_{\mathcal{T}_N}^{\intercal}
       Q^{\prime\prime} - \vect{g} - \alpha
       \vectgr{\Phi}_{\mu}^{\prime\intercal}Q }_{\Real^N}^2
  \\
  &&& \hphantom{ \arg\min_{ Q^{\prime\prime} \in \mathcal{H}
      \,\given\, Q^{\prime\prime} =
      P_{\vectgr{\Phi}_{\mathcal{T}_N}} Q^{\prime\prime}} }
      + \sigma \norm{ Q^{\prime\prime} -
      P_{\vectgr{\Phi}_{\mathcal{T}_N}} Q }^2_{\mathcal{H}}
      \notag \\
  & = && \vectgr{\Phi}_{\mathcal{T}_N} \vect{c}_{\star} \,,
\end{alignat*}
where $\vect{c}_{\star}$ satisfies
\begin{alignat}{2}
  &&& \negphantom{ {} = {} } \arg\min_{ \vect{c}\in
      \Real^N }
      \norm{ \vectgr{\Phi}_{\mathcal{T}_N}^{\intercal}
      \vectgr{\Phi}_{\mathcal{T}_N} \vect{c} -
      \vect{g} - \alpha
      \vectgr{\Phi}_{\mu}^{\prime\intercal} Q }_{\Real^N}^2
      \notag\\
  &&& \negphantom{ {} = {} } \hphantom{ \arg\min_{
      \vect{c}\in \Real^N } } + \sigma \norm{
      \vectgr{\Phi}_{\mathcal{T}_N} \vect{c} -
      P_{\vectgr{\Phi}_{\mathcal{T}_N}} Q
      }^2_{\mathcal{H}} \label{argmin.c} \\
  & {} = {}
    && \arg\min_{ \vect{c}\in
       \Real^N } \overbrace{ \norm{ \vect{K}_{\mathcal{T}_N}
       \vect{c} - \vect{g} - \alpha
      \vectgr{\Phi}_{\mu}^{\prime\intercal}Q }_{\Real^N}^2
       }^{ \mathcal{L}(\vectgr{\gamma}, \vectgr{\Upsilon} ) }
      \notag\\
  &&& \negphantom{ {} = {} } \hphantom{ \arg\min_{
      \vect{c}\in \Real^N } } + \sigma \underbrace{
      ( \vect{c} - \vect{K}_{\mathcal{T}_N}^{\dagger}
      \vectgr{\Phi}_{\mathcal{T}_N}^{\intercal} Q
      )^{\intercal} \vect{K}_{\mathcal{T}_N} ( \vect{c} -
      \vect{K}_{\mathcal{T}_N}^{\dagger}
      \vectgr{\Phi}_{\mathcal{T}_N}^{\intercal} Q)}_{
      \mathcal{R}(\vectgr{\gamma}, \vectgr{\Upsilon} ) }
      \notag \\
  & {} \ni {}
    && \vect{c}_{\star} \notag\\
  & {} \coloneqq {}
    && ( \vect{K}_{\mathcal{T}_N} + \sigma \vect{I}_N )^{-1}
       \bigl( \vect{g} + \sigma
       \vect{K}_{\mathcal{T}_N}^{\dagger}
       \vectgr{\Phi}_{\mathcal{T}_N}^{\intercal} Q + \alpha
       \vectgr{\Phi}_{\mu}^{\prime \intercal}Q \bigr) \notag\\
  & =
    && ( \vect{K}_{\mathcal{T}_N} + \sigma \vect{I}_N )^{-1}
       \vect{g} \notag\\
  &&& + ( \vect{K}_{\mathcal{T}_N} + \sigma \vect{I}_N
      )^{-1}
      [ \sigma \vect{K}_{\mathcal{T}_N}^{\dagger} \,,
      \alpha\vect{I}_N ]
      \begin{bmatrix}
        \vectgr{\Phi}_{\mathcal{T}_N}^{\intercal}Q \\
        \vectgr{\Phi}_{\mu}^{\prime \intercal}Q
      \end{bmatrix}
      \notag\\
  & =
    && ( \vect{K}_{\mathcal{T}_N} + \sigma \vect{I}_N )^{-1}
       \vect{g} \notag \\
  &&& + \alpha ( \vect{K}_{\mathcal{T}_N} + \sigma
      \vect{I}_N )^{-1} [ (\sigma/\alpha)
      \vect{K}_{\mathcal{T}_N}^{\dagger} \,, \vect{I}_N ]
      \vectgr{\Phi}_{\mu_{\star}}^{\textnormal{av}
      \intercal} Q \notag \\
  & =
    && \vectgr{\gamma}_{\star} + \alpha
       \vectgr{\Upsilon}_{\star}
       \vectgr{\Phi}_{\mu_{\star}}^{\textnormal{av}
       \intercal} Q \,, \notag
\end{alignat}
and where
$\vectgr{\gamma}_{\star} \coloneqq (
\vect{K}_{\mathcal{T}_N} + \sigma \vect{I}_N )^{-1}
\vect{g}$ and
$\vectgr{\Upsilon}_{\star} \coloneqq (
\vect{K}_{\mathcal{T}_N} + \sigma \vect{I}_N )^{-1} [
(\sigma/\alpha) \vect{K}_{\mathcal{T}_N}^{\dagger} \,,
\vect{I}_N ]$.

Under the light of~\eqref{LSPE.regularizer}, the previous
findings are summarized as follows:
$(\vectgr{\gamma}_{\star}, \vectgr{\Upsilon}_{\star})$
satisfies~\eqref{T.yields.several.maps.task} iff
$\vect{c}_{\star} = \vectgr{\gamma}_{\star} + \alpha
\vectgr{\Upsilon}_{\star}
\vectgr{\Phi}_{\mu_{\star}}^{\textnormal{av} \intercal} Q$
is one of the minimizers in~\eqref{argmin.c} iff
$T_{\textnormal{LSPE}, \mu}(Q) =
\vectgr{\Phi}_{\mathcal{T}_N} \vect{c}_{\star} =
T_{\mu_{\star}}(Q)$. These equivalences establish the claim
of \cref{prop:T.yields.LSPE}. As an additional remark,
notice that $T_{\textnormal{LSPE}, \mu}(Q) = Q$
in~\eqref{Tlspe.solution} yields~\eqref{fix.LSPE}.

The proofs of \Cref{prop:T.yields.BR,prop:T.inthispaper}
follow similar steps with the proof of
\cref{prop:T.yields.LSPE} and are thus skipped.

\section{Proof of~\cref{thm:new.Bellman.maps.nonexp}}%
\label[appendix]{app:thm:new.Bellman.maps.nonexp}

First, the following lemma and its proof are in order.

\begin{lemma}\label{proof-nonexpansive-lemma}
  For any $Q_1,Q_2\in\mathcal{H}$,
  $\vect{s}\in\mathfrak{S}$, there exists $\delta \neq 0$
  s.t.\ for any
  $\epsilon \in ( 0, \sqrt{\lvert\delta\rvert})$ an
  $a^{\prime} \in\mathfrak{A}$ can be always selected s.t.\
  \begin{align}
    \bigl[ \inf\nolimits_{a\in\mathfrak{A}}Q_1(\vect{s},a)
    & - \inf\nolimits_{a\in\mathfrak{A}}Q_2(\vect{s},a)
      \bigr]^2 \notag\\
    & \leq \left[ Q_1(\vect{s},a^\prime) -
      Q_2(\vect{s},a^\prime) \right]^2 + \epsilon
      \,. \label{epsilon-threshold}
  \end{align}
\end{lemma}

\begin{IEEEproof}
  Define
  $\delta \coloneqq \inf_{a\in\mathfrak{A}}Q_1(\vect{s},a) -
  \inf_{a\in\mathfrak{A}}Q_2(\vect{s},a)$. Whenever
  $\delta = 0$, \eqref{epsilon-threshold} holds true
  trivially. Consider now the case where $\delta > 0$, take
  any $\epsilon \in (0, \sqrt{\delta})$, and define
  $\epsilon^{\prime} \coloneqq \delta - \sqrt{\delta^2 -
    \epsilon} > 0$ so that
  $\epsilon = 2\epsilon^{\prime} \delta -
  \epsilon^{\prime}{}^2$ and
  $\delta - \epsilon^{\prime} >0$. Then,
  $\exists a^\prime\in\mathfrak{A}$ s.t.\
  $Q_2(\vect{s}, a^{\prime}) \leq \inf_{a\in\mathfrak{A}}
  Q_2(\vect{s}, a) + \epsilon^{\prime}$. This result
  together with
  $\inf_{a\in\mathfrak{A}} Q_1(\vect{s}, a) \leq
  Q_1(\vect{s}, a^{\prime})$ suggest:
  \begin{align*}
    Q_1(\vect{s}, a^{\prime}) - Q_2(\vect{s}, a^{\prime}) \geq
    \inf_{a\in\mathfrak{A}}Q_1(\vect{s}, a) -
    \inf_{a\in\mathfrak{A}} Q_2(\vect{s}, a) -
    \epsilon^{\prime}\,,
  \end{align*}
  where the right-hand side is positive, and
  \begin{align*}
    & \left[ Q_1(\vect{s},a^\prime)-Q_2(\vect{s},a^\prime)
      \right]^2 \\
    & \geq \left[ \inf\nolimits_{a\in\mathfrak{A}} Q_1(\vect{s}, a) -
      \inf\nolimits_{a\in\mathfrak{A}}Q_2(\vect{s}, a)
      \right]^2 - 2\epsilon^{\prime} \delta +
      \epsilon^{\prime}{}^2 \\
    & = \left[ \inf\nolimits_{a\in\mathfrak{A}} Q_1(\vect{s}, a) -
      \inf\nolimits_{a\in\mathfrak{A}}Q_2(\vect{s}, a) \right]^2 - \epsilon
      \,,
  \end{align*}
  which establishes~\eqref{epsilon-threshold}. The proof for
  the case $\delta < 0$ follows exactly the previous steps
  by using $-\delta$ instead of $\delta$ and by
  interchanging $Q_1$ with $Q_2$.
\end{IEEEproof}

Property~\eqref{Lipschitz.mu} is established as follows:
$\forall Q_1,Q_2\in\mathcal{H}$,
\begin{align*}
  & \norm{T_{\mu} (Q_1) - T_{\mu}(Q_2)}_{\mathcal{H}}^2 \\
  & = \alpha^2\innerp{
    \vectgr{\Psi}\vectgr{\Phi}^{\textnormal{av}}_{\mu}{}^{\intercal}
    (Q_1-Q_2) }{
    \vectgr{\Psi}\vectgr{\Phi}^{\textnormal{av}}_{\mu}{}^{\intercal}
    (Q_1-Q_2)}_{\mathcal{H}} \\
  & = \alpha^2\innerp{
    \vectgr{\Phi}^{\textnormal{av}}_{\mu}{}^{\intercal}
    (Q_1-Q_2) }{ \vectgr{\Psi}^{\intercal}
    \vectgr{\Psi}\vectgr{\Phi}^{\textnormal{av}}_{\mu}{}^{\intercal}
    (Q_1-Q_2)} \\
  & \leq \alpha^2 \norm{\vectgr{\Psi}^\intercal
    \vectgr{\Psi}}_2 \innerp{
    Q_1-Q_2 }{ \vectgr{\Phi}^{\textnormal{av}}_{\mu}
    \vectgr{\Phi}^{\textnormal{av}}_{\mu}{}^{\intercal}
    (Q_1-Q_2)}_{\mathcal{H}} \\
  & \leq \alpha^2 \norm{ \vect{K}_{\Psi} }_2 \cdot \norm{
    \vectgr{\Phi}^{\textnormal{av}}_{\mu}
    \vectgr{\Phi}^{\textnormal{av}}_{\mu}{}^{\intercal} }_2
    \cdot \norm{ Q_1-Q_2 }^2_{\mathcal{H}} \\
  & \leq \beta^2 \norm{Q_1-Q_2}^2_{\mathcal{H}} \,,
\end{align*}
where observation
$\norm{ \vectgr{\Phi}^{\textnormal{av}}_{\mu}
  \vectgr{\Phi}^{\textnormal{av}}_{\mu}{}^{\intercal} }_2 =
\norm{ \vectgr{\Phi}^{\textnormal{av}}_{\mu}{}^{\intercal}
  \vectgr{\Phi}^{\textnormal{av}}_{\mu} }_2 = \norm{
  \vect{K}_{\mu}^{\textnormal{av}} }_2$ and~\eqref{beta}
were used to obtain the last inequality.

The proof of~\eqref{Lipschitz.optimal} follows. For any
$Q_1, Q_2 \in \mathcal{H}$,
\begin{align}
  & \norm{T(Q_1) - T(Q_2)}_{\mathcal{H}}^2 \notag \\
  & = \alpha^2 \norm{ \vectgr{\Psi}
    (\vecinf_{\mathcal{S}_{\textnormal{av}}}(Q_1) -
    \vecinf_{\mathcal{S}_{\textnormal{av}}}(Q_2)
    )}_{\mathcal{H}}^2 \notag\\
  & \leq \alpha^2 \norm{ \vect{K}_{\Psi} }_2 \, \norm{
    \vecinf_{\mathcal{S}_{\textnormal{av}}}(Q_1) -
    \vecinf_{\mathcal{S}_{\textnormal{av}}}(Q_2)
    }^2 \notag\\
  & = \alpha^2 \norm{ \vect{K}_{\Psi} }_2
    \sum_{i=1}^{N_{\textnormal{av}}}
    \bigl[ \inf_{a_i \in\mathfrak{A}}
    Q_1(\vect{s}_i^{\textnormal{av}}, a_i) -
    \inf_{a_i\in\mathfrak{A}}
    Q_2(\vect{s}_i^{\textnormal{av}}, a_i) \bigr]^2
    \,. \label{T.Lip.step.1}
\end{align}
According to~\cref{proof-nonexpansive-lemma}, there exist
$\{ \delta_i \neq 0 \}_{i=1}^{N_{\textnormal{av}}}$ s.t.\
for any
$\epsilon\in (0, \min_{i\in \{ 1, \ldots,
  N_{\textnormal{av}}\}}\{ \sqrt{\lvert \delta_i \rvert} \}
)$ actions $\{ a_i^{\prime} \}_{i=1}^{N_{\textnormal{av}}}$
can be selected so that~\eqref{T.Lip.step.1} yields
\begin{alignat}{2}
  &&& \negphantom{ {} \leq {} } \norm{T(Q_1) -
      T(Q_2)}_{\mathcal{H}}^2 \notag \\
  & {} \leq {}
    && \alpha^2 \norm{ \vect{K}_{\Psi} }_2
       \sum_{i=1}^{N_{\textnormal{av}}}
    \bigl[ Q_1(\vect{s}_i^{\textnormal{av}}, a_i^{\prime}) -
       Q_2(\vect{s}_i^{\textnormal{av}},
       a_i^{\prime}) \bigr]^2 \notag \\
  &&& + \epsilon \alpha^2 N_{\textnormal{av}} \norm{
      \vect{K}_{\Psi} }_2 \,. \label{T.Lip.step.2}
\end{alignat}
Take now any stationary policy
$\mu^{\prime} \in \mathcal{M}$ s.t.\
$\mu^{\prime}( \vect{s}_i^{\textnormal{av}} ) =
a_i^{\prime}$, $\forall i$. Then, by
$\vectgr{\Phi}_{\mu^{\prime}}^{\textnormal{av}} \coloneqq [
\varphi(\vect{s}_1^{\textnormal{av}}, \mu^{\prime}(
\vect{s}_1^{\textnormal{av}} ) ), \ldots,
\varphi(\vect{s}_{N_{\textnormal{av}}}^{\textnormal{av}},
\mu^{\prime}(
\vect{s}_{N_{\textnormal{av}}}^{\textnormal{av}} ) )]$,
\begin{align*}
  & \sum\nolimits_{i=1}^{N_{\textnormal{av}}}
    \bigl[ Q_1(\vect{s}_i^{\textnormal{av}}, a_i^{\prime}) -
    Q_2(\vect{s}_i^{\textnormal{av}},
    a_i^{\prime}) \bigr]^2 \\
  & = \sum\nolimits_{i=1}^{N_{\textnormal{av}}} \bigl[
    Q_1(\vect{s}_i^{\textnormal{av}},
    \mu^{\prime}(\vect{s}_i^{\textnormal{av}}) ) -
    Q_2(\vect{s}_i^{\textnormal{av}},
    \mu^{\prime}(\vect{s}_i^{\textnormal{av}})) \bigr]^2 \\
  & = \norm{
    \vectgr{\Phi}_{\mu^{\prime}}^{\textnormal{av}}{}^{\intercal}
    (Q_1 - Q_2) }^2 \\
  & =
    \innerp{\vectgr{\Phi}_{\mu^{\prime}}^{\textnormal{av}}{}^{\intercal}
    (Q_1 - Q_2)}
    {\vectgr{\Phi}_{\mu^{\prime}}^{\textnormal{av}}{}^{\intercal}
    (Q_1 - Q_2)} \\
  & = \innerp{Q_1 - Q_2}
    { \vectgr{\Phi}_{\mu^{\prime}}^{\textnormal{av}}
    \vectgr{\Phi}_{\mu^{\prime}}^{\textnormal{av}}{}^{\intercal}
    (Q_1 - Q_2)}_{\mathcal{H}} \\
  & \leq \norm{ \vect{K}_{\mu^{\prime}}^{\textnormal{av}}
    }_2 \, \norm{ Q_1 - Q_2 }_{\mathcal{H}}^2 \,,
\end{align*}
so that~\eqref{T.Lip.step.2} results in
\begin{align*}
  & \norm{T(Q_1) - T(Q_2)}_{\mathcal{H}}^2 \notag \\
  & \leq
    \alpha^2 \norm{ \vect{K}_{\Psi} }_2
    \sup_{\mu^{\prime}\in \mathcal{H}} \norm{
    \vect{K}_{\mu^{\prime}}^{\textnormal{av}} }_2 \, \norm{
    Q_1 - Q_2 }_{\mathcal{H}}^2
    + \epsilon \alpha^2 N_{\textnormal{av}} \norm{
    \vect{K}_{\Psi} }_2 \\
  & = \beta^2 \norm{ Q_1 - Q_2 }_{\mathcal{H}}^2
    + \epsilon \alpha^2 N_{\textnormal{av}} \norm{
    \vect{K}_{\Psi} }_2 \,.
\end{align*}
Since $\epsilon$ can be made arbitrarily small, this last
inequality establishes~\eqref{Lipschitz.optimal}.

\section{Proof of~\cref{thm:new.Bellman.maps.consistency}}%
\label[appendix]{app:thm:new.Bellman.maps.consistency}

The proof is built on arguments
of~\cite{fukumizu2004dimensionality, Song:ICML:09,
  Song:AISTATS:10}. In particular, following~\cite[\S3,
\S4.1]{Song:AISTATS:10} and for an arbitrarily fixed policy
$\mu(\cdot) \in \mathcal{M}$, define the linear covariance
operators
$\Sigma_{zz}, \Sigma_{s^{\prime} z}^{\mu}, \Sigma_{
  s^{\prime} \given z }^{\mu} \colon \mathcal{H} \to
\mathcal{H}$ by
\begin{subequations}\label{Sigma.def}
  \begin{alignat}{2}
    \Sigma_{zz} (Q)
    & {} \coloneqq {}
    && \expect_{\vect{z}}
       \{\innerp{\varphi(\vect{z})}{Q}_{\mathcal{H}} \,
       \varphi(\vect{z})\}\,, \label{Sigma.zz}\\
    \Sigma_{s^{\prime} z}^{\mu} (Q)
    & \coloneqq
    &&
       \expect_{ (\vect{s}^{\prime}, \vect{z}) } \{\innerp{
       \varphi(\vect{s}^\prime, \mu
       (\vect{s}^\prime))}{Q}_{\mathcal{H}} \,\varphi(\vect{z})
       \}\,, \label{Sigma.sprimez} \\
    \Sigma_{ s^{\prime} \given z }^{\mu} (Q)
    & \coloneqq
    && \Sigma_{zz}^{-1}
       \Sigma_{s^{\prime} z}^{\mu} (Q)
       \,, \label{Sigma.conditional.map}
  \end{alignat}
\end{subequations}
where $\expect_{ \vect{z} }\{\cdot\}$ denotes expectation
with respect to (w.r.t.) the $\sigma$-subalgebra generated
by $\vect{z}$~\cite{Williams:probability:91}, similarly for
$\expect_{ (\vect{s}^{\prime}, \vect{z}) }\{ \cdot\}$, and
$\Sigma_{zz}^{-1}$ stands for the inverse of
$\Sigma_{zz}$. It is important to stress here that the
expectation symbols in~\eqref{Sigma.def} are considered in
the following sense. In~\eqref{Sigma.zz}, for example, for
arbitrarily fixed $Q\in \mathcal{H}$, $\Sigma_{zz} (Q)$
stands for the unique point of $\mathcal{H}$ which
satisfies, according to the Riesz representation
theorem~\cite{Conway},
$\innerp{\Sigma_{zz} (Q)} {h} = L_{zz} ( h )$,
$\forall h\in \mathcal{H}$, where $L_{zz} (\cdot)$ is the
linear continuous operator defined by
$L_{zz} (\cdot) \colon \mathcal{H} \to \Real \colon h
\mapsto L_{zz} ( h ) \coloneqq \expect_{\vect{z}} \Set{
  \innerp{ \innerp{Q} { \varphi(\vect{z}) }_{\mathcal{H}}
    \cdot \varphi(\vect{z})} {h}_{\mathcal{H}} }$, and where
expectation is taken here in the usual
sense~\cite{Williams:probability:91}. Operators~\eqref{Sigma.sprimez}
and~\eqref{Sigma.conditional.map} are defined in a similar
way.

Recall here also the definitions for the minimum
$\sigma_{\min} ( \mathcal{A} )$ and maximum
$\sigma_{\max} ( \mathcal{A} )$ spectral values of a linear
bounded and self-adjoint operator
$\mathcal{A} \colon \mathcal{H} \to
\mathcal{H}$~\cite[Thm.~9.2-3]{kreyszig:91}:
\begin{subequations}
  \begin{alignat}{2}
    \sigma_{\min} ( \mathcal{A} )
    & {} \coloneqq {}
    && \inf\nolimits_{ \norm{h}_{\mathcal{H}} = 1 } \innerp{h}
       {\mathcal{A} (h)}_{\mathcal{H}}
       \,, \label{sigma.min.def} \\
    \sigma_{\max} ( \mathcal{A} )
    & {} \coloneqq {}
    && \sup\nolimits_{ \norm{h}_{\mathcal{H}} = 1 } \innerp{h}
       {\mathcal{A} (h)}_{\mathcal{H}} = \norm{ \mathcal{A} }
       \,. \label{sigma.max.def}
  \end{alignat}
\end{subequations}

It can be verified
by~\cite[Thm.~2]{fukumizu2004dimensionality} and
\cite[\S4.1]{Song:AISTATS:10} that
$\expect_{ \vect{s}^\prime \given \vect{z} } \{
Q(\vect{s}^{\prime}, \mu(\vect{s}^{\prime})) \} = \innerp{
  \Sigma_{ s^{\prime} \given z }^{\mu\, *} (
  \varphi(\vect{z}) ) }{ Q }_{\mathcal{H}}$, where $*$
denotes the adjoint of a linear
operator~\cite{Conway}. Hence, by the reproducing property
of the kernel in $\mathcal{H}$,
$\expect_{ \vect{s}^\prime \given \vect{z} } \{
Q(\vect{s}^{\prime}, \mu(\vect{s}^{\prime})) \} =
\innerp{\varphi( \vect{z} )}{ \Sigma_{ s^{\prime} \given z
  }^{\mu} (Q) }_{\mathcal{H}} = \Sigma_{ s^{\prime} \given z
}^{\mu} (Q) (\vect{z})$, and the classical
B-Maps~\eqref{classical.Bellman.maps} take the following
equivalent form:
\begin{subequations}\label{classical.Bellman.alternative}
  \begin{alignat}{2}
    (T_{\mu}^{\diamond} Q)(\vect{z})
    & {} \coloneqq {}
    && g(\vect{z}) + \alpha \Sigma_{
       s^{\prime} \given z }^{\mu} (Q)
       (\vect{z})
       \,, \label{classical.Bellman.mu.alternative}\\
    (T^{\diamond} Q)(\vect{z})
    & \coloneqq
    && g(\vect{z}) + \alpha \Sigma_{ s^{\prime}
       \given z }^{\mu_Q} (Q) (\vect{z})
       \,, \label{classical.Bellman.inf.alternative}
  \end{alignat}
\end{subequations}
where the stationary policy $\mu_Q(\cdot)$ is defined as
in~\eqref{classical.Bellman.min}. Along the lines of
\cref{ass:T.inthispaper} and~\cite[(7)]{Song:AISTATS:10},
define
\begin{align}
  \hat{\Sigma}_{ s^{\prime} \given z }^{\mu}
  & \coloneqq \hat{\Sigma}_{ s^{\prime} \given z }^{\mu} (N)
    \notag \\
  & \coloneqq
    \vectgr{\Phi}_{\mathcal{T}_N} ( \vect{K}_{\mathcal{T}_N} +
    N \sigma^{\prime}_N \vect{I}_N )^{-1}
    \vectgr{\Phi}_{\mu}^{\textnormal{av}\intercal} \notag \\
  & = \tfrac{1}{\sqrt{N}} \vectgr{\Phi}_{\mathcal{T}_N}
    \left( \tfrac{1}{N} \vect{K}_{\mathcal{T}_N} +
    \sigma^{\prime}_N \vect{I}_N \right)^{-1}
    \tfrac{1}{\sqrt{N}}
    \vectgr{\Phi}_{\mu}^{\textnormal{av}\intercal}
    \,, \label{hat.classical.Bellman.mu.alternative}
\end{align}
so that~\eqref{new.Bellman.mu.special} is recast as
\begin{align}
  (T_{\mu} Q) (\vect{z})
  & \coloneqq g(\vect{z}) + \alpha \hat{\Sigma}_{ s^{\prime}
    \given z }^{\mu} (Q) (\vect{z})
    \,. \label{new.Bellman.mu.alternative}
\end{align}

The following theorem asserts that the previous quantity is
a consistent finite-sample estimate
of~\eqref{Sigma.conditional.map}.

\begin{thm}\label{thm:consistency.Gamma}
  \cref{ass:HS} means that
  \begin{align*}
    \sum\nolimits_{i=1}^{+\infty} \norm{ \Sigma_{zz}^{-3/2}
    \Sigma_{ s^{\prime}z }^{\mu} e_i}_{\mathcal{H}}^2 <
    +\infty,
  \end{align*}
  for a countable orthonormal basis $(e_i)_{i=1}^{+\infty}$
  of $\mathcal{H}$~\cite[p.~267]{Conway}. Under also
  \cref{ass:reg.coeff.sigma},
  $\Plim_{N\to \infty} \norm{ \Sigma_{ s^{\prime} \given z
    }^{\mu} - \hat{\Sigma}_{ s^{\prime} \given z }^{\mu} (N)
  } = 0$.
\end{thm}

\begin{IEEEproof}
  \cite[Thm.~1]{Song:AISTATS:10} yields
  $\Plim_{N\to \infty} \norm{ \Sigma_{ s^{\prime} \given z
    }^{\mu} - \hat{\Sigma}_{ s^{\prime} \given z }^{\mu} (N)
  }_{\textnormal{HS}} = 0$, where
  $\norm{\cdot}_{\textnormal{HS}}$ stands for the
  Hilbert-Schmidt norm of an
  operator~\cite[p.~267]{Conway}. Consequently, the fact
  $\norm{\cdot} \leq
  \norm{\cdot}_{\textnormal{HS}}$~\cite[p.~267]{Conway}
  establishes the claim of the theorem.
\end{IEEEproof}

The claim of \cref{thm:new.Bellman.maps.consistency} that
$T_{\mu}$ and $T$ are contractions follows directly by
\eqref{Lipschitz.inequality} and assumption
$\beta(N) \leq \beta_{\infty} < 1$, a.s. Now, the triangle
inequality and \cref{ass:consistency.fix} suggest%
\begin{subequations}\label{contraction.proof}%
  \begin{alignat}{2}
    &&& \negphantom{ {} \leq {} } \norm{T_{\mu}^{\diamond}
        (Q_1) - T_{\mu}^{\diamond}
        (Q_2)}_{\mathcal{H}} \\
    & {} \leq {}
      && \norm{T_{\mu}^{\diamond} (Q_1) - T_{\mu}
         (Q_1)}_{\mathcal{H}} +  \norm{
         T_{\mu}^{\diamond} (Q_2) - T_{\mu}
         (Q_2)}_{\mathcal{H}} \notag \\
    &&& + \norm{T_{\mu} (Q_1) - T_{\mu} (Q_2)}_{\mathcal{H}}
        \notag \\
    & {} \leq {}
      && \alpha \norm{ \Sigma_{ s^{\prime} \given z
         }^{\mu} - \hat{\Sigma}_{ s^{\prime} \given z }^{\mu} (N)
         }\, \norm{ Q_1
         }_{\mathcal{H}} \label{contraction.proof.i} \\
    &&& + \alpha \norm{ \Sigma_{ s^{\prime} \given z
         }^{\mu} - \hat{\Sigma}_{ s^{\prime} \given z
        }^{\mu} (N) }\, \norm{ Q_2
        }_{\mathcal{H}} \label{contraction.proof.ii} \\
    &&& + \beta_{\infty} \norm{Q_1 - Q_2}_{\mathcal{H}} \,.
  \end{alignat}
\end{subequations}
By applying $\Plim_{N\to \infty}$ to
\eqref{contraction.proof} and by
\cref{thm:consistency.Gamma}, it can be verified that
$\norm{T_{\mu}^{\diamond} (Q_1) - T_{\mu}^{\diamond}
  (Q_2)}_{\mathcal{H}} \leq \beta_{\infty} \norm{Q_1 -
  Q_2}_{\mathcal{H}}$, $\forall Q_1, Q_2\in \mathcal{H}$,
a.s. The claim
$\norm{T^{\diamond} (Q_1) - T^{\diamond}
  (Q_2)}_{\mathcal{H}} \leq \beta_{\infty} \norm{Q_1 -
  Q_2}_{\mathcal{H}}$, $\forall Q_1, Q_2\in \mathcal{H}$,
a.s., can be established in a similar way to
\eqref{contraction.proof}, but with $\mu_{Q_1}$ and
$\mu_{Q_2}$, whose definitions are provided
below~\eqref{classical.Bellman.min}, taking the place of
$\mu$ in \eqref{contraction.proof.i} and
\eqref{contraction.proof.ii}, respectively.

Consider now the fixed points $Q_{\mu}^{\diamond}$ and
$Q_{\mu}$ of $T_{\mu}^{\diamond}$ and $T_{\mu}$,
respectively. Notice now that
\begin{alignat}{2}
  &&& \negphantom{ {} = {} } \norm{Q_{\mu}^{\diamond} -
      Q_{\mu}}_{\mathcal{H}} \notag \\
  & {} = {}
    && \norm{ T_{\mu}^{\diamond} (Q_{\mu}^{\diamond}) -
       T_{\mu}(Q_{\mu}) }_{\mathcal{H}} \notag \\
  & \leq
    && \norm{T_{\mu}^{\diamond} (Q_{\mu}^{\diamond})
    - T_{\mu} (Q_{\mu}^{\diamond})}_{\mathcal{H}} + \norm{
       T_{\mu} (Q_{\mu}^{\diamond}) -
       T_{\mu}(Q_{\mu})}_{\mathcal{H}} \notag \\
  & \leq
    && \alpha \norm{ ( \Sigma_{ s^{\prime} \given z
         }^{\mu} - \hat{\Sigma}_{ s^{\prime} \given z
        }^{\mu} (N) ) (Q_{\mu}^{\diamond}) }_\mathcal{H} +
       \beta \norm{Q_{\mu}^\diamond - Q_{\mu}}_{\mathcal{H}}
       \notag\\
  & \leq
    && \alpha \norm{ \Sigma_{ s^{\prime} \given z
         }^{\mu} - \hat{\Sigma}_{ s^{\prime} \given z
        }^{\mu} (N) } \, \norm{ Q_{\mu}^{\diamond}
       }_\mathcal{H} + \beta_{\infty} \norm{Q_{\mu}^\diamond
       -
       Q_{\mu}}_{\mathcal{H}}\,, \label{convergence.prob.i}
\end{alignat}
which yields
\begin{align*}
  & \norm{Q_{\mu}^{\diamond} - Q_{\mu}(N)}_{\mathcal{H}}
    \leq \frac{\alpha \norm{ Q_{\mu}^{\diamond}
    }_\mathcal{H} }{1-\beta_{\infty}} \norm{ \Sigma_{
    s^{\prime} \given z }^{\mu} - \hat{\Sigma}_{ s^{\prime}
    \given z }^{\mu} (N)} \,,
\end{align*}
that establishes in turn~\eqref{consistency.mu} by
\cref{thm:consistency.Gamma}.

Notice now that $Q_*^{\diamond} - Q_* =
T^{\diamond}(Q_*^{\diamond}) - T(Q_*) =
T^{\diamond}(Q_*^{\diamond}) - T(Q_*^{\diamond}) +
T(Q_*^{\diamond}) - T(Q_*)$, that
\begin{align*}
  T^{\diamond}(Q_*^{\diamond}) - T(Q_*^{\diamond})
  = ( \Sigma_{
  s^{\prime} \given z }^{ \mu_{Q_*^{\diamond}} } -
  \hat{\Sigma}_{ s^{\prime} \given z }^{
  \mu_{Q_*^{\diamond}} } (N) ) ( Q_*^{\diamond} ) \,,
\end{align*}
and follow steps like those in \eqref{convergence.prob.i} to
establish \eqref{consistency.*} by
\cref{thm:consistency.Gamma}.

\section{Proof of \cref{thm:PI-bound}}%
\label[appendix]{app:PI-bound}

The following discussion is performed for any $\omega$
chosen arbitrarily from $E^{(\epsilon)}$ of
\cref{ass:liminf.event}, after possibly excluding from
$E^{(\epsilon)}$ the union of zero-probability events which
appear via the qualifier ``a.s.'' in
\Cref{ass:beta.infty.online,%
  ass:Qn.close.Qmun,%
  ass:Tdiammu.minus.Tdiam,%
  ass:Id-Tdiamond.bound}. By the definition of
$E^{(\epsilon)}$, $\omega \in E^{(\epsilon)}$ implies that
there exists a sufficiently large $n_0$ s.t.\ for any
$n\geq n_0$, there exists a sufficiently large $N[n]$ with
$\omega \in E_{n,N[n]}^{(\epsilon)}$.

By \cref{ass:consistency.fix}, $T_{\mu_n}^{\diamond}$ is
$\beta_{\infty}$-Lipschitz continuous, and thus a
contraction for all sufficiently large $n$ by
\cref{ass:beta.infty.online} and the discussion
after~\eqref{contraction.proof}. Recall then the
Banach-Picard fixed-point theorem~\cite{hb.plc.book}, which
guarantees that $\forall Q\in \mathcal{H}$,
$\lim_{K\to \infty} (T_{\mu_{n+1}}^{\diamond})^K(Q) =
Q_{\mu_{n+1}}^{\diamond}$, with $Q_{\mu_{n+1}}^{\diamond}$
being the unique fixed point of
$T_{\mu_{n+1}}^{\diamond}$.

\begin{lemma}\label{proof-PI-bound-1}
  For all sufficiently large $n$, a.s.,
  \begin{align*}
    \norm{Q_{\mu_{n+1}}^{\diamond} - T_{\mu_{n+1}}^{\diamond}
    (Q_{\mu_n}^\diamond)}_{\mathcal{H}} \leq
    \frac{\Delta_2}{1-\beta_\infty} \,.
  \end{align*}
\end{lemma}

\begin{IEEEproof}
  For any $k\in \IntegerPP$,
  \begin{align*}
    & \norm{(T_{\mu_{n+1}}^{\diamond})^k(Q_{\mu_n}^\diamond) -
      (T_{\mu_{n+1}}^{\diamond})^{k-1}(Q_{\mu_n}^\diamond)}_{\mathcal{H}}
    \\
    & \leq \beta_\infty \norm{(T_{\mu_{n+1}}^\diamond)^{k-1}
      (Q_{\mu_n}^\diamond) - (T_{\mu_{n+1}}^\diamond)^{k-2}
      (Q_{\mu_n}^\diamond)}_{\mathcal{H}} \\
    & \leq \beta_\infty^{k-1} \norm{(T_{\mu_{n+1}}^\diamond
      - \Id) (Q_{\mu_n}^\diamond) }_\mathcal{H} \leq
      \beta_\infty^{k-1} \Delta_2 \,,
  \end{align*}
  by \cref{ass:Id-Tdiamond.bound}. Hence, for any
  $K\in \IntegerPP$,
  \begin{align}
    & \norm{(T_{\mu_{n+1}}^\diamond)^K (Q_{\mu_n}^\diamond) -
      T_{\mu_{n+1}}^\diamond(Q_{\mu_n}^\diamond)}_{\mathcal{H}}
      \notag\\
    & \leq \sum\nolimits_{k=1}^K
      \norm{(T_{\mu_{n+1}}^\diamond)^k (Q_{\mu_n}^\diamond)
      - (T_{\mu_{n+1}}^\diamond)^{k-1}
      (Q_{\mu_n}^\diamond)}_{\mathcal{H}} \notag\\
    & \leq \sum\nolimits_{k=1}^K \beta_\infty^{k-1} \Delta_2
      \leq \Delta_2
      \sum \nolimits_{k=0}^{\infty} \beta_\infty^{k} =
      \tfrac{\Delta_2}{1-\beta_\infty}
      \,. \label{bound.Id-T.1}
  \end{align}
  Since
  $\lim_{K\to \infty}(T_{\mu_{n+1}}^\diamond)^K
  (Q_{\mu_n}^\diamond) = Q_{\mu_{n+1}}^\diamond$, the
  application of $\lim_{K\to \infty}$
  to~\eqref{bound.Id-T.1} establishes
  \cref{proof-PI-bound-1}.
\end{IEEEproof}

\begin{lemma}\label{proof-PI-bound-2}
  For all sufficiently large $n$,
  \begin{alignat*}{2}
    &&& \negphantom{ {} \leq {} }
        \norm{Q_{\mu_{n+1}}^\diamond -
        Q_{*}^\diamond}_{\mathcal{H}} \\
    & {} \leq {}
      && \beta_\infty\norm{Q_{\mu_{n}}^\diamond -
      Q_{*}^\diamond}_{\mathcal{H}} + 2 \beta_\infty
      (\Delta_0 + \epsilon) \\
    &&& + \Delta_1 + \frac{\Delta_2}{1-\beta_\infty} \,.
  \end{alignat*}
\end{lemma}

\begin{IEEEproof}
  By \cref{ass:beta.infty.online} and by following again the
  discussion after~\eqref{contraction.proof}, it can be
  verified that $T^\diamond$ is a
  $\beta_{\infty}$-contraction. Observe also by
  \cref{ass:PI-bound} and \cref{proof-PI-bound-1} that
  \begin{alignat*}{2}
    &&& \hspace{-20pt}
        \norm{Q_{\mu_{n+1}}^\diamond-Q_*^\diamond}_{\mathcal{H}}
    \\
    & {} \leq {}
      && \norm{T^\diamond(Q_n)-Q_*^\diamond}_{\mathcal{H}} +
         \norm{Q_{\mu_{n+1}}^\diamond-T^\diamond(Q_n)}_{\mathcal{H}}
    \\
    & =
      && \norm{T^\diamond(Q_n) - T^\diamond(Q_*^\diamond)
         }_{\mathcal{H}} +
         \norm{Q_{\mu_{n+1}}^\diamond-T^\diamond(Q_n)}_{\mathcal{H}}
    \\
    & \leq
      && \beta_\infty\norm{Q_n-Q_*^\diamond}_{\mathcal{H}} +
         \norm{Q_{\mu_{n+1}}^\diamond -
         T_{\mu_{n+1}}^\diamond(Q_n)}_{\mathcal{H}} \\
    &&& + \norm{T_{\mu_{n+1}}^\diamond(Q_n) -
        T^\diamond(Q_n)}_{\mathcal{H}}\\
    & \leq
      && \beta_{\infty} \norm{Q_{\mu_n}^{\diamond} -
         Q_*^\diamond }_{\mathcal{H}} +
         \beta_{\infty} \norm{Q_n -
         Q_{\mu_n}(N[n])}_{\mathcal{H}} \\
    &&& + \beta_{\infty} \norm{Q_{\mu_n}(N[n]) -
         Q_{\mu_n}^\diamond}_{\mathcal{H}} \\
    &&& + \norm{Q_{\mu_{n+1}}^\diamond -
        T_{\mu_{n+1}}^\diamond(Q_{\mu_n}^\diamond)}_{\mathcal{H}} \\
    &&& + \norm{T_{\mu_{n+1}}^\diamond(Q_{\mu_n}^\diamond) -
        T_{\mu_{n+1}}^\diamond(Q_n)}_{\mathcal{H}}\\
    &&& + \norm{T_{\mu_{n+1}}^\diamond(Q_n) -
        T^\diamond(Q_n)}_{\mathcal{H}} \\
    & \leq
      && \beta_{\infty} \norm{Q_{\mu_n}^{\diamond} -
         Q_*^\diamond }_{\mathcal{H}} +
         \beta_{\infty} (\Delta_0 + \epsilon) \\
    &&& + \tfrac{\Delta_2}{1-\beta_{\infty}} +
        \beta_{\infty} \norm{Q_{\mu_n}^\diamond -
        Q_n}_{\mathcal{H}} + \Delta_1 \\
    & \leq
      && \beta_{\infty} \norm{Q_{\mu_n}^{\diamond} -
         Q_*^\diamond }_{\mathcal{H}} +
         \beta_{\infty} (\Delta_0+\epsilon) \\
    &&& + \tfrac{\Delta_2}{1-\beta_{\infty}}
        + \beta_{\infty} \norm{Q_{\mu_n}(N[n]) -
        Q_n}_{\mathcal{H}} \\
    &&& + \beta_{\infty} \norm{Q_{\mu_n}^\diamond -
        Q_{\mu_n}(N[n])}_{\mathcal{H}} + \Delta_1 \\
    & \leq
      && \beta_{\infty} \norm{Q_{\mu_n}^{\diamond} -
         Q_*^\diamond }_{\mathcal{H}} +
         2 \beta_{\infty} (\Delta_0+\epsilon) + \Delta_1
         + \tfrac{\Delta_2}{1-\beta_{\infty}} \,,
  \end{alignat*}
  which establishes \cref{proof-PI-bound-2}.
\end{IEEEproof}

For a sufficiently large $n$, the application of
\cref{proof-PI-bound-2} recursively for $K$ times yields
\begin{alignat*}{2}
  &&& \hspace{-20pt} \norm{Q_{\mu_{n+K}}^{\diamond} -
      Q_*^{\diamond}}_{\mathcal{H}} \\
  & {} \leq {} && \beta_{\infty}^K
                  \norm{Q_{\mu_n}^{\diamond} -
                  Q_*^{\diamond}}_{\mathcal{H}} \\
  &&& + \sum\nolimits_{k=0}^{K-1} \beta_\infty^{k} \left(
      2\beta_\infty (\Delta_0 + \epsilon) + \Delta_1 + \tfrac{
      \Delta_2}{1-\beta_\infty} \right) \\
  & \leq && \beta_{\infty}^K
            \norm{Q_{\mu_n}^\diamond -
            Q_*^\diamond}_{\mathcal{H}} \\
  &&& + \underbrace{\tfrac{1}{1-\beta_\infty} \left(
      2\beta_\infty (\Delta_0 + \epsilon) + \Delta_1 +
      \tfrac{\Delta_2}{1- \beta_\infty}
      \right)}_{\Delta^{\prime}} \,,
\end{alignat*}
and because of $\beta_{\infty} < 1$,
\begin{align*}
  \limsup_{n \to \infty} \norm{Q_{\mu_{n}}^\diamond -
  Q_*^\diamond}_{\mathcal{H}} = \limsup_{K \to \infty}
  \norm{Q_{\mu_{n+K}}^\diamond - Q_*^\diamond}_{\mathcal{H}}
  \leq \Delta^{\prime} \,.
\end{align*}

Now, the triangle inequality suggests
\begin{alignat*}{2}
  &&& \negphantom{ {} \leq {} } \norm{Q_{n} -
      Q_*^{\diamond}}_{\mathcal{H}} \\
  & {} \leq {}
    && \norm{Q_n - Q_{\mu_n}(N[n])}_{\mathcal{H}} + \norm{
       Q_{\mu_n}(N[n]) - Q_{\mu_n}^{\diamond}}_{\mathcal{H}} \\
  &&& + \norm{Q_{\mu_n}^{\diamond} -
      Q_*^{\diamond}}_{\mathcal{H}} \\
  & \leq
    && \Delta_0 + \epsilon + \norm{Q_{\mu_n}^{\diamond} -
      Q_*^{\diamond}}_{\mathcal{H}}\,,
\end{alignat*}
and an application of $\limsup_{n\to \infty}$ to the
previous inequality yields
$\limsup_{n\to \infty} \norm{Q_{n} -
  Q_*^{\diamond}}_{\mathcal{H}} \leq \Delta_0 + \epsilon +
\Delta^{\prime}$, which establishes \cref{thm:PI-bound}.

\section{Proof of \cref{thm:evaluation}}%
\label[appendix]{app:thm:evaluation}

First, recall operators~\eqref{Sigma.conditional.map}
and~\eqref{hat.classical.Bellman.mu.alternative}. Define
then
\begin{align}
  \xi_n \coloneqq ( \hat{\Sigma}_{ s^{\prime} \given z
  }^{\mu_n} (N) - \Sigma_{ s^{\prime} \given z
  }^{\mu_n} )^* (\varphi(\vect{z}_{n-1})) \,, \label{xi.def}
\end{align}
where superscript $*$ over a bounded linear operator denotes
its adjoint operator~\cite{Conway}.

The linear covariance operators
$\Sigma_{zz}^{(n)}, \Sigma_{\xi z}^{(n)}, \Sigma_{\xi
  \xi}^{(n)} \colon \mathcal{H} \to \mathcal{H}$ are
introduced next; $\forall Q\in \mathcal{H}$,
\begin{subequations}\label{prop:E.loss.covariace.maps}
  \begin{alignat}{3}
    \Sigma_{zz}^{(n)} (Q)
    & {} \coloneqq {}
    && \expect \{
    && \innerp{\varphi(\vect{z}_{n})}{Q}_{\mathcal{H}}
       \cdot \varphi(\vect{z}_{n}) \}
       \,, \\
    \Sigma_{\xi z}^{(n)} (Q)
    & \coloneqq
    && \expect \{
    && \innerp{\varphi(\vect{z}_{n-1})}{Q}_{\mathcal{H}}
       \cdot \xi_n \} \,, \notag \\
    & =
    && \expect \{
    && \innerp{\varphi(\vect{z}_{n-1})}{Q}_{\mathcal{H}}
       \notag \\
    &&&&& \cdot ( \hat{\Sigma}_{ s^{\prime} \given z
          }^{\mu_n} (N) - \Sigma_{ s^{\prime} \given z
          }^{\mu_n} )^* (\varphi(\vect{z}_{n-1})) \}
          \,, \label{Sigma.xiz} \\
    \Sigma_{\xi\xi}^{(n)} (Q)
    & \coloneqq
    && \expect \{
    && \innerp{\xi_n}{Q}_{\mathcal{H}} \cdot
       \xi_n \} \notag\\
    & =
    && \expect \{
    && \innerp{ \varphi(\vect{z}_{n-1}) }
       { ( \hat{\Sigma}_{ s^{\prime}
       \given z }^{\mu_n} (N) - \Sigma_{ s^{\prime} \given z
       }^{\mu_n} ) (Q) }_{\mathcal{H}} \notag \\
    &&&&& \cdot ( \hat{\Sigma}_{ s^{\prime} \given z
          }^{\mu_n} (N) - \Sigma_{ s^{\prime} \given z
          }^{\mu_n} )^* (\varphi(\vect{z}_{n-1})) \} \,,
          \label{Sigma.xixi}
  \end{alignat}
\end{subequations}
where expectations in~\eqref{prop:E.loss.covariace.maps} are
considered along the lines of~\eqref{Sigma.def}. It can be
verified that $\Sigma_{zz}^{(n)}, \Sigma_{\xi\xi}^{(n)}$ are
self adjoint, \ie,
$\Sigma_{zz}^{(n)}{}^* = \Sigma_{zz}^{(n)}$ and
$\Sigma_{\xi\xi}^{(n)}{}^* = \Sigma_{\xi\xi}^{(n)}$.

\begin{prop}\label{prop:E.loss}
  With $\mathcal{L}_{\mu_n}^{(n)} [\vect{z}_{n-1}] (\cdot)$
  defined in~\eqref{loss.hyperplane}, its expected loss
  $G_{\mu_n} (\cdot) \coloneqq \expect
  \{\mathcal{L}_{\mu_n}^{(n)} [ \vect{z}_{n-1} ] (\cdot) \}$
  takes the following form for all sufficiently large $n$:
  \begin{alignat}{2}
    G_{\mu_n} (Q)
    & {} = {} && \tfrac{1}{2} \innerp{Q}
                 {\mathcal{A}_{\mu_n} (Q)}_{\mathcal{H}} +
                 \innerp{Q}{\mathcal{B}_{\mu_n}
                 (g)}_{\mathcal{H}} \notag \\
    &&& + \tfrac{1}{2}
        \innerp{g}{\Sigma_{zz}(g)}_{\mathcal{H}} \,, \quad
        \forall Q\in \mathcal{H}\,, \label{G(Q)-def}
  \end{alignat}
  where the linear
  $\mathcal{A}_{\mu_n}, \mathcal{B}_{\mu_n} \colon
  \mathcal{H} \to \mathcal{H}$ are defined by
  \begin{subequations}\label{prop:E.loss.maps.quadratic}
    \begin{alignat}{2}
      \mathcal{A}_{\mu_n}
      & {} \coloneqq {}
      && (\alpha \Sigma_{ s^{\prime} \given z }^{\mu_n} - \Id)^*
         \Sigma_{zz} (\alpha \Sigma_{ s^{\prime} \given z
         }^{\mu_n} - \Id) + \alpha^2
         \Sigma_{\xi\xi} \notag \\
      &&& + \alpha \Sigma_{\xi z} (\alpha \Sigma_{
          s^{\prime} \given z }^{\mu_n} - \Id) + \alpha(\alpha
          \Sigma_{ s^{\prime} \given z }^{\mu_n} - \Id)^*
          \Sigma_{\xi z}^* \label{def.map.A} \,, \\
      \mathcal{B}_{\mu_n}
      & \coloneqq
      && (\alpha \Sigma_{ s^{\prime} \given z }^{\mu_n} -
         \Id)^* \Sigma_{zz} + \alpha\Sigma_{\xi z}
         \,. \label{def.map.B}
    \end{alignat}
  \end{subequations}
\end{prop}

\begin{IEEEproof}
  Loss~\eqref{loss.hyperplane} takes the following form:
  $\forall Q\in \mathcal{H}$,
  \begin{alignat}{3}
    \mathcal{L}_{\mu_n}^{(n)} [\vect{z}_{n-1}] (Q)
    & {} = {} \tfrac{1}{2}
    &&&& \negphantom{ \Bigl[ }
         \innerp{T_{\mu_n}^{(n)}(Q) -
         Q}{\varphi( \vect{z}_{n-1} )}_{\mathcal{H}}^2 \notag
    \\
    & {} = {}
      \tfrac{1}{2}
    && \Bigl[
    && \innerp{(\alpha \Sigma_{ s^{\prime} \given z
       }^{\mu_n} - \Id)(Q)}
       {\varphi(\vect{z}_{n-1})}_{\mathcal{H}} \notag \\
    &&&&& + \alpha \innerp{\xi_n}{Q}_{\mathcal{H}}
          + \innerp{g}
          {\varphi(\vect{z}_{n-1})}_{\mathcal{H}} \Bigr]^2
          \notag \\
    & {} = {} &&&& \negphantom{ \tfrac{1}{2} \Bigl[ }
                   \textnormal{term}_1 + \textnormal{term}_2 +
                   \textnormal{term}_3 \,, \label{Loss_n}
  \end{alignat}
  where
  \begin{alignat*}{2}
    &&& \negphantom{{} \coloneqq {} \tfrac{1}{2} \Bigl[}
        \textnormal{term}_1 \notag\\
    & {} \coloneqq {}
      \tfrac{1}{2} \Bigl[
      && \innerp{(\alpha \Sigma_{ s^{\prime} \given z }^{\mu_n}
         - \Id)
         (Q)}{\varphi(\vect{z}_{n-1})}_{\mathcal{H}}^2
         + \alpha^2 \innerp{\xi_n}{Q}_{\mathcal{H}}^2 \notag \\
    &&& + 2 \alpha \innerp{(\alpha \Sigma_{ s^{\prime}
        \given z }^{\mu_n} - \Id)(Q)}
        {\varphi(\vect{z}_{n-1})}_{\mathcal{H}}
        \cdot \innerp{\xi_n}
        {Q}_{\mathcal{H}} \Bigr] \,, \\
    &&& \negphantom{{} \coloneqq {} \tfrac{1}{2} \Bigl[}
        \textnormal{term}_2 \notag\\
    & {} \coloneqq {}
      && \negphantom{\tfrac{1}{2} \Bigl[} \innerp{g}
         {\varphi(\vect{z}_{n-1})}_{\mathcal{H}} \cdot
         \innerp{(\alpha \Sigma_{ s^{\prime} \given z
         }^{\mu_n} - \Id) (Q)}
         {\varphi(\vect{z}_{n-1})}_{\mathcal{H}} \notag \\
    &&& \negphantom{\tfrac{1}{2} \Bigl[} + \alpha
        \innerp{g}
        {\varphi(\vect{z}_{n-1})}_{\mathcal{H}}
        \cdot \innerp{\xi_n}
        {Q}_{\mathcal{H}} \,, \\
    &&& \negphantom{ {} \coloneqq {} \tfrac{1}{2} \Bigl[ }
        \textnormal{term}_3 {} \coloneqq \tfrac{1}{2}
        \innerp{g} {\varphi(\vect{z}_{n-1})}_{\mathcal{H}}^2
        \,.
  \end{alignat*}

  A closer look at $\textnormal{term}_1$
  via~\eqref{Sigma.zz} suggests that
  \begin{alignat*}{2}
    &&& \negphantom{ {} = {} } \expect \{ \innerp{(\alpha
        \Sigma_{ s^{\prime} \given z }^{\mu_n} - \Id) (Q)}
        {\varphi(\vect{z}_{n-1})}_{\mathcal{H}}^2 \} \\
    & {} = {}
      && \expect \{ \langle (\alpha
         \Sigma_{ s^{\prime} \given z }^{\mu_n} - \Id)(Q)
         \given \innerp{(\alpha \Sigma_{ s^{\prime} \given z
         }^{\mu_n} - \Id)(Q)}
         {\varphi(\vect{z}_{n-1})}_{\mathcal{H}} \\
    &&& \phantom{ \expect_{\vect{z}_{n-1}} \{ } \cdot
        \varphi(\vect{z}_{n-1}) \rangle_{\mathcal{H}} \} \\
    & {} = {}
      && \langle (\alpha \Sigma_{ s^{\prime} \given z
         }^{\mu_n} - \Id)(Q) \given
         \expect \{ \innerp{(\alpha \Sigma_{ s^{\prime}
         \given z }^{\mu_n} - \Id)(Q)}
         {\varphi(\vect{z}_{n-1})}_{\mathcal{H}} \\
    &&& \phantom{ \langle (\alpha \Sigma_{ s^{\prime}
         \given z }^{\mu_n} - \Id)(Q) \given
        \expect_{\vect{z}_{n-1}}
        \{ } \cdot \varphi(\vect{z}_{n-1}) \}
        \rangle_{\mathcal{H}} \\
    & {} = {}
      && \innerp{Q}{(\alpha \Sigma_{ s^{\prime}
         \given z }^{\mu_n} - \Id)^*
         \Sigma_{zz} (\alpha \Sigma_{ s^{\prime}
         \given z }^{\mu_n} - \Id) (Q)}_{\mathcal{H}} \,,
  \end{alignat*}
  that
  \begin{align*}
    \alpha^2 \expect \{\innerp{\xi_n}
    {Q}_{\mathcal{H}}^2\}
    & = \alpha^2 \innerp{Q} {\expect \{ \innerp{\xi_n}
      {Q}_{\mathcal{H}} \xi_n \}}_{\mathcal{H}} \\
    & = \innerp{Q}
      {\alpha^2 \Sigma_{\xi\xi} (Q)}_{\mathcal{H}} \,,
  \end{align*}
  and
  \begin{alignat*}{2}
    &&& \negphantom{ {} = {} } 2 \alpha
        \expect \{\innerp{(\alpha \Sigma_{ s^{\prime}
         \given z }^{\mu_n} -
        \Id)(Q)} {\varphi(\vect{z}_{n-1})}_{\mathcal{H}}
        \innerp{\xi_n} {Q}_{\mathcal{H}}\} \\
    & {} = {}
      && 2 \alpha \innerp{Q} {\expect
         \{\innerp{(\alpha \Sigma_{ s^{\prime}
         \given z }^{\mu_n} - \Id)(Q)}
         {\varphi(\vect{z}_{n-1})}_{\mathcal{H}}\,
         \xi_n \}}_{\mathcal{H}} \\
    & =
      && \alpha \innerp{Q}{\Sigma_{\xi z} (\alpha \Sigma_{
         s^{\prime} \given z }^{\mu_n} - \Id)(Q)}_{\mathcal{H}} \\
    &&& + \alpha \innerp{Q} {\Sigma_{\xi z} (\alpha \Sigma_{
         s^{\prime} \given z }^{\mu_n} - \Id)
        (Q)}_{\mathcal{H}} \\
    & =
      && \alpha \innerp{Q} {\Sigma_{\xi z}
         (\alpha \Sigma_{
         s^{\prime} \given z }^{\mu_n} - \Id) (Q)}_{\mathcal{H}}
    \\
    &&& + \alpha \innerp{Q}{(\alpha \Sigma_{
         s^{\prime} \given z }^{\mu_n} -
        \Id)^* \Sigma_{\xi z}^*(Q)}_{\mathcal{H}} \\
    & =
      && \innerp{Q} { ( \alpha \Sigma_{\xi z}
         (\alpha \Sigma_{
         s^{\prime} \given z }^{\mu_n} - \Id) + \alpha
         ( \alpha \Sigma_{
         s^{\prime} \given z }^{\mu_n} - \Id)^*
         \Sigma_{\xi z}^*) (Q)}_{\mathcal{H}} \,.
  \end{alignat*}
  Hence,
  $\expect\{ \textnormal{term}_1 \} = (1/2)
  \innerp{Q}{\mathcal{A}_{\mu_n}
    (Q)}_{\mathcal{H}}$. Moreover,
  \begin{alignat*}{2}
    &&& \negphantom{ {} = {} } \expect\{ \textnormal{term}_2 \} \\
    & {} = {}
      && \expect
         \{ \innerp{g}
         {\varphi(\vect{z}_{n-1})}_{\mathcal{H}} \cdot
         \innerp{(\alpha \Sigma_{ s^{\prime}
         \given z }^{\mu_n} - \Id) (Q)}
         {\varphi(\vect{z}_{n-1})}_{\mathcal{H}} \\
    &&& \phantom{\expect \{} + \alpha \innerp{g}
        {\varphi(\vect{z}_{n-1})}_{\mathcal{H}} \cdot
        \innerp{\xi_n}{Q}_{\mathcal{H}}\} \\
    & =
      && \innerp{(\alpha \Sigma_{ s^{\prime}
         \given z }^{\mu_n} - \Id)(Q)}
         {\expect \{\innerp{g}
         {\varphi(\vect{z}_{n-1})}_{\mathcal{H}} \,
         \varphi(\vect{z}_{n-1}) \} }_{\mathcal{H}} \\
    &&& + \alpha \innerp{Q} {\expect \{ \innerp{g}
        {\varphi(\vect{z}_{n-1})}_{\mathcal{H}} \, \xi_n \}
        }_{\mathcal{H}} \\
    & {} = {}
      && \innerp{Q}{(\alpha \Sigma_{ s^{\prime}
         \given z }^{\mu_n} - \Id)^*
         \Sigma_{zz} (g)}_{\mathcal{H}} +
         \alpha \innerp{Q}{\Sigma_{\xi z} (g)}_{\mathcal{H}}
    \\
    & {} = {}
      && \innerp{Q}{\mathcal{B}_{\mu_n} (g)}_{\mathcal{H}}
         \,,
  \end{alignat*}
  and finally,
  \begin{align*}
    \expect\{ \textnormal{term}_3 \}
    & = \tfrac{1}{2} \expect \{ \innerp{g}
      {\varphi(\vect{z}_{n-1})}_{\mathcal{H}}^2 \} \\
    & = \tfrac{1}{2} \innerp{g}
      {\expect \{ \innerp{g}
      {\varphi(\vect{z}_{n-1})}_{\mathcal{H}}
      \varphi (\vect{z}_{n-1})\}}_{\mathcal{H}}\\
    & = \tfrac{1}{2} \innerp{g} {\Sigma_{zz}
      (g)}_{\mathcal{H}} \,,
  \end{align*}
  which completes the proof of \cref{prop:E.loss}.
\end{IEEEproof}

It is worth noting here that by the adopted assumptions and
the observation
$\mathcal{A}_{\mu_n} = \mathcal{A}_{\mu_n}^*$, operator
$\mathcal{A}_{\mu_n}$ turns out to be bounded linear and
self-adjoint.

\begin{lemma}\label{lemma:strongly-convex}
  The expected loss $G_{\mu_n}(\cdot)$~\eqref{G(Q)-def} is
  $\sigma_{\min} (\mathcal{A}_{\mu_n})$-strongly convex for
  all sufficiently large $n$.
\end{lemma}

\begin{IEEEproof}
  Verify that $\forall Q_1, Q_2 \in \mathcal{H}$,
  $\forall \gamma \in (0,1)$,
  \begin{alignat*}{2}
    &&& \negphantom{ {} = {} } \gamma G_{\mu_n} (Q_1) + (1 -
        \gamma) G_{\mu_n}(Q_2) - G_{\mu_n}( \gamma Q_1 + (1 -
        \gamma) Q_2 ) \\
    & {} = {}
      && \tfrac{1}{2} \gamma (1 - \gamma) [
         \innerp{Q_1}{\mathcal{A}_{\mu_n} (Q_1)}_{\mathcal{H}} +
         \innerp{Q_2}{\mathcal{A}_{\mu_n}
         (Q_2)}_{\mathcal{H}} \\
    &&& \phantom{ \tfrac{1}{2} \gamma (1 - \gamma) [ } - 2
        \innerp{Q_1}{\mathcal{A}_{\mu_n}
        (Q_2)}_{\mathcal{H}} ] \\
    & = && \tfrac{1}{2} \gamma ( 1 - \gamma)
           \innerp{Q_1-Q_2}{\mathcal{A}_{\mu_n}
           (Q_1 - Q_2)}_{\mathcal{H}}\\
    & \geq && \tfrac{1}{2} \gamma (1 - \gamma)\,
              \sigma_{\min} (\mathcal{A}_{\mu_n}) \,
              \norm{Q_1 - Q_2}_{\mathcal{H}} \,,
  \end{alignat*}
  and recall that $\sigma_{\min} (\mathcal{A}_{\mu_n})$ is
  assumed to be positive for all sufficiently large $n$.
\end{IEEEproof}

Given the assertion of \cref{lemma:strongly-convex}, define
the minimizer
\begin{align}
  \breve{Q}_{\mu_n}^{\diamond} \coloneqq
  \argmin\nolimits_{ Q\in\mathcal{H} } G_{\mu_n} (Q)
  \,, \label{breve.Q.def}
\end{align}
which is well defined and unique because of the coercivity
and strongly convexity of $G_{\mu_n}$~\cite{hb.plc.book}.

\begin{lemma}\label{lemma:a-b}
  For any $h_1, h_2 \in \mathcal{H}$ and any
  $Q\in \mathcal{H}$,
  \begin{align*}
    \nabla \left( \innerp{ \cdot }{\innerp{h_1}{ \cdot
    }_{\mathcal{H}}\, h_2}_{\mathcal{H}} \right) (Q) =
    \innerp{Q}{h_1}_{\mathcal{H}}\, h_2 +
    \innerp{Q}{h_2}_{\mathcal{H}}\, h_1 \,,
  \end{align*}
  where $\nabla$ stands for the Fr\'{e}chet
  gradient~\cite{hb.plc.book}.
\end{lemma}

\begin{IEEEproof}
  Notice that for any $q\in \mathcal{H}$,
  \begin{alignat*}{2}
    &&& \negphantom{ {} = {} } \innerp{Q+q}
        {\innerp{h_1}{Q+q}_{\mathcal{H}}\, h_2}_{\mathcal{H}} -
        \innerp{Q}{\innerp{h_1}{Q}_{\mathcal{H}}\,
        h_2}_{\mathcal{H}} \\
    & {} = {}
      && \innerp{q}
         {\innerp{h_1}{Q}_{\mathcal{H}}\, h_2}_{\mathcal{H}} +
         \innerp{Q} {\innerp{h_1}{q}_{\mathcal{H}}\,
         h_2}_{\mathcal{H}} \\
    &&& + \innerp{q}
        {\innerp{h_1}{q}_{\mathcal{H}}\, h_2}_{\mathcal{H}} \\
    & =
      && \innerp{q}
         { \innerp{Q}{h_1}_{\mathcal{H}}\, h_2 +
         \innerp{Q}{h_2}_{\mathcal{H}}\, h_1 }_{\mathcal{H}}
    \\
    &&& + \innerp{q} {\innerp{h_1}{q}_{\mathcal{H}}\,
         h_2}_{\mathcal{H}} \,,
  \end{alignat*}
  that
  $\lim_{0\neq \norm{q}_{\mathcal{H}} \to 0} \innerp{q}
  {\innerp{h_1}{q}_{\mathcal{H}}\, h_2}_{\mathcal{H}} /
  \norm{q}_{\mathcal{H}} = 0$, and recall the definition of
  the Fr\'{e}chet gradient~\cite{hb.plc.book} to establish
  \cref{lemma:a-b}.
\end{IEEEproof}

\begin{lemma}\label{lemma:gradient.G}\mbox{}
  \begin{lemlist}

  \item\label{lemma:differensial} For all sufficiently large
    $n$ and for any $Q\in \mathcal{F}_{n-1}$, a.s.,
    \begin{align*}
      \nabla G_{\mu_n} (Q) = \expect_{ \given
      \mathcal{F}_{n-1} } \{ \nabla \mathcal{L}_{\mu_n}^{(n)}
      [\vect{z}_{n-1}] (Q) \} \,,
    \end{align*}
    where $\expect_{ \given \mathcal{F}_{n-1} } \{ \cdot \}$
    stands for the conditional expectation, conditioned on
    the filtration $\mathcal{F}_{n-1}$.

  \item\label{lemma:Lipschitz.gradient} $\nabla G_{\mu_n}$
    is $\norm{ \mathcal{A}_{\mu_n} }$-Lipschitz continuous.

  \end{lemlist}
\end{lemma}

\begin{IEEEproof}
  Because of~\eqref{G(Q)-def},
  $\nabla G_{\mu_n}(Q) = \mathcal{A}_{\mu_n} (Q) +
  \mathcal{B}_{\mu_n} (g)$. By~\eqref{Loss_n},
  $\nabla \mathcal{L}_{\mu_n}^{(n)} [\vect{z}_{n-1}] (Q) =
  \nabla \textnormal{term}_1 + \nabla
  \textnormal{term}_2$. Following the lines of the proof of
  \cref{prop:E.loss}, notice by \cref{lemma:a-b} that
  \begin{align}
    & \expect_{ \given
      \mathcal{F}_{n-1} } \{ \nabla \langle (\alpha
      \Sigma_{ s^{\prime} \given z }^{\mu_n} - \Id)(Q)
      \given \innerp{(\alpha
      \Sigma_{ s^{\prime} \given z }^{\mu_n} - \Id)(Q)}
      {\varphi(\vect{z}_{n-1})}_{\mathcal{H}} \notag \\
    & \phantom{ = \expect_{ \given
      \mathcal{F}_{n-1} } \{ } \cdot
      \varphi(\vect{z}_{n-1}) \rangle_{\mathcal{H}} \}
    \notag \\
    & = \expect_{ \given
      \mathcal{F}_{n-1} } \{ \nabla \langle Q \given \innerp{Q}
      {(\alpha \Sigma_{ s^{\prime} \given z }^{\mu_n} - \Id)^*
      \varphi(\vect{z}_{n-1})}_{\mathcal{H}} \notag \\
    & \phantom{ = \expect_{ \given \mathcal{F}_{n-1} \{ } }
      \cdot (\alpha \Sigma_{ s^{\prime} \given z }^{\mu_n} - \Id)^*
      \varphi(\vect{z}_{n-1}) \rangle_{\mathcal{H}} \}
      \notag \\
    & = \expect_{ \given
      \mathcal{F}_{n-1} } \{ 2 \innerp{Q}{(\alpha
      \Sigma_{ s^{\prime} \given z }^{\mu_n} - \Id)^*
      \varphi( \vect{z}_{n-1}
      )}_{\mathcal{H}} (\alpha \Sigma_{ s^{\prime} \given z
      }^{\mu_n} - \Id)^* \notag \\
    & \phantom{ = \expect_{ \given \mathcal{F}_{n-1} } \{ }
      \cdot \varphi( \vect{z}_{n-1} ) \} \notag \\
    & = 2 (\alpha \Sigma_{ s^{\prime} \given z }^{\mu_n} -
      \Id)^* \notag \\
    & \phantom{ = \expect_{ \given \mathcal{F}_{n-1} } \{ }
      \cdot \expect_{ \given \mathcal{F}_{n-1} }
      \{\innerp{(\alpha \Sigma_{ s^{\prime} \given z
      }^{\mu_n} - \Id)(Q)}{\varphi( \vect{z}_{n-1}
      )}_{\mathcal{H}} \varphi( \vect{z}_{n-1} )\} \notag \\
    & = 2 (\alpha \Sigma_{ s^{\prime} \given z }^{\mu_n} -
      \Id)^* \notag \\
    & \phantom{ = \expect_{ \given \mathcal{F}_{n-1} } \{ }
      \cdot \expect \{\innerp{(\alpha \Sigma_{ s^{\prime}
      \given z }^{\mu_n} - \Id)(Q)}{\varphi( \vect{z}_{n-1}
      )}_{\mathcal{H}}
      \varphi( \vect{z}_{n-1} )\} \label{use.independency} \\
    & = 2 (\alpha \Sigma_{ s^{\prime} \given z }^{\mu_n} - \Id)^*
      \Sigma_{zz} (\alpha \Sigma_{ s^{\prime} \given z
      }^{\mu_n} - \Id)(Q) \,, \notag
  \end{align}
  where \cref{ass:independency} was used
  in~\eqref{use.independency}. Furthermore,
  \begin{align*}
    & \alpha^2 \expect_{ \given \mathcal{F}_{n-1} } \{ \nabla
    \innerp{Q} {\innerp{Q}{\xi_n}_{\mathcal{H}}
    \xi_n}_{\mathcal{H}} \} \\
    & = \alpha^2 \expect_{ \given \mathcal{F}_{n-1} } \{ 2
      \innerp{Q}{\xi_n}_{\mathcal{H}} \xi_n\} \\
    & = \alpha^2 \expect \{ 2
      \innerp{Q}{\xi_n}_{\mathcal{H}} \xi_n\} \\
    & = 2 \alpha^2 \Sigma_{\xi\xi}(Q) \,,
  \end{align*}
  and
  \begin{alignat*}{2}
    &&& \negphantom{ {} = {} } \alpha
        \expect_{ \given \mathcal{F}_{n-1} } \{ \nabla
        \innerp{Q} {\innerp{(\alpha \Sigma_{ s^{\prime}
        \given z }^{\mu_n} - \Id)(Q)} {\varphi(
        \vect{z}_{n-1} )}_{\mathcal{H}} \xi_n}_{\mathcal{H}}\} \\
    & {} = {}
      && \alpha
         \expect_{ \given \mathcal{F}_{n-1} } \{ \nabla
         \innerp{Q} {\innerp{Q} { (\alpha
         \Sigma_{ s^{\prime} \given z }^{\mu_n} - \Id)^*
         \varphi( \vect{z}_{n-1} )}_{\mathcal{H}}
         \xi_n}_{\mathcal{H}}\} \\
    & =
      && \alpha \expect_{ \given \mathcal{F}_{n-1} } \{
         \innerp{Q}{ (\alpha \Sigma_{ s^{\prime} \given z
         }^{\mu_n} - \Id)^* (\varphi(
         \vect{z}_{n-1} ))}_{\mathcal{H}} \xi_n \\
    &&& \phantom{ \alpha \expect_{ \given \mathcal{F}_{n-1}
        } \{ } + \innerp{Q}{\xi_n}_{\mathcal{H}} ( \alpha
        \Sigma_{ s^{\prime} \given z }^{\mu_n} - \Id)^* (\varphi(
        \vect{z}_{n-1} ))\} \\
    & =
      && \alpha \expect_{ \given \mathcal{F}_{n-1} } \{
         \innerp{ (\alpha \Sigma_{ s^{\prime} \given z
         }^{\mu_n} - \Id)(Q)} {\varphi(
         \vect{z}_{n-1} )}_{\mathcal{H}} \xi_n \} \\
    &&& \phantom{ \alpha \expect_{ \given \mathcal{F}_{n-1}
        } \{ } + \alpha (\alpha \Sigma_{ s^{\prime} \given z
        }^{\mu_n} - \Id)
        \expect_{ \given \mathcal{F}_{n-1} }
        \{\innerp{Q}{\xi_n}_{\mathcal{H}} \varphi(
        \vect{z}_{n-1} )\} \\
    & =
      && \alpha \expect \{
         \innerp{ (\alpha \Sigma_{ s^{\prime} \given z
         }^{\mu_n} - \Id)(Q)} {\varphi(
         \vect{z}_{n-1} )}_{\mathcal{H}} \xi_n \} \\
    &&& \phantom{ \alpha \expect \{ } + \alpha (\alpha
        \Sigma_{ s^{\prime} \given z }^{\mu_n} - \Id)
        \expect \{\innerp{Q}{\xi_n}_{\mathcal{H}} \varphi(
        \vect{z}_{n-1} )\} \\
    & =
      && \left( \alpha
         \Sigma_{\xi z} (\alpha \Sigma_{ s^{\prime} \given z
         }^{\mu_n} - \Id)
         + \alpha (\alpha \Sigma_{ s^{\prime} \given z
         }^{\mu_n} - \Id)^* \Sigma_{\xi z}^* \right) (Q) \,,
  \end{alignat*}
  where \cref{ass:independency} was used again as
  in~\eqref{use.independency} to replace conditional
  expectations by $\expect\{ \cdot \}$. The previous
  derivations suggest
  $\expect_{ \given \mathcal{F}_{n-1} } \{ \nabla
  \textnormal{term}_1 \} =
  \mathcal{A}_{\mu_n}(Q)$. Moreover,
  \begin{alignat*}{2}
    &&& \negphantom{ {} = {} } \expect_{ \given
        \mathcal{F}_{n-1} } \{ \nabla \textnormal{term}_2 \}
    \\
    & {} = {}
      && \expect_{ \given \mathcal{F}_{n-1} } \{ \nabla
         \innerp{g} {\varphi( \vect{z}_{n-1}
         )}_{\mathcal{H}} \innerp{ (\alpha \Sigma_{
         s^{\prime} \given z }^{\mu_n} - \Id)(Q)} {\varphi(
         \vect{z}_{n-1} )}_{\mathcal{H}}
    \\
    &&& \phantom{ \expect_{ \given \mathcal{F}_{n-1}
        } \{ } + \alpha \nabla \innerp{g}
        {\varphi( \vect{z}_{n-1} )}_{\mathcal{H}}
        \innerp{\xi_n}{Q}_{\mathcal{H}}\} \\
    & =
      && (\alpha \Sigma_{ s^{\prime} \given z
         }^{\mu_n} - \Id)^* \expect_{
         \given \mathcal{F}_{n-1} }
         \{\innerp{g} {\varphi( \vect{z}_{n-1} )}_{\mathcal{H}}
         \varphi( \vect{z}_{n-1} ) \} \\
    &&& \phantom{ \expect_{ \given \mathcal{F}_{n-1}
        } \{ } + \alpha  \expect_{ \given \mathcal{F}_{n-1}
        } \{ \innerp{g} {\varphi( \vect{z}_{n-1} )} \xi_n \}
    \\
    & =
      && (\alpha \Sigma_{ s^{\prime} \given z
         }^{\mu_n} - \Id)^* \expect
         \{\innerp{g} {\varphi( \vect{z}_{n-1} )}_{\mathcal{H}}
         \varphi( \vect{z}_{n-1} ) \} \\
    &&& \phantom{ \expect_{ \given \mathcal{F}_{n-1}
        } \{ } + \alpha  \expect \{ \innerp{g} {\varphi(
        \vect{z}_{n-1} )} \xi_n \} \\
    & = && (\alpha \Sigma_{ s^{\prime} \given z
         }^{\mu_n} - \Id)^* \Sigma_{zz}
           (g) + \alpha \Sigma_{\xi z} (g) =
           \mathcal{B}_{\mu_n} (g) \,.
  \end{alignat*}

  Gathering all of the previous results,
  $\expect_{ \given \mathcal{F}_{n-1} } \{ \nabla
  \textnormal{term}_1 + \nabla \textnormal{term}_2 \} =
  \mathcal{A}_{\mu_n}(Q) + \mathcal{B}_{\mu_n} (g) = \nabla
  G_{\mu_n}(Q)$, which establishes the proof of
  \cref{lemma:differensial}.

  The proof of \cref{lemma:Lipschitz.gradient} follows
  directly from the observation that $\forall Q_1, Q_2\in
  \mathcal{H}$,
  \begin{align*}
    \norm{ \nabla G_{\mu_n}(Q_1) - \nabla G_{\mu_n}(Q_2)
    }_{\mathcal{H}}
    & = \norm{ \mathcal{A}_{\mu_n}(Q_1) -
      \mathcal{A}_{\mu_n}(Q_2) }_{\mathcal{H}} \\
    & \leq \norm{ \mathcal{A}_{\mu_n} } \, \norm{ Q_1 - Q_2
      }_{\mathcal{H}} \,,
  \end{align*}
  where it can be also observed
  by~\eqref{prop:E.loss.maps.quadratic} that
  \begin{align*}
    \norm{ \mathcal{A}_{\mu_n} } {} \leq {}
    & \norm{ \Sigma_{zz} }\,
      \norm{ \alpha \Sigma_{ s^{\prime} \given z
      }^{\mu_n} - \Id}^2 + \alpha^2
      \norm{ \Sigma_{\xi\xi} } \\
    & + 2\alpha \norm{ \Sigma_{\xi z} }\, \norm{ \alpha
      \Sigma_{ s^{\prime} \given z
      }^{\mu_n} - \Id} \,.
  \end{align*}
\end{IEEEproof}

\begin{lemma}\label{lemma:differencial-bound}
  There exist $c_1, c_2 \in \RealPP$ s.t.\ for all
  sufficiently large $n$ and for any
  $Q \in \mathcal{F}_{n-1}$, a.s.,
  \begin{align*}
    \expect_{ \given \mathcal{F}_{n-1} } \{ \norm{
    \nabla \mathcal{L}_{\mu_n}^{(n)} [\vect{z}_{n-1}] (Q) -
    \nabla G(Q) }_{\mathcal{H}}^2 \} \leq c_1
    \norm{Q}_{\mathcal{H}}^2 + c_2 \,.
  \end{align*}
\end{lemma}

\begin{IEEEproof}
  Because of \cref{lemma:differensial},
  \begin{align*}
    & \expect_{ \given \mathcal{F}_{n-1} } \{ \norm{
      \nabla \mathcal{L}_{\mu_n}^{(n)} [\vect{z}_{n-1}] (Q)
      - \nabla G(Q) }_{\mathcal{H}}^2 \} \notag\\
    & = \expect_{ \given \mathcal{F}_{n-1} } \{ \norm{
      \nabla \mathcal{L}_{\mu_n}^{(n)}
      [\vect{z}_{n-1}] (Q) }_{\mathcal{H}}^2 \} -
      \norm{ \nabla G(Q)
      }_\mathcal{H}^2 \\
    & \leq \expect_{ \given \mathcal{F}_{n-1} } \{
      \norm{ \nabla \mathcal{L}_{\mu_n}^{(n)}
      [\vect{z}_{n-1}] (Q) }_{\mathcal{H}}^2 \} \,.
  \end{align*}

  By following the proof of \cref{lemma:differensial}, it
  can be readily verified that $\nabla
  \mathcal{L}_{\mu_n}^{(n)} [\vect{z}_{n-1}] (Q) =
  \mathcal{A}_{\mu_n}^{(n)} [ \vect{z}_{n-1} ] (Q) +
  \mathcal{B}_{\mu_n}^{(n)} [ \vect{z}_{n-1} ] (g)$, where
  the mappings
  \begin{alignat*}{2}
    \mathcal{A}_{\mu_n}^{(n)} [ \vect{z}_{n-1} ] (\cdot)
    & {} \coloneqq {}
    && \innerp{\cdot}{(\alpha \Sigma_{ s^{\prime} \given z
         }^{\mu_n} - \Id)^* \varphi
       (\vect{z}_{n-1})}_{\mathcal{H}} \\
    &&& \hspace{30pt} \cdot (\alpha \Sigma_{ s^{\prime} \given z
         }^{\mu_n} - \Id)^* \varphi(\vect{z}_{n-1}) \\
    &&& + \alpha \innerp{\cdot}{(\alpha \Sigma_{ s^{\prime} \given z
         }^{\mu_n} - \Id)^* \varphi
        (\vect{z}_{n-1})}_{\mathcal{H}} \xi_n\\
    &&& + \alpha \innerp{\cdot} {\xi_n}_{\mathcal{H}}
        (\alpha \Sigma_{ s^{\prime} \given z }^{\mu_n} - \Id)^*
        \varphi (\vect{z}_{n-1}) \\
    &&& + \alpha^2 \innerp{\cdot} {\xi_n}_{\mathcal{H}} \xi_n
        \,, \\
    \mathcal{B}_{\mu_n}^{(n)} [ \vect{z}_{n-1} ] (\cdot)
    & \coloneqq
    && (\alpha \Sigma_{ s^{\prime} \given z
         }^{\mu_n} - \Id)^* \innerp{\cdot} { \varphi
       (\vect{z}_{n-1})}_{\mathcal{H}} \varphi
       (\vect{z}_{n-1}) \\
    &&& + \alpha \innerp{\cdot} {\varphi
       (\vect{z}_{n-1})}_{\mathcal{H}} \xi_n \,.
  \end{alignat*}

  Notice now by \cref{ass:T.mu.diamond.contraction} that for
  any $Q\in \mathcal{H}$,
  \begin{align*}
    \norm{ \Sigma_{ s^{\prime} \given z }^{\mu_n} (Q)
    }_{\mathcal{H}}
    & = \norm{ \Sigma_{ s^{\prime} \given z }^{\mu_n} (Q) -
      \Sigma_{ s^{\prime} \given z }^{\mu_n} (0) }_{\mathcal{H}} \\
    & = \tfrac{1}{\alpha} \norm{ T_{\mu_n}^{\diamond} (Q) -
      T_{\mu_n}^{\diamond}(0) }_{\mathcal{H}} \\
    & \leq \tfrac{1}{\alpha} \beta_{\infty} \norm{ Q - 0
      }_{\mathcal{H}} = \tfrac{1}{\alpha} \beta_{\infty}
      \norm{ Q }_{\mathcal{H}} \,,
  \end{align*}
  which suggests that
  $\norm{ \Sigma_{ s^{\prime} \given z }^{\mu_n} } \leq
  \beta_{\infty} / \alpha$. This implies in turn
  $\norm{\alpha \Sigma_{ s^{\prime} \given z }^{\mu_n} -
    \Id}_\mathcal{H} \leq \alpha \norm{ \Sigma_{ s^{\prime}
      \given z }^{\mu_n} } + \norm{\Id} \leq \alpha
  \beta_{\infty} / \alpha + 1 \leq \beta_{\infty} +
  1$. Moreover, the reproducing property of the kernel
  $\kappa$ yields
  $\norm{\varphi (\vect{z}_{n-1})}_{\mathcal{H}}^2 = \kappa(
  \vect{z}_{n-1}, \vect{z}_{n-1}) \leq B_{\kappa}$. Notice
  also that
  \begin{alignat*}{2}
    &&& \negphantom{ {} \leq {} } \norm{
        \mathcal{A}_{\mu_n}^{(n)} [ \vect{z}_{n-1} ]
        (Q)}_{\mathcal{H}} \\
    & {} \leq {}
      && \Bigl[ \norm{\alpha \Sigma_{ s^{\prime} \given z
         }^{\mu_n} - \Id}^2
         \norm{\varphi (\vect{z}_{n-1})}_{\mathcal{H}}^2 \\
    &&& + 2\alpha \norm{\alpha \Sigma_{ s^{\prime} \given z
        }^{\mu_n} - \Id}
        \, \norm{ \varphi (\vect{z}_{n-1}) }_{\mathcal{H}}
        \, \norm{\xi_n}_{\mathcal{H}} \\
    &&& + \alpha^2 \norm{ \xi_n }_{\mathcal{H}}^2 \Bigr]
        \norm{Q}_{\mathcal{H}} \\
    & \leq
      && \Bigl[ (\beta_{\infty} + 1)^2
         B_{\kappa} + 2\alpha (\beta_{\infty} + 1) \sqrt{
         B_{\kappa} } \norm{ \xi_n }_{\mathcal{H}} \\
    &&& + \alpha^2 \norm{ \xi_n
        }_{\mathcal{H}}^2 \Bigr] \norm{Q}_{\mathcal{H}} \,, \\
    &&& \negphantom{ {} \leq {} }
        \norm{\mathcal{B}_{\mu_n}^{(n)} [ \vect{z}_{n-1} ]
        (g)}_{\mathcal{H}} \\
    & \leq
      && \norm{\alpha \Sigma_{ s^{\prime} \given z }^{\mu_n}
         - \Id}
         \, \norm{ \varphi
         (\vect{z}_{n-1}) }_{\mathcal{H}}^2 + \alpha
         \norm{\xi_n}_{\mathcal{H}}\, \norm{g}_{\mathcal{H}} \\
    & \leq
      && (\beta_{\infty} + 1) B_{\kappa} + \alpha \norm{
         \xi_n }_{\mathcal{H}}\, \norm{g}_{\mathcal{H}}  \,.
  \end{alignat*}

  Observe that function
  $(\cdot)^{4/i} \colon \Real \to \Real$ is convex
  $\forall i\in \{1, 2, 3\}$, and recall Jensen's inequality
  for conditional expectation~\cite{Williams:probability:91}
  to verify that
  $( \expect_{ \given \mathcal{F}_{n-1} } \{ \norm{ \xi_n
  }_{\mathcal{H}}^i \} )^{4/i} \leq \expect_{ \given
    \mathcal{F}_{n-1} } \{ \norm{ \xi_n
  }_{\mathcal{H}}^{i\cdot 4/i} \} = \expect \{ \norm{ \xi_n
  }_{\mathcal{H}}^4 \} = \mathfrak{m}_{\xi}^{(4)}$ for all
  sufficiently large $n$, $\forall i\in \{1, 2, 3\}$. These
  arguments suggest that there exist
  $\{ \varrho_i \}_{i=0}^4 \subset \RealP$ s.t.\
  \begin{alignat*}{2}
    &&& \negphantom{ {} \leq {} } \expect_{ \given
        \mathcal{F}_{n-1} } \{ \norm{
        \mathcal{A}_{\mu_n}^{(n)} [ \vect{z}_{n-1} ] (Q)
        }_{\mathcal{H}}^2 \} \\
    & {} \leq {}
      && \expect_{ \given
         \mathcal{F}_{n-1} } \Bigl\{ \Bigl[ (\beta_{\infty} +
         1)^2 B_{\kappa} + 2\alpha (\beta_{\infty} + 1)
         \sqrt{ B_{\kappa} } \norm{ \xi_n }_{\mathcal{H}} \\
    &&& \phantom{ \expect_{ \given
        \mathcal{F}_{n-1} } \Bigl\{ } + \alpha^2 \norm{
        \xi_n }_{\mathcal{H}}^2 \Bigr]^2 \Bigr\}
        \norm{Q}_{\mathcal{H}}^2 \\
    & =
      && \expect_{ \given
         \mathcal{F}_{n-1} } \left\{ \sum\nolimits_{i=0}^4
         \varrho_i \norm{\xi_n}_{\mathcal{H}}^i \right\}
         \norm{Q}_{\mathcal{H}}^2 \\
    & \leq
      && \left[ \sum\nolimits_{i=0}^4
         \varrho_i \left( \expect_{ \given
         \mathcal{F}_{n-1} } \left\{
         \norm{\xi_n}_{\mathcal{H}}^4 \right\} \right)^{i/4}
         \right] \norm{Q}_{\mathcal{H}}^2 \\
    & =
      && \underbrace{ \left( \sum\nolimits_{i=0}^4
         \varrho_i ( \mathfrak{m}_{\xi}^{(4)} )^{i/4} \right)
         }_{ c_1^{\prime} }
         \norm{Q}_{\mathcal{H}}^2 \leq
         c_1^{\prime} \norm{Q}_{\mathcal{H}}^2
         \,, \\
    &&& \negphantom{ {} \leq {} } \expect_{ \given
        \mathcal{F}_{n-1} } \{
        \norm{\mathcal{B}_{\mu_n}^{(n)} [ \vect{z}_{n-1} ]
        (g)}_{\mathcal{H}}^2 \} \\
    & \leq
      && (\beta_{\infty} + 1)^2 B_{\kappa}^2 \\
    &&& + 2\alpha
        (\beta_{\infty} + 1) B_{\kappa}
        \norm{g}_{\mathcal{H}}\, \expect_{ \given
        \mathcal{F}_{n-1} } \{ \norm{ \xi_n
        }_{\mathcal{H}} \}  \\
    &&& + \alpha^2 \norm{g}_{\mathcal{H}}^2\, \expect_{
        \given \mathcal{F}_{n-1} } \{ \norm{ \xi_n
        }_{\mathcal{H}}^2 \}  \\
    & \leq
      && (\beta_{\infty} + 1)^2 B_{\kappa}^2 \\
    &&& + 2\alpha
        (\beta_{\infty} + 1) B_{\kappa}
        \norm{g}_{\mathcal{H}} \, ( \expect_{ \given
        \mathcal{F}_{n-1} } \{ \norm{ \xi_n
        }_{\mathcal{H}}^4 \} )^{1/4}  \\
    &&& + \alpha^2 \norm{g}_{\mathcal{H}}^2 \, ( \expect_{
        \given \mathcal{F}_{n-1} } \{ \norm{ \xi_n
        }_{\mathcal{H}}^4 \} )^{2/4} \\
    & =
      && (\beta_{\infty} + 1)^2 B_{\kappa}^2 + 2\alpha
         (\beta_{\infty} + 1) B_{\kappa}
         \norm{g}_{\mathcal{H}} \, ( \mathfrak{m}_{\xi}^{(4)}
         )^{1/4} \\
    &&& + \alpha^2 \norm{g}_{\mathcal{H}}^2 \, (
        \mathfrak{m}_{\xi}^{(4)} )^{1/2} \\
    & {} \eqqcolon {} && c_2^{\prime} \,.
  \end{alignat*}

  Consequently,
  \begin{alignat*}{2}
    \negphantom{ {} \leq {} } \expect_{ \given
    \mathcal{F}_{n-1} } \{ \norm{ \nabla
    \mathcal{L}_{\mu_n}^{(n)} [\vect{z}_{n-1}] (Q)
    }_{\mathcal{H}}^2 \}
    & {} \leq {}
    && 2 c_1^{\prime}{}^2
       \norm{Q}_{\mathcal{H}}^2 + 2 c_2^{\prime}{}^2 \\
    & =
    && c_1 \norm{Q}_{\mathcal{H}}^2 + c_2 \,,
  \end{alignat*}
  where $c_1 \coloneqq 2 c_1^{\prime}{}^2$ and $c_2
  \coloneqq 2 c_2^{\prime}{}^2$. This completes the proof.
\end{IEEEproof}

An inspection of \cite[Lemma~3.1]{sayed2014adaptation},
under the light of \Cref{lemma:strongly-convex,%
  lemma:gradient.G,%
  lemma:differencial-bound}, suggests that for any
sufficiently small step size $\eta$, or more specifically,
for any
\begin{align*}
  \eta < \frac{2 \sigma_{\min}( \mathcal{A}_{\mu_n} ) } {
  \sigma_{\max}^2 ( \mathcal{A}_{\mu_n} ) + 2 c_1 } \,,
\end{align*}
there exists $c_3 \in \RealPP$ s.t.\
\begin{align}
  \limsup \nolimits_{n \to \infty}\, \expect\{ \norm{ Q_n -
  \breve{Q}_{\mu_n}^{\diamond} }_{\mathcal{H}}^2 \} \leq c_3
  \eta \,. \label{Qn.to.Qmin}
\end{align}

Now, because $T_{\mu_n}^{\diamond}$ is a contraction, its
fixed point $Q_{\mu_n}^{\diamond}$ is unique and satisfies
$(\alpha \Sigma_{ s^{\prime} \given z }^{\mu_n} - \Id)
(Q_{\mu_n}^{\diamond}) = -g$. This implies that the linear
$(\alpha \Sigma_{ s^{\prime} \given z }^{\mu_n} - \Id)$ is
non-singular, because any $Q$ in the null space of
$(\alpha \Sigma_{ s^{\prime} \given z }^{\mu_n} - \Id)$
satisfies
$(\alpha \Sigma_{ s^{\prime} \given z }^{\mu_n} - \Id)
(Q_{\mu_n}^{\diamond} - Q) = -g$, and thus
$Q = 0$~\cite[Thm.~2.6-10(a)]{kreyszig:91}. Therefore,
$(\alpha \Sigma_{ s^{\prime} \given z }^{\mu_n} - \Id)^*
(\alpha \Sigma_{ s^{\prime} \given z }^{\mu_n} - \Id)$ is
also non-singular. Consequently, with
$\sigma_{\min} (\cdot)$ defined by~\eqref{sigma.min.def},
$\sigma_{\min}( (\alpha \Sigma_{ s^{\prime} \given z
}^{\mu_n} - \Id)^* (\alpha \Sigma_{ s^{\prime} \given z
}^{\mu_n} - \Id) ) > 0$, because otherwise, and
by~\eqref{sigma.min.def}, there would exist a sequence
$(h_k)_{k\in \IntegerP} \subset \mathcal{H}$, with
$\norm{ h_k }_{\mathcal{H}} = 1$, s.t.\
$\lim_{k \to \infty} \norm{ (\alpha \Sigma_{ s^{\prime}
    \given z }^{\mu_n} - \Id) h_k}_{\mathcal{H}}^2 = 0
\Leftrightarrow \lim_{k \to \infty} ( h_k^{\prime} \coloneqq
(\alpha \Sigma_{ s^{\prime} \given z }^{\mu_n} - \Id) h_k ) =
0 \Leftrightarrow \lim_{k \to \infty} ( h_k = (\alpha \Sigma_{
  s^{\prime} \given z }^{\mu_n} - \Id)^{-1} h_k^{\prime} ) = 0
\Rightarrow 0 = \lim_{k \to \infty} \norm{h_k}_{\mathcal{H}}
= 1$, which is absurd. Moreover, notice that because of
\cref{lemma:strongly-convex}, $G_{\mu_n}$ is
coercive~\cite{hb.plc.book}, and there exists thus
$B_{\breve{q}} \in \RealPP$ s.t.\
$\norm{ \breve{Q}_{\mu_n} }_{\mathcal{H}} \leq
B_{\breve{q}}$, for all sufficiently large $n$, via
\cref{ass:stationary.policy}.

Notice that there exists $\mathfrak{m}_{\xi}^{(2)} \in
\RealPP$ s.t.\
\begin{subequations}\label{e-E+V}
  \begin{alignat}{2}
    \norm{\Sigma_{\xi\xi}}
    & {} = {}
    && \sup_{\norm{h}_{\mathcal{H}} = 1}
       \innerp{h} {\expect_{\xi_n}
       \{\innerp{h}{\xi_n}_{\mathcal{H}} \, \xi_n
       \} }_{\mathcal{H}} \notag \\
    & =
    && \sup_{\norm{h}_{\mathcal{H}} = 1} \expect_{\xi_n}
       \{ \innerp{h} {\innerp{h}{\xi_n}_{\mathcal{H}} \, \xi_n
       }_{\mathcal{H}} \} \notag \\
    & =
    && \expect \{ \norm{ \xi_n }_{\mathcal{H}}^2 \}
       \eqqcolon \mathfrak{m}_{\xi}^{(2)} \,, \label{e-E} \\
    \intertext{and}
    \norm{\Sigma_{\xi z}}
    & =
    && \sup_{\norm{h}_{\mathcal{H}} = 1} \innerp{h} {
       \expect_{\vect{z}_{n-1}, \xi_n}
       \{\innerp{h}{ \varphi(\vect{z}_{n-1}) }_{\mathcal{H}}\,
        \xi_n \}}_{\mathcal{H}} \notag\\
    & =
    && \sup_{\norm{h}_{\mathcal{H}} = 1}
       \expect_{\vect{z}_{n-1}, \xi_n} \{ \innerp{h} {
       \innerp{h}{ \varphi(\vect{z}_{n-1}) }_{\mathcal{H}}\,
        \xi_n }_{\mathcal{H}} \} \notag\\
    & \leq
    && \sup_{\norm{h}_{\mathcal{H}} = 1}
       \norm{h}_{\mathcal{H}}^2 \expect_{\vect{z}_{n-1}, \xi_n}
       \{ \norm{ \xi_n }_{\mathcal{H}} \, \norm{ \varphi(
       \vect{z}_{n-1} ) }_{\mathcal{H}} \} \notag\\
    & \leq
    && \sqrt{B_{\kappa}}
       \expect \{ \norm{ \xi_n }_{\mathcal{H}} \}
       \leq ( B_{\kappa} \mathfrak{m}_{\xi}^{(2)} )^{1/2}
       \,, \label{e-V}
  \end{alignat}
\end{subequations}
where Jensen's inequality
$\expect \{ \norm{\xi_n}_{\mathcal{H}} \} \leq (
\mathfrak{m}_{\xi}^{(2)})^{1/2}$~\cite{Williams:probability:91},
propelled by the convexity of the function $(\cdot)^2$, was
used in \eqref{e-V}. Moreover, define
\begin{alignat*}{2}
  \tilde{\mathcal{A}}_{\mu_n}
  & {} \coloneqq {}
  && \mathcal{A}_{\mu_n} - (\alpha \Sigma_{ s^{\prime}
     \given z }^{\mu_n} -
     \Id)^* \Sigma_{zz} ( \alpha \Sigma_{ s^{\prime} \given
     z }^{\mu_n} - \Id ) \\
  & =
  && \alpha \Sigma_{\xi z} (\alpha \Sigma_{ s^{\prime}
     \given z }^{\mu_n} - \Id) + \alpha(\alpha \Sigma_{
     s^{\prime} \given z }^{\mu_n} - \Id)^* \Sigma_{\xi
     z}^* + \alpha^2 \Sigma_{\xi\xi} \,, \\
  \tilde{\mathcal{B}}_{\mu_n}
  & \coloneqq
  && \mathcal{B}_{\mu_n} - (\alpha \Sigma_{ s^{\prime}
     \given z }^{\mu_n} - \Id)^* \Sigma_{zz} =
     \alpha\Sigma_{\xi z} \,.
\end{alignat*}
Via~\eqref{e-E+V},
\begin{alignat*}{2}
  \norm{ \tilde{\mathcal{A}}_{\mu_n} }
  & {} \leq {}
  && 2\alpha ( B_{\kappa} \mathfrak{m}_{\xi}^{(2)} )^{1/2}
     \norm{(\alpha \Sigma_{ s^{\prime} \given z }^{\mu_n} -
     \Id)} + \alpha^2 \mathfrak{m}_{\xi}^{(2)} \\
  & \leq
  && 2\alpha ( B_{\kappa} \mathfrak{m}_{\xi}^{(2)} )^{1/2}
     ( \beta_{\infty} + 1 ) + \alpha^2
     \mathfrak{m}_{\xi}^{(2)} \,, \\
  \norm{\tilde{\mathcal{B}}_{\mu_n} }
  & \leq
  && \alpha ( B_{\kappa} \mathfrak{m}_{\xi}^{(2)} )^{1/2}
     \,.
\end{alignat*}

Recall $(\alpha \Sigma_{ s^{\prime} \given z }^{\mu_n} - \Id)
(Q_{\mu_n}^{\diamond}) + g = 0$, and observe
$\nabla G (\breve{Q}_{\mu_n}^{\diamond}) = 0$, because of
the convexity of $G_{\mu_n} (\cdot)$. Hence,
\begin{alignat*}{2}
    0 & {} = {}
    && (\alpha \Sigma_{ s^{\prime} \given z }^{\mu_n} -
       \Id)^* \Sigma_{zz} (\alpha \Sigma_{ s^{\prime} \given
       z }^{\mu_n} - \Id) (Q_{\mu_n}^{\diamond}) \\
      &&& + (\alpha \Sigma_{ s^{\prime} \given z }^{\mu_n} -
          \Id)^* \Sigma_{zz}(g) \,, \\
    0 & =
    && \mathcal{A}_{\mu_n} ( \breve{Q}_{\mu_n}^{\diamond}) +
       \mathcal{B}_{\mu_n} (g) \,,
\end{alignat*}
which lead in turn to
\begin{alignat*}{2}
  0 & {} = {}
  && \lVert
     (\alpha \Sigma_{ s^{\prime} \given z }^{\mu_n} - \Id)^*
     \Sigma_{zz} (\alpha \Sigma_{ s^{\prime} \given z
     }^{\mu_n} - \Id) (Q_{\mu_n}^{\diamond}) \\
    &&& \phantom{ \lVert } + (\alpha \Sigma_{ s^{\prime}
        \given z }^{\mu_n} - \Id)^*
        \Sigma_{zz}(g) - [ \mathcal{A}_{\mu_n}(
        \breve{Q}_{\mu_n}^{\diamond}) + \mathcal{B}(g) ]
        \rVert_{\mathcal{H}} \\
    & = && \lVert (\alpha \Sigma_{ s^{\prime} \given z
           }^{\mu_n} - \Id)^*
           \Sigma_{zz} (\alpha \Sigma_{ s^{\prime} \given z
           }^{\mu_n} - \Id) (Q_{\mu_n}^\diamond -
           \breve{Q}_{\mu_n}^{\diamond}) \\
    &&& \phantom{ \lVert } - [ \tilde{\mathcal{A}}_{\mu_n}
        (\breve{Q}_{\mu_n}^{\diamond}) +
        \tilde{\mathcal{B}}_{\mu_n} (g) ] \rVert_{\mathcal{H}} \\
    & \geq
  && \norm{(\alpha \Sigma_{ s^{\prime} \given z }^{\mu_n} - \Id)^*
     \Sigma_{zz} (\alpha \Sigma_{ s^{\prime} \given z
     }^{\mu_n} - \Id) (Q_{\mu_n}^{\diamond} -
     \breve{Q}_{\mu_n}^{\diamond})}_{\mathcal{H}} \\
    &&& \phantom{ \lVert } - \norm{ \tilde{\mathcal{A}}_{\mu_n}
        (\breve{Q}_{\mu_n}^{\diamond}) +
        \tilde{\mathcal{B}}_{\mu_n} (g) }_{\mathcal{H}} \\
    & \geq
  && \sigma_{\min} ( (\alpha \Sigma_{ s^{\prime} \given z
     }^{\mu_n} - \Id)^* (\alpha \Sigma_{ s^{\prime} \given z
     }^{\mu_n} - \Id) )\, \sigma_{\min} ( \Sigma_{zz} ) \\
    &&& \phantom{ \lVert } \cdot \norm{Q_{\mu_n}^{\diamond}
        - \breve{Q}_{\mu_n}^{\diamond}}_{\mathcal{H}} - \norm{
        \tilde{\mathcal{A}}_{\mu_n}}\,
        \norm{\breve{Q}_{\mu_n}^{\diamond}}_{\mathcal{H}} -
        \norm{ \tilde{\mathcal{B}}_{\mu_n}}\,
        \norm{g}_{\mathcal{H}} \,.
\end{alignat*}
The previous discussion suggests that there exists $c_4
\in \RealPP$ s.t.\
\begin{alignat*}{2}
  &&& \negphantom{ {} \leq {} } \norm{Q_{\mu_n}^{\diamond} -
      \breve{Q}_{\mu_n}^{\diamond}}_{\mathcal{H}} \\
  & {} \leq {}
    && \frac{\norm{ \tilde{\mathcal{A}}_{\mu_n}}\,
       \norm{\breve{Q}_{\mu_n}^{\diamond}}_{\mathcal{H}} +
       \norm{ \tilde{\mathcal{B}}_{\mu_n}}\,
       \norm{g}_{\mathcal{H}}}{\sigma_{\min} ( (\alpha
       \Sigma_{ s^{\prime} \given z }^{\mu_n} - \Id)^*
       (\alpha \Sigma_{ s^{\prime} \given z }^{\mu_n} - \Id)
       )\, \sigma_{\min} ( \Sigma_{zz} )} \\
  & \leq
    && \frac{ \left[ 2\alpha \sqrt{B_{\kappa}}
       ( \beta_{\infty} + 1 ) + \alpha^2
       (\mathfrak{m}_{\xi}^{(2)})^{1/2} \right] B_{\breve{q}}
       }{\sigma_{\min} ( (\alpha \Sigma_{ s^{\prime} \given
       z }^{\mu_n} - \Id)^* (\alpha \Sigma_{ s^{\prime}
       \given z }^{\mu_n} - \Id))\,
       \sigma_{\min} ( \Sigma_{zz} )}
       (\mathfrak{m}_{\xi}^{(2)})^{1/2} \\
  &&& + \frac{ \alpha \sqrt{B_{\kappa}}
      \norm{g}_{\mathcal{H}} }{\sigma_{\min} ( (\alpha
      \Sigma_{ s^{\prime} \given z }^{\mu_n} - \Id)^*
      (\alpha
      \Sigma_{ s^{\prime} \given z }^{\mu_n} - \Id) )\,
      \sigma_{\min} ( \Sigma_{zz} )}
      (\mathfrak{m}_{\xi}^{(2)})^{1/2} \\
  & \leq && c_4 (\mathfrak{m}_{\xi}^{(2)})^{1/2} \,.
\end{alignat*}
Notice also that
\begin{alignat}{2}
  \mathfrak{m}_{\xi}^{(2)}
  & {} = {}
    && \expect\{ \norm{ \xi_n }_{\mathcal{H}}^2 \} \notag \\
  & =
    && \expect\{ \langle ( \hat{\Sigma}_{ s^{\prime}
       \given z }^{\mu_n} - \Sigma_{ s^{\prime} \given z
       }^{\mu_n} )^* (\varphi(\vect{z}_{n-1})) \notag \\
  &&& \phantom{ \expect\{ } \given ( \hat{\Sigma}_{
      s^{\prime} \given z }^{\mu_n} - \Sigma_{
      s^{\prime} \given z }^{\mu_n} )^*
      (\varphi(\vect{z}_{n-1})) \rangle_{\mathcal{H}} \}
      \notag \\
  & =
    && \expect\{ \langle \varphi(\vect{z}_{n-1}) \notag \\
  &&& \phantom{ \expect\{ }
      \given ( \hat{\Sigma}_{
      s^{\prime} \given z }^{\mu_n} - \Sigma_{
      s^{\prime} \given z }^{\mu_n} ) ( \hat{\Sigma}_{
      s^{\prime} \given z }^{\mu_n} - \Sigma_{
      s^{\prime} \given z }^{\mu_n} )^*
      (\varphi(\vect{z}_{n-1})) \rangle_{\mathcal{H}} \}
      \notag \\
  & \leq
    && \expect\{ \sigma_{\max} ( ( \hat{\Sigma}_{
       s^{\prime} \given z }^{\mu_n} - \Sigma_{
       s^{\prime} \given z }^{\mu_n} ) ( \hat{\Sigma}_{
       s^{\prime} \given z }^{\mu_n} - \Sigma_{
       s^{\prime} \given z }^{\mu_n} )^* ) \notag \\
  &&& \phantom{ \expect\{ } \cdot \kappa( \vect{z}_{n-1},
      \vect{z}_{n-1}) \} \notag \\
  & \leq
    && B_{\kappa} \expect\{ \norm{ ( \hat{\Sigma}_{
       s^{\prime} \given z }^{\mu_n} - \Sigma_{
       s^{\prime} \given z }^{\mu_n} ) ( \hat{\Sigma}_{
       s^{\prime} \given z }^{\mu_n} - \Sigma_{
       s^{\prime} \given z }^{\mu_n} )^* } \} \notag \\
  & =
    && B_{\kappa} \expect\{ \norm{ \hat{\Sigma}_{
       s^{\prime} \given z }^{\mu_n}( N[n] ) - \Sigma_{
       s^{\prime} \given z }^{\mu_n} }^2 \} \,,
\end{alignat}
so that
\begin{alignat}{2}
  &&& \negphantom{ {} = {} } \limsup_{ n\to\infty }
      \expect\{ \norm{Q_{\mu_n}^{\diamond} -
      \breve{Q}_{\mu_n}^{\diamond}}_{\mathcal{H}}^2
      \} \notag \\
  & {} = {}
    && \limsup_{ n\to\infty }\,
       \norm{Q_{\mu_n}^{\diamond} -
       \breve{Q}_{\mu_n}^{\diamond}}_{\mathcal{H}}^2 \notag
  \\
  & \leq
    && c_4^2 \mathfrak{m}_{\xi}^{(2)} \notag \\
  & \leq
    && c_4^2 B_{\kappa} \limsup_{ n\to\infty } \expect\{
       \norm{ \hat{\Sigma}_{ s^{\prime} \given z }^{\mu_n}(
       N[n] ) - \Sigma_{ s^{\prime} \given z }^{\mu_n} }^2
       \} \,. \label{Qdiam.to.Qmin}
\end{alignat}

Combine now~\eqref{Qn.to.Qmin} with~\eqref{Qdiam.to.Qmin} to
obtain
\begin{alignat*}{2}
  &&& \negphantom{ {} \leq {} } \limsup \nolimits_{
      n\to\infty } \, \expect\{ \norm{ Q_n -
      Q_{\mu_n}^{\diamond} }_{\mathcal{H}}^2 \} \notag \\
  & {} \leq {}
  && 2 \limsup \nolimits_{ n\to\infty } \, \expect\{ \norm{
     Q_n - \breve{Q}_{\mu_n}^{\diamond} }_{\mathcal{H}}^2 \}
     \notag \\
  &&& + 2 \limsup \nolimits_{ n\to\infty } \, \expect\{
      \norm{ Q_{\mu_n}^{\diamond} -
      \breve{Q}_{\mu_n}^{\diamond} }_{\mathcal{H}}^2 \}
      \notag \\
  & \leq
  && 2c_3 \eta + 2 c_4^2 B_{\kappa} \limsup_{ n\to\infty }
     \expect\{ \norm{ \hat{\Sigma}_{ s^{\prime} \given z
     }^{\mu_n}( N[n] ) - \Sigma_{ s^{\prime} \given z
     }^{\mu_n} }^2 \} \,,
\intertext{and hence,}
  &&& \negphantom{ {} \leq {} } \limsup \nolimits_{
      n\to\infty } \, \expect\{ \norm{ Q_n -
      Q_*^{\diamond} }_{\mathcal{H}}^2 \} \\
  & {} \leq {}
    && 2 \limsup \nolimits_{
       n\to\infty } \, \expect\{ \norm{ Q_n -
       Q_{\mu_n}^{\diamond} }_{\mathcal{H}}^2 \} \\
  &&& + 2 \limsup \nolimits_{
      n\to\infty } \, \expect\{ \norm{ Q_{\mu_n}^{\diamond}
      - Q_*^{\diamond} }_{\mathcal{H}}^2 \} \\
  & \leq
    && 4c_3 \eta + 4c_4^2 B_{\kappa} \limsup_{n\to \infty}\,
       \expect\{ \norm{ \hat{\Sigma}_{ s^{\prime} \given z
       }^{\mu_n}( N[n] ) - \Sigma_{ s^{\prime} \given z
       }^{\mu_n} }^2 \} \\
  &&& + 2 \limsup \nolimits_{
      n\to\infty } \, \expect\{ \norm{ Q_{\mu_n}^{\diamond}
      - Q_*^{\diamond} }_{\mathcal{H}}^2 \} \,,
\end{alignat*}
which establishes \cref{thm:evaluation} with
$\Delta_4 \coloneqq 4c_3$ and
$\Delta_5 \coloneqq 4 c_4^2 B_{\kappa}$.

\section{Proof of \cref{thm:optimal.regret}}%
\label[appendix]{app:optimal.regret}

\begin{lemma}\label{proof-concave-inequality}
  There exist $c_1 \in \RealPP$ and $c_2\in \RealP$ s.t.\
  the following inequalities hold true for all
  $m \in \{ n - M_{\textnormal{av}} + 1, \ldots, n \}$ and
  for all sufficiently large $n$, a.s.,
  \begin{align*}
    c_1 (y_m - \vectgr{\theta}_{n+1}^{\intercal} \vect{x}_m )^2 +
    c_2
    & \leq \log\, \lvert y_m -
      \vectgr{\theta}_{n+1}^{\intercal} \vect{x}_m
      \rvert^2 \\
    & \leq 1 + ( y_m -
      \vectgr{\theta}_{n+1}^{\intercal} \vect{x}_m )^2 \,.
  \end{align*}
\end{lemma}

\begin{IEEEproof}
  By the concavity of $\log (\cdot)$,
  $\log \varpi \leq 1 + \varpi$,
  $\forall \varpi\in \RealPP$. Hence,
  $\log\, \lvert y_m -
  \vectgr{\theta}_{n+1}^{\intercal} \vect{x}_m
  \rvert^2 \leq 1 + (y_m -
  \vectgr{\theta}_{n+1}^{\intercal}
  \vect{x}_m)^2$. The concavity of $\log (\cdot)$
  suggests also that
  $\forall \varpi\in ( \Delta_6^2, \Delta_7^2 )$,
  \begin{align*}
    \log \varpi
    & \geq \frac{ \log \Delta_7^2 - \log \Delta_6^2 } {
      \Delta_7^2 - \Delta_6^2 } (\varpi -
      \Delta_6^2 ) + \log \Delta_6^2 =
      c_1 \varpi + c_2 \,,
  \end{align*}
  where
  $c_1 \coloneqq ( \log \Delta_7^2 - \log \Delta_6^2 ) / (
  \Delta_7^2 - \Delta_6^2 )$ and
  $c_2 \coloneqq \log \Delta_6^2 - \Delta_6^2 ( \log
  \Delta_7^2 - \log \Delta_6^2 ) / ( \Delta_7^2 - \Delta_6^2
  ) $. Substituting $\varpi$ in the previous inequality by
  $\lvert y_m - \vectgr{\theta}_{n+1}^{\intercal} \vect{x}_m
  \rvert^2$ establishes \cref{proof-concave-inequality}.
\end{IEEEproof}

Recall now the data model in \cref{sec:RobAdaFilt} to verify
that
$y_m - \vectgr{\theta}_{n+1}^{\intercal} \vect{x}_m =
\vectgr{\theta}_*^{\intercal} \vect{x}_m + o_m -
\vectgr{\theta}_{n+1}^{\intercal} \vect{x}_m = (
\vectgr{\theta}_* - \vectgr{\theta}_{n+1} )^{\intercal}
\vect{x}_m + o_m$. Moreover,
recall~\eqref{def.sn} and~\eqref{one.step.loss.value} to
verify that the chosen one-step loss satisfies
\begin{align*}
  g( \vect{s}_n, \mu (\vect{s}_n) )
  & = g( \vect{s}_n, a_n ) = g( \vect{z}_n )\\
  & = \tfrac{1}{M_{\textnormal{av}}} \sum_{m = n -
    M_{\textnormal{av}} + 1}^{n}
    \log \frac{\lvert y_m -
    \vectgr{\theta}_{n+1}^{\intercal} \vect{x}_m \rvert^2}
    {\norm{\vect{x}_m }_2^2} \,.
\end{align*}
Observe then via \cref{proof-concave-inequality} that
\begin{subequations}\label{g.bounds}
  \begin{alignat}{2}
    &&& \negphantom{ M_{\textnormal{av}} }
        M_{\textnormal{av}} + \sum \nolimits_{m = n -
        M_{\textnormal{av}} + 1}^{n} ( y_m -
        \vectgr{\theta}_{n+1}^{\intercal} \vect{x}_m )^2
        \notag \\
    &&& - \sum \nolimits_{m = n - M_{\textnormal{av}} +
        1}^{n} \log\, \norm{\vect{x}_m
        }_2^2 \label{g.upper.bound} \\
    & {} \geq {}
      && \sum \nolimits_{m = n - M_{\textnormal{av}} + 1}^{n}
         \log\, \lvert y_m -
         \vectgr{\theta}_{n+1}^{\intercal} \vect{x}_m
         \rvert^2 \notag \\
    &&& - \sum \nolimits_{m = n - M_{\textnormal{av}} +
        1}^{n} \log\, \norm{\vect{x}_m }_2^2 \notag \\
    & =
      && \sum \nolimits_{m = n - M_{\textnormal{av}} +
         1}^{n}
         \log \frac{\lvert y_m -
         \vectgr{\theta}_{n+1}^{\intercal} \vect{x}_m \rvert
         ^2} {\norm{\vect{x}_m }_2^2} \notag \\
    & =
      && M_{\textnormal{av}} g( \vect{s}_n,
         \mu (\vect{s}_n) ) \label{g.in.between} \\
    & \geq
      && c_1 \sum
         \nolimits_{m = n - M_{\textnormal{av}} +
         1}^{n} [ (\vectgr{\theta}_* -
         \vectgr{\theta}_{n+1} )^{\intercal} \vect{x}_m +
         o_m ]^2 \notag \\
    &&& + M_{\textnormal{av}} c_2 - \sum
        \nolimits_{m = n - M_{\textnormal{av}} +
        1}^{n} \log\, \norm{\vect{x}_m}_2^2 \notag \\
    & =
      && c_1 \sum
         \nolimits_{m = n - M_{\textnormal{av}} + 1}^{n}
         (\vectgr{\theta}_* -
         \vectgr{\theta}_{n+1} )^{\intercal}
         \vect{x}_m \vect{x}_m^{\intercal}
         (\vectgr{\theta}_* - \vectgr{\theta}_{n+1} ) \notag
    \\
    &&& + 2c_1 \sum
        \nolimits_{m = n - M_{\textnormal{av}} + 1}^{n}
        (\vectgr{\theta}_* -
        \vectgr{\theta}_{n+1} )^{\intercal}
        \vect{x}_m o_m \notag \\
    &&& + c_1 \sum
        \nolimits_{m = n - M_{\textnormal{av}} + 1}^{n}
        o_m^2 + M_{\textnormal{av}} c_2 \notag \\
    &&& - \sum
        \nolimits_{m = n - M_{\textnormal{av}} + 1}^{n}
        \log\, \norm{\vect{x}_m}_2^2 \label{g.lower.bound} \,.
  \end{alignat}
\end{subequations}
Notice also by
\Cref{ass:independency.signals,ass:stationary.x} that there
exists $c_3 \in \Real$ s.t.\
$\expect_{ \given \vectgr{\theta}_{n+1} } \{ \log\,
\norm{\vect{x}_m }_2^2 \} = \expect \{ \log\,
\norm{\vect{x}_m }_2^2 \} \geq c_3$, for all sufficiently
large $m$, a.s. Observe also that
$\expect\{ \norm{ \vect{x}_m }_2^2 \} = \expect\{ \trace
(\vect{x}_m \vect{x}_m^{\intercal}) \} = \trace ( \expect\{
\vect{x}_m \vect{x}_m^{\intercal} \} ) = \trace (
\Sigma_{xx} )$. Hence, by \eqref{g.bounds}, a.s.,
\begin{alignat*}{2}
  &&& \negphantom{ M_{\textnormal{av}} } M_{\textnormal{av}}
      + \sum \nolimits_{m = n - M_{\textnormal{av}} + 1}^{n}
      \expect_{ \given \vectgr{\theta}_{n+1} } \{ ( y_m -
      \vectgr{\theta}_{n+1}^{\intercal} \vect{x}_m )^2
      \} \\
  &&& - M_{\textnormal{av}} c_3 \\
  & {} \geq {}
    && M_{\textnormal{av}}
       + \sum \nolimits_{m = n - M_{\textnormal{av}} + 1}^{n}
       \expect_{ \given \vectgr{\theta}_{n+1} } \{ ( y_m -
       \vectgr{\theta}_{n+1}^{\intercal} \vect{x}_m )^2
       \} \\
  &&& - \sum \nolimits_{m = n - M_{\textnormal{av}} +
      1}^{n} \expect_{ \given \vectgr{\theta}_{n+1} }
      \{ \log\, \norm{\vect{x}_m }_2^2 \} \\
  & \geq
    && M_{\textnormal{av}} \expect_{ \given
       \vectgr{\theta}_{n+1} } \{ g( \vect{s}_n,
       \mu (\vect{s}_n) ) \} \\
  & \geq
    && c_1 \sum
       \nolimits_{m = n - M_{\textnormal{av}} + 1}^{n}
       (\vectgr{\theta}_* -
       \vectgr{\theta}_{n+1} )^{\intercal}
       \expect_{ \given \vectgr{\theta}_{n+1} } \{
       \vect{x}_m \vect{x}_m^{\intercal} \}
       (\vectgr{\theta}_* - \vectgr{\theta}_{n+1} ) \\
  &&& + 2c_1 \sum
      \nolimits_{m = n - M_{\textnormal{av}} + 1}^{n}
      (\vectgr{\theta}_* -
      \vectgr{\theta}_{n+1} )^{\intercal}
      \expect_{ \given \vectgr{\theta}_{n+1} } \{
      \vect{x}_m o_m \} \\
  &&& + c_1 \sum
      \nolimits_{m = n - M_{\textnormal{av}} + 1}^{n}
      \expect_{ \given \vectgr{\theta}_{n+1} } \{ o_m^2 \}
      + M_{\textnormal{av}} c_2 \\
  &&& - \sum
      \nolimits_{m = n - M_{\textnormal{av}} + 1}^{n}
      \expect_{ \given \vectgr{\theta}_{n+1} } \{ \log\,
      \norm{\vect{x}_m}_2^2 \} \\
  & \geq
    && c_1 \sum
       \nolimits_{m = n - M_{\textnormal{av}} + 1}^{n}
       (\vectgr{\theta}_* -
       \vectgr{\theta}_{n+1} )^{\intercal}
       \expect\{
       \vect{x}_m \vect{x}_m^{\intercal} \}
       (\vectgr{\theta}_* - \vectgr{\theta}_{n+1} ) \\
  &&& + 2c_1 \sum
      \nolimits_{m = n - M_{\textnormal{av}} + 1}^{n}
      (\vectgr{\theta}_* -
      \vectgr{\theta}_{n+1} )^{\intercal}
      \expect \{ \vect{x}_m \} \expect \{ o_m \} \\
  &&& + c_1 \sum
      \nolimits_{m = n - M_{\textnormal{av}} + 1}^{n}
      \expect \{ o_m^2 \}
      + M_{\textnormal{av}} c_2 \\
  &&& - \sum
      \nolimits_{m = n - M_{\textnormal{av}} + 1}^{n}
      \log \expect_{ \given \vectgr{\theta}_{n+1} } \{
      \norm{\vect{x}_m}_2^2 \} \\
  & \geq
    && c_1 \lambda_{\min} (\Sigma_{xx})
       \sum\nolimits_{m = n - M_{\textnormal{av}} + 1}^{n}
       \norm{ \vectgr{\theta}_* - \vectgr{\theta}_{n+1}
       }_2^2 \\
  &&& + c_1 M_{\textnormal{av}} \sigma_o^2 +
      M_{\textnormal{av}} c_2 \\
  &&& - \sum \nolimits_{m = n - M_{\textnormal{av}} +
      1}^{n} \log \expect \{ \norm{\vect{x}_m}_2^2 \} \\
  & =
    && c_1 M_{\textnormal{av}} \lambda_{\min}
       (\Sigma_{xx}) \norm{ \vectgr{\theta}_* -
       \vectgr{\theta}_{n+1} }_2^2
       + c_1 M_{\textnormal{av}} \sigma_o^2 \\
  &&& + M_{\textnormal{av}} c_2 -
      M_{\textnormal{av}} \log \trace( \Sigma_{xx} ) \,,
\end{alignat*}
which yield in turn
\begin{alignat*}{2}
  &&& \negphantom{ {} \geq {} } M_{\textnormal{av}}( 1 - c_3
      ) + M_{\textnormal{av}} \Delta_7^2 \\
  & {} \geq {}
    && M_{\textnormal{av}} + \sum
       \nolimits_{m = n - M_{\textnormal{av}} + 1}^{n}
       \expect \{ ( y_m -
       \vectgr{\theta}_{n+1}^{\intercal} \vect{x}_m )^2
       \} \\
  & \geq
    && M_{\textnormal{av}} \expect \{ g( \vect{s}_n,
       \mu (\vect{s}_n) ) \} \\
  & \geq
    && c_1 M_{\textnormal{av}} \lambda_{\min}
       (\Sigma_{xx}) \expect\{ \norm{ \vectgr{\theta}_* -
       \vectgr{\theta}_{n+1} }_2^2 \}
       + c_1 M_{\textnormal{av}} \sigma_o^2 \\
  &&& + M_{\textnormal{av}} c_2 -
      M_{\textnormal{av}} \log \trace( \Sigma_{xx} ) \,,
\end{alignat*}
or equivalently,
\begin{subequations}
  \begin{alignat}{2}
    &&& \negphantom{ {} \geq {} } 1 - c_3 +
        \Delta_7^2 \label{expect.g.upper} \\
    & {} \geq {}
      && \expect \{ g( \vect{s}_n,
         \mu (\vect{s}_n) ) \} \notag \\
    & \geq
      && c_1 \lambda_{\min}
         (\Sigma_{xx}) \expect\{ \norm{ \vectgr{\theta}_* -
         \vectgr{\theta}_{n+1} }_2^2 \}
         + c_1 \sigma_o^2 \notag \\
    &&& + c_2 - \log \trace( \Sigma_{xx} )
        \,. \label{expect.g.lower}
  \end{alignat}
\end{subequations}

Recall now that
$Q_{\mu}^{\diamond} = T_{\mu}^{\diamond} (
Q_{\mu}^{\diamond} )$. Hence,
by~\eqref{classical.Bellman.mu}, $\forall n\geq n_0$,
\begin{align*}
  & Q_{\mu}^{\diamond} ( \vect{s}_n, \mu( \vect{s}_n ) ) \\
  & = T_{\mu}^{\diamond} ( Q_{\mu}^{\diamond} ) (
    \vect{s}_n, \mu( \vect{s}_n ) ) \\
  & = g( \vect{s}_n, \mu( \vect{s}_n ) ) + \alpha \expect_{
    \vect{s}_{n+1} \given \vect{s}_n} \{
    Q_{\mu}^{\diamond} ( \vect{s}_{n+1}, \mu( \vect{s}_{n+1}
    ) ) \} \,.
\end{align*}
It can be directly verified by this last recursion and
induction that for any $K\in \IntegerPP$, a.s.,
\begin{alignat*}{2}
  &&& \negphantom{ {} = {} } Q_{\mu}^{\diamond} (
      \vect{s}_{n_0}, \mu( \vect{s}_{n_0} ) ) \\
  & {} = {}
    && g( \vect{s}_{n_0}, \mu( \vect{s}_{n_0} ) )
       + \sum\nolimits_{ \nu = n_0 + 1 }^{n_0 + K - 1}
       \alpha^{\nu - n_0} \expect_{
       \vect{s}_{\nu} \given \vect{s}_{n_0} }  \{
       g ( \vect{s}_{\nu}, \mu( \vect{s}_{\nu} ) ) \} \\
  &&& + \alpha^K \expect_{
      \vect{s}_{n_0 + K} \given
      \vect{s}_{n_0} } \{ Q_{\mu}^{\diamond} (
      \vect{s}_{n_0+K}, \mu(
      \vect{s}_{n_0+K} ) ) \} \,,
\end{alignat*}
and hence,
\begin{alignat}{2}
  \expect\{ Q_{\mu}^{\diamond} (
  \vect{s}_{n_0}, \mu( \vect{s}_{n_0} ) ) \}
  & {} = {}
  && \sum\nolimits_{ \nu = n_0  }^{n_0 + K - 1}
     \alpha^{\nu - n_0} \expect \{ g ( \vect{s}_{\nu}, \mu(
     \vect{s}_{\nu} ) ) \} \notag \\
  &&& + \alpha^K \expect \{ Q_{\mu}^{\diamond} (
      \vect{s}_{n_0+K}, \mu( \vect{s}_{n_0+K} ) ) \}
      \,. \label{Qmu.is.sum.g.K}
\end{alignat}

By~\eqref{expect.g.upper}, $\forall K\in \IntegerPP$,
\begin{alignat*}{2}
  &&& \negphantom{ {} \leq {} } \sum\nolimits_{ \nu = n_0
      }^{n_0 + K - 1} \alpha^{\nu - n_0} \expect \{ g (
      \vect{s}_{\nu}, \mu( \vect{s}_{\nu} ) ) \} \\
  & {} \leq {}
    && ( 1 - c_3 + \Delta_7^2 ) \sum\nolimits_{ \nu = n_0
       }^{ n_0 + K - 1 } \alpha^{\nu - n_0} \\
  & \leq
    && ( 1 - c_3 + \Delta_7^2 ) \sum\nolimits_{ \nu = n_0
       }^{ +\infty } \alpha^{\nu - n_0} =
       \frac{ 1 }{1 - \alpha} ( 1 - c_3 + \Delta_7^2 ) \,,
\end{alignat*}
so that
$\sum_{ \nu = n_0 }^{+\infty} \alpha^{\nu - n_0} \expect \{
g ( \vect{s}_{\nu}, \mu( \vect{s}_{\nu} ) ) \} < +
\infty$. Thus, by applying $\limsup_{ K \to \infty}$
to~\eqref{Qmu.is.sum.g.K} and by recalling
\cref{ass:expect.Qmu.upper.bound},
\begin{align*}
  \expect\{ Q_{\mu}^{\diamond} (
  \vect{s}_{n_0}, \mu( \vect{s}_{n_0} ) ) \} =
  \sum\nolimits_{ n = n_0 }^{+\infty} \alpha^{n - n_0}
  \expect \{ g ( \vect{s}_{n}, \mu( \vect{s}_{n} ) ) \}
  \,.
\end{align*}
Hence, by~\eqref{expect.g.lower},
\begin{alignat}{2}
  &&& \negphantom{ {} \geq {} }
      \expect \{ Q_{\mu}^{\diamond}( \vect{s}_{n_0},
      \mu(\vect{s}_{n_0}) ) \} {} = {}
      \sum \nolimits_{n = n_0}^{+\infty} \alpha^{n - n_0}
      \expect\{ g( \vect{s}_n, \mu (\vect{s}_n) ) \} \notag
  \\
  & {} \geq {}
    && c_1 \lambda_{\min}
       (\Sigma_{xx})  \sum
       \nolimits_{n = n_0}^{+\infty} \alpha^{n - n_0}
       \expect\{ \norm{ \vectgr{\theta}_* -
       \vectgr{\theta}_{n+1} }_2^2 \} \notag \\
  &&& + \sum \nolimits_{n = n_0}^{+\infty} \alpha^{n - n_0}
      [ c_1 \sigma_o^2 + c_2 - \log \trace( \Sigma_{xx} ) ]
      \notag \\
  & =
    && c_1 \lambda_{\min}
       (\Sigma_{xx})  \frac{1}{\alpha^{n_0+1}} \sum
       \nolimits_{n = n_0 + 1}^{+\infty} \alpha^n
       \expect\{ \norm{ \vectgr{\theta}_* -
       \vectgr{\theta}_{n} }_2^2 \} \notag \\
  &&& + \frac{1}{1 - \alpha} [ c_1 \sigma_o^2 + c_2 - \log
      \trace( \Sigma_{xx} ) ] \,, \notag
      \intertext{which yields}
  &&& \negphantom{ {} \leq {} }
      \sum
      \nolimits_{n= n_0 + 1}^{\infty} \alpha^n \expect\{
      \norm{ \vectgr{\theta}_* - \vectgr{\theta}_{n} (
      \mu( \vect{s}_{n-1} ) ) }_2^2 \} \notag \\
  & {} \leq {}
    && \frac{\alpha^{n_0+1}}{c_1 \lambda_{\min}
       (\Sigma_{xx})} \expect \{
       Q_{\mu}^{\diamond}( \vect{s}_{n_0},
       \mu(\vect{s}_{n_0}) ) \} \notag \\
  &&& + \frac{1}{c_1 \lambda_{\min}
      (\Sigma_{xx})} \cdot \frac{\alpha^{n_0+1} }{ 1 - \alpha } [
      \log \trace ( \Sigma_{xx} )  - c_1 \sigma_o^2 - c_2 ]
      \,. \notag 
\end{alignat}
Observe that the way to update $\vectgr{\theta}_n$,
$\forall n\geq n_0$, was not specified throughout the
previous analysis. As such, set
$\vectgr{\theta}_n \coloneqq \vectgr{\theta}_{n_0}$,
$\forall n\geq n_0$, $\Delta_8 \coloneqq c_1$,
$\Delta_9 \coloneqq c_2$ in the last inequality, and finally
substitute $n_0$ by $n$ to establish
\cref{thm:optimal.regret}.

\end{document}